\documentclass[10pt,journal,compsoc]{IEEEtran}
\ifCLASSOPTIONcompsoc
  \usepackage[nocompress]{cite}
\else
  \usepackage{cite}
\fi
\ifCLASSINFOpdf
\else
\fi
\usepackage{graphicx}
\usepackage{amsmath,amssymb,amsfonts,marvosym}
\usepackage{array}
\usepackage{multirow}
\usepackage{subcaption} % TODO EZ: I used subcaption because subfig is incompatible with hyperref. Can we do this?
\newcommand\MYhyperrefoptions{bookmarks=true,bookmarksnumbered=true,
pdfpagemode={UseOutlines},plainpages=false,pdfpagelabels=true,
colorlinks=true,linkcolor={black},citecolor={black},urlcolor={black},
pdftitle={FDGram Draft for TSE},%<!CHANGE!
pdfsubject={Typesetting},%<!CHANGE!
pdfauthor={Lukas Kirschner},%<!CHANGE!
pdfkeywords={Computer Society, IEEEtran, journal, LaTeX, paper,
             template}}%<^!CHANGE!
\ifCLASSINFOpdf
\usepackage[\MYhyperrefoptions,pdftex]{hyperref}
\else
\usepackage[\MYhyperrefoptions,breaklinks=true,dvips]{hyperref}
\usepackage{breakurl}
\fi
\usepackage{algorithmic}
\usepackage{textcomp}
\usepackage[usenames,dvipsnames]{xcolor}
\usepackage{listings}
\usepackage{xspace}
\usepackage{multirow}
\usepackage{booktabs}
\usepackage{colortbl}
\usepackage{tikz}
\usepackage{pgfplots}
\usepackage{pgfplotstable}
\usepackage[linguistics]{forest}
\usepackage{balance}
\usepackage{framed}
\usepackage{ragged2e}
 
\makeatletter
\def\endthebibliography{%
  \def\@noitemerr{\@latex@warning{Empty `thebibliography' environment}}%
  \endlist
}
\makeatother

\newenvironment{result}{\begin{framed}\centering\it}{\end{framed}}

\newcommand{\revise}[1]{\textcolor{black}{#1}}
\newcommand{\rev}[1]{\textcolor{black}{#1}}
\newcommand{\lrevise}[1]{\textcolor{black}{#1}}

\newcommand{\approach}{\textsc{FdLoop}\xspace} % DirGram DGFuzz DirGram(Fuzz) FdLoop FeedGram(m) %LoopGram %LoopGen %FDGen %Fd %FdGram
\newcommand{\evogfuzz}{EvoGFuzz\xspace}
\newcommand{\dynamosa}{DynaMOSA\xspace}
\newcommand{\evosuite}{EvoSuite\xspace}
\makeatletter

\newcommand\TODOcaption[1]{%
    \edef\todo@caption{#1}}
\newcommand\TODOtitle[1]{%
    \edef\todo@title{#1}}
\makeatother
\usetikzlibrary{calc,positioning,arrows,matrix,fit,shapes,decorations.pathmorphing,snakes,tikzmark}
\pgfdeclarelayer{background}
\pgfsetlayers{background,main}

\def\BibTeX{{\rm B\kern-.05em{\sc i\kern-.025em b}\kern-.08em
    T\kern-.1667em\lower.7ex\hbox{E}\kern-.125emX}}

\lstnewenvironment{cfglisting}{\lstset{basicstyle=\ttfamily\scriptsize,numbers=none,mathescape=true,basewidth=0.5em,escapechar=!}}{}

\definecolor{kulerA}{RGB}{243,243,243}%BOX BACKGROUNDS (modified)
\definecolor{kulerB}{RGB}{143,141,141}
\definecolor{kulerC}{RGB}{8,175,238}
\definecolor{kulerD}{RGB}{50,150,50}%HIGHLIGHTED TEXT
\definecolor{kulerE}{RGB}{14,13,14}%HEADLINES

\def\rqone{\textbf{RQ1 Effectiveness:} How effective is our approach (\approach) in targeting specific testing goal(s), i.e., a single testing goal and multiple testing goals? How does \approach's generated test suite compare to the initial seed inputs in achieving the targeted goals(s)?
}
\def\rqtwo{\textbf{RQ2 Comparison to the state-of-the-art:} How does our approach (\approach) compare to 
state-of-the-art methods in achieving a targeted testing goal? How effective is \approach in comparison to the following baselines -- %:namely the following 
(a) %the initial seed inputs , (b) 
typical grammar-based test generators (i.e., random, probabilistic, and inverted baselines), \revise{ (b) 
a state-of-the-art (SOTA) evolutionary grammar-based baseline (i.e.,  EvoGFuzz), 
and (c) 
a SOTA non-grammar-based evolutionary baseline (i.e.,  \dynamosa)?}
}

\def\rqthree{\textbf{RQ3 Ablation Study:} What is the contribution of each test feedback in achieving specific testing goal(s)? What is the contribution of our design choices (i.e., input mutation and grammar mutation) to the effectiveness of \approach ?}
\def\rqfour{\textbf{RQ4 Sensitivity Analysis:} How sensitive is our approach to the following parameter settings -- the number of initial seeds, the number of generations,  the number of generated inputs,  \revise{and varying  random seed values?}}
\newlength\boxw
\newlength\boxh
\newlength\nsep
\newlength\dmarg

\newcommand\bdcaptionbot[3][]{\node[bdcaptionbot] at (#2.south) (#2-caption) {#3};}
\makeatletter
\newcommand\multiconnect@nodes[3]{\edef\mc@c{#1}%
\draw[draw=none] (#2) -- %
    node[shape=coordinate,pos=.4] (\mc@c 21) {}%
    node[shape=coordinate,pos=.6] (\mc@c 22) {}%
    node[shape=coordinate,pos=.3] (\mc@c 31) {}%
    node[shape=coordinate,pos=.5] (\mc@c 32) {}%
    node[shape=coordinate,pos=.7] (\mc@c 33) {}%
    node[shape=coordinate,pos=.2] (\mc@c 41) {}%
    node[shape=coordinate,pos=.4] (\mc@c 42) {}%
    node[shape=coordinate,pos=.6] (\mc@c 43) {}%
    node[shape=coordinate,pos=.8] (\mc@c 44) {}%
(#3);%
}
\newcommand\multiconnectnodesouth[1]{\multiconnect@nodes{#1-south}{#1.south west}{#1.south east}}
\newcommand\multiconnectnodenorth[1]{\multiconnect@nodes{#1-north}{#1.north west}{#1.north east}}
\newcommand\multiconnectnodeeast[1]{\multiconnect@nodes{#1-east}{#1.north east}{#1.south east}}
\newcommand\multiconnectnodewest[1]{\multiconnect@nodes{#1-west}{#1.north west}{#1.south west}}
 
\makeatother

\tikzset{%
blockdiagramlines/.style={draw,stroke=black,line width=1.2pt},
blockdiagramarrow/.style={blockdiagramlines,->},
blockdiagramdashedarrow/.style={blockdiagramarrow,dashed},
blockdiagramfont/.style={font=\scriptsize},
blockdiagramannot/.style={blockdiagramlines,text=black,fill=white,align=center,blockdiagramfont},
blockdiagramblock/.style={blockdiagramannot,minimum width=1.8cm,minimum height=1.3cm,rounded corners=2pt},
blockdiagrammicroblock/.style={blockdiagramannot,font=\tiny,minimum width=1.5cm,minimum height=.5cm,rounded corners=1pt},
blockdiagramboxa/.style={draw=green!50!black,dashed,line width=1pt},
blockdiagramcaption/.style={blockdiagramfont,inner sep=1.5pt,align=center},
blockdiagramcaptiona/.style={blockdiagramcaption,text=green!50!black,font=\scriptsize\sffamily},
blockdiagramboxb/.style={blockdiagramboxa,draw=blue!50!black},
blockdiagramcaptionb/.style={blockdiagramcaptiona,text=blue!50!black},
blockdiagramboxc/.style={blockdiagramboxa,draw=red!50!black},
blockdiagramcaptionc/.style={blockdiagramcaptiona,text=red!50!black},
blockdiagramarrowcaption/.style={blockdiagramcaptiona,text=black},
pics/numbering/.style args={#1}{%
    code={
        \node[draw=black,shape=circle,fill=white,text=black,font=\normalfont\scriptsize,inner sep=1pt,text width=8pt,text height=6pt,align=center,outer sep=0] (-number) {#1};
    }
},
microtree/.style={every node/.style={font=\tiny,draw=none,inner sep=2pt,outer sep=0},level 1/.style={sibling distance={6mm}},level 2/.style={sibling distance=3mm},sibling distance=4mm,level distance=4mm},
bdinnertext/.style={font=\footnotesize\sffamily\linespread{0.8}\selectfont,align=center},
bdlinewidth/.style={line width=1pt,draw=black},
bdline/.style={bdlinewidth,rounded corners=1pt},
bdcaption/.style={bdinnertext,font=\scriptsize\sffamily\linespread{0.8}\selectfont,anchor=south,text width=1.3\boxw},
bdcaptionbot/.style={bdcaption,anchor=north},
bdblock/.style={bdinnertext,bdline,minimum width=\boxw,minimum height=\boxh,text width=\boxw,inner sep=1mm,outer sep=0},
bdslimblock/.style={bdblock,minimum height=.45cm,font=\tiny\sffamily},
bdarrow/.style={bdline,-latex},
bddashedarrow/.style={bdarrow,dashed,draw=gray},
bddashedbox/.style={bdline,dashed,opacity=0.4,draw=#1!60!white,fill=#1!30!white},
bddashedcaption/.style={bdinnertext,opacity=0.8,text=#1!90!white},
bdcode/.style={bdinnertext,font=\ttfamily\linespread{0.8}\selectfont}
}

\newcommand\annotprob[2][kulerD]{{\tikz[baseline=(annotprob.base)]{\node[font=\tiny\sffamily\itshape,text=#1,draw=#1,rectangle,inner sep=0.5mm,outer sep=1mm](annotprob){#2};}\hspace{.5mm}}} % Annotations for CFG

\begin{document}
\title{
%Feedback-driven 
Directed 
Grammar-Based Test Generation
%(\approach)
%Conference Paper Title*\\
}

%TODO Change authors from example!
\author{Lukas~Kirschner %~\IEEEmembership{Member,~IEEE,}
        and Ezekiel~Soremekun%~\IEEEmembership{Member,~IEEE,}
%        and~Andreas~Zeller,%~\IEEEmembership{Member,~IEEE}% <-this % stops a space
\IEEEcompsocitemizethanks{\IEEEcompsocthanksitem 
L. Kirschner is with %CISPA Helmholtz Center for Information Security 
%\revise{
Saarland University and University of Luxembourg. %}
%\todo{no affiliation?}
%M. Shell was with the Department
%of Electrical and Computer Engineering, Georgia Institute of Technology, Atlanta,
%GA, 30332.
\protect\\
E-mail: kirschlu@googlemail.com
%see http://www.michaelshell.org/contact.html
\IEEEcompsocthanksitem E. Soremekun is with 
%Royal Holloway,  University of London. 
\rev{Singapore University of Technology and Design
\protect\\
E-mail: ezekiel\_soremekun@sutd.edu.sg
}}% <-this % stops a space
\thanks{Manuscript received April XXX, 202X; revised August XX, 202X.}}


%TODO Change paper headers!
\markboth{
IEEE Transactions on Software Engineering, %(TSE), 
%Journal of \LaTeX\ Class Files,
~Vol.~XX, No.~XX, August~202X
}%
{
Kirschner \MakeLowercase{\textit{et al.}}: FDGRAM: Feedback-driven Grammar-Based Test Generation for IEEE Transactions on Software Engineering (TSE)
%Shell \MakeLowercase{\textit{et al.}}: Bare Advanced Demo of IEEEtran.cls for IEEE Computer Society Journals
}



\IEEEtitleabstractindextext{%
\justify{\begin{abstract} 
\textbf{Context}: 
To effectively test complex software, 
it is important to generate 
\textit{goal-specific inputs}, i.e., inputs that 
%contain specific input structures that 
achieve a specific testing goal.  
For instance, developers may intend to target one or more testing goal(s) during testing -- generate complex inputs or trigger new or error-prone behaviors. 
\textbf{Problem:} However, most state-of-the-art test generators %but 
are not designed to 
\textit{%can not 
target specific goals. } 
Notably,  grammar-based test generators, which  
%ion techniques generate 
(randomly) produce \textit{syntactically valid inputs} via an 
%by employing the 
input specification (i.e., grammar)
%, 
% as a producer. 
%typically,
% these techniques 
%generate valid %test 
%inputs which 
have a low probability of 
%may or may \textit{not} 
achieving an arbitrary testing goal. 
%-at-hand. %a specific testing goal . 
%However, such 
%These
%Even though d
%the goal-at-hand.} 
% (e.g., program failure)}
%, complex inputs or high execution time)}. 
%\par
%\textit{
%To achieve specific testing goal(s), i
%However,  
\textbf{Aim}: %} 
This work addresses this challenge by proposing an  
%we propose an 
automated test generation 
approach (called \approach) 
which iteratively learns relevant input properties from existing inputs 
%and evolves 
%from existing inputs 
to 
%evolve and 
drive the generation of goal-specific inputs. 
%\par
\textbf{Method}: 
The main idea of our approach is to leverage \textit{test feedback} to generate \textit{goal-specific inputs} via 
a combination of \textit{evolutionary testing} and \textit{grammar learning}. 
\approach automatically learns a mapping between input structures and a specific testing goal, such mappings allow to generate inputs that target the goal-at-hand. 
Given a testing goal, \approach iteratively selects, evolves and learn the input distribution of goal-specific test inputs via test feedback and a probabilistic grammar. 
%\par 
%In our evaluation 
We concretize \approach for 
%demonstrate that \approach effectively achieves each of our 
%the following four 
four testing goals, namely 
unique code coverage, input-to-code complexity, program failures (exceptions) and long execution time.  
We evaluate \approach using three (3) well-known input formats (JSON, CSS and JavaScript) and 20 
%large, 
open-source software. 
\textbf{Results}: 
%Our evaluation results show that \approach is effective in achieving each testing goal and multiple goals. 
%achieves each testing goal in few (\todo{How do we define ``achieving'' a testing goal? Our average coverage converged at about 10 generations}) generations. 
%\revise{and scales to multiple testing goals}. 
%We found that 
\revise{
In most (86\%) settings, 
\approach outperforms all five tested baselines namely the baseline grammar-based test generators (random,  probabilistic and inverse-probabilistic methods),  \evogfuzz and \dynamosa. 
% in most (86\%) settings.  
\approach is (up to) 
twice (2X) as effective as the best baseline (\evogfuzz) in inducing erroneous behaviors. 
%up to 
%\todo{89\%} more effective 
%and quicker 
%than the baseline grammar-based test generators (i.e., 
%initial seed inputs, 
%random, probabilistic and inverse-probabilistic methods). 
%. 
%We also show that \approach 
%and i
%It outperforms the closest state-of-the-art 
%evolutionary testing 
%approaches,  \evogfuzz and \dynamosa, by up to \todo{77\%} and \todo{X\%}, respectively.  
}
%Further results show that the inputs generated by \approach are \revise{similar to human-written inputs and outperform the initial seed inputs}. 
In addition, %\revise{
%we conduct an \textit{ablation study} to evaluate the contribution of our design choices. We 
we show that the main components of \approach (i.e., \revise{input mutator, grammar mutator and test feedbacks}) contribute positively to its effectiveness.
% of our approach. 
%Generally, %we demonstrate that 
We also observed that \approach is effective across varying parameter settings -- the number of initial seed inputs, the number of generated inputs,  \revise{the number of input generations and varying random seed values}.
%sets and sizes of seed inputs, and
%Besides, it % reveals \checknumber{X} 
%new bugs and \checknumber{X} existing bugs in \checknumber{X} software systems. %}
%generated \checknumber{X} files that reveal \checknumber{X} bugs in \checknumber{X} software systems.
%\par
\textbf{Implications}:
%\revise{These results imply 
Finally, our evaluation demonstrates that \approach 
effectively achieves 
%is effective for 
%in aiding developers to 
%targeting a 
single testing goals (revealing erroneous behaviors,  generating complex inputs,  or 
%producing inputs with %that induce %leading to 
inducing long execution time) 
%Finally, we illustrate that 
%\approach 
%outperforms the state-of-the-art grammar-based test generators and it 
and scales to 
%several single and 
multiple testing goals.  
%Finally,  \approach generates inputs that are similar to real-world seed inputs but outperform them.
%}
\end{abstract}}

}

\maketitle

% is not selected <OR> if conference mode is selected - because compsoc
% conference papers position the abstract like regular (non-compsoc)
\IEEEdisplaynontitleabstractindextext

% \fi
\IEEEpeerreviewmaketitle

\ifCLASSOPTIONcompsoc
\IEEEraisesectionheading{\section{Introduction}\label{sec:introduction}}
\else
\section{Introduction}
\label{sec:introduction}
\fi

%TODO Change this text

%\todo{check all tables and figures are referenced}

%\IEEEPARstart{C}{omplex} s
\IEEEPARstart{S}{oftware} systems often need to process highly structured inputs 
 which
%  Structured inputs 
% as input.  These inputs 
are difficult and highly improbable to generate via random test generation,  especially without a knowledge of the input \revise{specification~\cite{hodovan2018grammarinator}. }
%or structured inputs,  the probability of generating a valid input
% (e.g., JSON) 
%is low . 
%\footnote{
%}
%\IEEEPARstart{T}{raditional} 
%\IEEEPARstart{G}{rammar-based} 
%To this end,  t
%Traditional g
Grammar-based test generators address this concern by leveraging
% the input specification, i.e., 
input grammars as a producer
%test generation techniques aim 
to generate 
\textit{syntactically valid inputs} for assessing software quality.
% of software systems. 
%~\todo{cite}.  
%Subsequently,  t
Researchers have proposed several grammar-based test generators,  
% have been proposed 
%to achieve this goal,~ \todo{cite}. S
some of which have been shown to be highly effective in 
%revealing faults,
%effectively
generating valid test inputs~\cite{8952419, hodovan2018grammarinator, 180229}.  
These techniques ensure that generated test inputs \textit{not only} exercise the input validation components of the software, but also assess the actual program logic.  
%To this end,  r

%However,  

\begin{figure*}[tbp!]
    \begin{subfigure}[l]{0.49\textwidth}
%>>>>>>> 01eff8c0d191689b1e9571c648105bf4d8a74a9f
    \centering
    \begin{minipage}{.5\textwidth}%
\begin{lstlisting}[basewidth=0.5em,language=Java,basicstyle=\ttfamily,keywordstyle=\bfseries\color{BlueViolet},numbers=left,stepnumber=1,numberstyle=\ttfamily\small\color{darkgray},numbersep=0.2cm,xleftmargin=0.5cm,emph={euclid,x,y,t},emphstyle=\bfseries\color{Bittersweet},escapeinside={(*}{*)}]
public int euclid(int x, int y) {
  if (x == 0) {
    return 1;
  }
  if (x < y) {
    int t = x;
    x = y;
    y = t;
  }
  if (x % y == 0) { (*\mbox{\Lightning \normalfont{Bug}}*)
    return y;
  } else {
    return euclid(y, x % y);
  }
}
\end{lstlisting}
\end{minipage}
\caption{\centering 
A sample Java function ``\texttt{euclid()}''  which computes the greatest common divisor of two positive given integers (adapted from \cite{exampleprogramurl})\revise{.} 
There is a bug in line 10 that causes a division-by-zero exception when ``y'' has the value 0.
To fix the bug, ``\lstinline|y == 0|'' needs to be added as condition in line 2.
}
\label{exampleprogram}% \\
\vspace{\baselineskip}
\end{subfigure}
\begin{minipage}{0.49\textwidth}
    \begin{subtable}[c]{1.0\textwidth}
        \centering\ttfamily
    \begin{tabular}{ll}
        euclid(36,20) &
        euclid(1,40) \\
        euclid(56,19) &
        euclid(5,307) \\
        euclid(92,81) &
        euclid(1032,45) \\
        euclid(19,23) &
        euclid(54,36) \\
    \end{tabular}
%<<<<<<< HEAD
%    \caption{Seed Files used to learn grammar probabilities in \autoref{examplegrammar}}\label{tab:seedfiles-example}
%\end{table}
%=======
    \subcaption{Seed \rev{Inputs} used to learn grammar probabilities in \autoref{examplegrammar}}\label{tab:seedfiles-example}
\end{subtable}
\\[2em]
\begin{subfigure}[l]{1.0\textwidth}
    \centering
    \begin{minipage}{0.65\textwidth}
\begin{cfglisting}
start = !\annotprob{1.0}! "euclid(" integer "," integer ")"
integer = !\annotprob{0.04}!digit | !\annotprob{0.96}!nzdigit number
number = !\annotprob{0.74}!digit | !\annotprob{0.26}!digit number
digit = !\annotprob{0.09}!"0" | !\annotprob{0.91}!nzdigit
nzdigit = !\annotprob{0.12}!"1" | !\annotprob{0.14}!"2" | !\annotprob{0.11}!"3"
          !\annotprob{0.14}!"4" | !\annotprob{0.13}!"5" | !\annotprob{0.10}!"6"
          !\annotprob{0.09}!"7" | !\annotprob{0.09}!"8" | !\annotprob{0.08}!"9"
\end{cfglisting}
\end{minipage}
\subcaption{\centering The learned probabilistic grammar for the ``\texttt{euclid()}'' function shown in \autoref{exampleprogram}}
\label{examplegrammar}
\end{subfigure}
\end{minipage}
\vspace{-\baselineskip}
\caption{\centering An example Java program together with seed inputs and a context-free grammar that describes the format of the inputs accepted by the program. The probabilities shown in the grammar are learned from the seed inputs.}
\label{fig:example-euclid}
\end{figure*}

Despite their success in generating valid inputs, current grammar-based test generators 
% they 
are not effective for \textit{achieving or maximizing arbitrary testing goals}. 
During testing,  developers often aim to achieve specific testing goal(s), beyond input validity. 
%of current grammar-based test generators, .   
For instance,  a developer may aim to generate arbitrarily complex inputs,  failure-inducing inputs,  inputs that achieve high input/code coverage or all of the aforementioned goals.  These testing goals are particularly more complex than achieving syntactic validity since they require 
%. This is due to their relation to 
unique program behaviors or input structures.  
% beyond  the input validity.  

%Besides, 
%To address this challenge, t
This work aims to 
\textit{automatically generate valid inputs that effectively target arbitrary testing goal(s)}.
% is challenging}. 
%learning the program behavior or input structure required to achieve an arbitrary goal is challenging. 
%In this work,  w
We address this challenge via a combination of grammar learning and evolutionary testing.  The main idea of our work is to leverage test feedback to learn the relationship between input structures and 
%the goal-specific inputs and 
the testing goal-at-hand. 
%The key insight is to l
%\approach
Our approach (\approach\footnote{\approach means ``\textbf{F}ee\textbf{d}back \textbf{Loop} for Directed Grammar-based Test Generation''}) 
leverages this relationship to drive test generation towards inputs that maximize the goal-at-hand.  
%In this work, we 

\rev{
We illustrate \approach using a motivating example (\autoref{fig:example-euclid}) and an algorithm  
%describing its components 
(\autoref{algorithm}).  For an arbitrary testing goal,  \approach
% iven a testing goal, FDLOOP 
iteratively selects, evolves and learns the input distribution of goal-specific inputs via test feedback and a probabilistic grammar. 
%\todo{Describe \autoref{algorithm}}
It first learns a probabilistic input grammar from sample seed inputs (\autoref{algorithm}: \autoref{alg:lrnPG}). 
We provide an example of the probabilistic grammar learning process in \autoref{fig:example-euclid},  with an example grammar (\autoref{examplegrammar}) and sample seed inputs (\autoref{tab:seedfiles-example}). 
\approach then leverages the learned input grammar as a producer to generate new inputs (\autoref{algorithm}: \autoref{alg:gen}).  Next (\autoref{algorithm}: \autoref{alg:instart} to \autoref{alg:inend}),  it iteratively evolves the generated inputs towards the goal by selecting relevant inputs (\autoref{algorithm}: \autoref{alg:sel}). 
%to the goal-at-hand . 
%evolving or 
Finally,  it mutates relevant inputs (\autoref{algorithm}: \autoref{alg:mut}) and mutates the learned grammar (\autoref{algorithm}: \autoref{alg:gmu}) to generate new inputs, until the goal is achieved. 
\autoref{approachbetter} further illustrates this evolution process 
%(parsing, mutation,  grammar learning  is further illustrated in  
with an example and   \autoref{approachoverview} shows a high-level workflow of \approach, with components mapped (color-coded) to \autoref{algorithm} steps.   
%e process of input parsing,  input mutation process  (\autoref{approachbetter}).  
}
In this work,  we concretize \approach for four testing goals, namely unique code coverage, input-to-code complexity (aka mappings), program failures (aka exceptions) and long execution time (aka run time). 
\revise{ %Most of these goals are common in testing except for 
``Input-to-code complexity'' is a mapping from features that are present in the test input (aka input features) to executed methods in the program}. 
We have selected these four testing goals because they have been targeted in previous work and they are relevant  
%to  developers 
for software testing practice~\cite{8952419, hodovan2018grammarinator, 180229, 9154602, eberlein2020evolutionary}.

As an example,  
%consider the example 
the \texttt{euclid()} program in \autoref{exampleprogram} shows a buggy program that computes 
the greatest common divisor (GCD) of two integers
 \texttt{x} and \texttt{y}.
 \rev{
Let us assume  the \texttt{euclid()} program \textit{only} accepts user inputs in the following format -- ``\texttt{euclid(x, y)}''. 
% e.g.,  as specified in its main method or entry point: 
\autoref{tab:seedfiles-example} and \autoref{examplegrammar} show examples of valid inputs and the expected input format for the \texttt{euclid()} program, respectively.}
%This input structure assumes t3
%This example demonstrates the 
%importance of grammar-based test generation: 
%limitation of random test generation.  Notably, t
The probability of generating a valid input for this program without an input specification is extremely low.   \autoref{tab:inputstable} shows sample test inputs generated via random test generators. 
%versus grammar based test generators for the  \texttt{euclid} program.  
As shown in \autoref{tab:inputstable}, the inputs generated by the random fuzzer can not exercise the actual program logic (\autoref{exampleprogram}) and achieves zero (0) code coverage.  %In addition,  
\autoref{tab:inputstable} also shows that the inputs 
generated by the grammar-based baselines for the \texttt{euclid()} program (in \autoref{exampleprogram}) using its input grammar (\autoref{examplegrammar}) are valid, exercise the program logic and achieve better coverage than the random fuzzer.  
%\rev{This demonstrates the importance and contribution of the input grammar. }

\rev{Despite achieving validity,  we note that 
%the baselines do not effectively target the testing goals at hand.} 
%\rev{
%Meanwhile,  b
%both 
the random test generator and other grammar-based test generators are unable to 
%Even though several inputs generated by the grammar-based baselines are syntactically valid and achieve better coverage than the random fuzzer,  they are unable to 
maximize any of the aforementioned four testing goals. 
%Meanwhile,  
\autoref{tab:inputstable} (rows \#8, \#9 and \#10) shows that only 
 \approach effectively maximizes each of these goals:
%\autoref{tab:inputstable} demonstrates that 
%(except for the inverse probabilistic fuzzer) 
For instance,  only \approach maximized exceptions (row \#9), mappings  and runtime (rows \#8 and \#9).  Likewise,  only \approach (and the inverse probabilistic fuzzer)
generated inputs that induce the \texttt{division  by zero} failure in line ten (10) of \autoref{exampleprogram}.}

To the best of our knowledge, this work presents the first grammar-based test generation technique that systematically targets arbitrary testing goal(s). 
%to  ..... 
Overall, this paper makes the following contributions:
\begin{itemize}
\item \textbf{\approach:} We present a directed grammar-based test generation approach that systematically targets and maximises an arbitrary testing goal (\autoref{sec:approach}).  \approach employs a synergistic combination of evolutionary testing and grammar learning to generate test suites that 
%allow developers to 
achieves 
%maximises 
%their intended 
testing goal(s).   

%\item \textbf{Single Testing Goal:} 

\item \textbf{Evaluation:} We present the experimental evaluation of our approach using 20 open source Java programs and three (3) popular input formats -- JSON, CSS and JavaScript  (\autoref{evaluationsetup}). 
%\item \textbf{Targeting Single and Multiple testing Goals:} 

\item \textbf{Testing Goals:}  Our results show that \approach is effective in achieving a \textit{single testing goal} for all four (4) examined goals. 
% (\autoref{evaluation}).  
We further demonstrate that our approach scales to \textit{multiple testing goals} and it 
% In addition, we show that \approach 
is tunable to\textit{ avoid and ignore specific goal(s)}
%  This is achieved via the weighted contribution of multipl goals, which is tunable by a developer (
(\autoref{evaluation}).  

\item \textbf{Comparison to the State-of-the-art:} \revise{We compare 
%the performance of 
\approach to five state-of-the-art (SOTA) baselines including three 
%(SOTA) 
grammar-based baselines and 
%the seed inputs,
two evolutionary 
%grammar-based t
test generators (\evogfuzz
%) 
and 
%the SOTA non-grammar-based evolutionary test generator (
\dynamosa).   
\approach 
%Results demonstrate that it 
outperforms all baselines
in most (86\%) settings. 
It is 
%up to 
twice (2X) as 
%\revise{ \todo{77\%} and \todo{X\%} more 
effective as the best baseline (\evogfuzz)
in triggering exceptions
%neous behaviors 
% and  \dynamosa,  respectively}
% by 
%  in achieving the testing goals 
(\autoref{evaluation}). }  

\item \textbf{Ablation and Sensitivity Study:} We demonstrate  
%via an ablation study 
that \approach 's components 
%indeed targets the intended testing goal(s), its components 
are necessary to effectively achieve testing goal(s) 
and \approach is effective across different parameter settings 
(\autoref{evaluation}). 
%(\textit{see}  \todo{XXX}).   

\end{itemize}

%The rest of 
This paper is organised as follows: Section \ref{sec:overview} provides an overview of \approach, and \autoref{sec:approach} presents a detailed description of our approach.  
In \autoref{evaluationsetup} and \autoref{evaluation}, we present the experimental setup and discuss our findings, respectively. In \autoref{threats}, we discuss the threats to validity and we present closely related work in \autoref{relatedwork}. Finally, we conclude with the discussion of future work in \autoref{sec:conclusion}.

\begin{table*}[tb]
    \centering
    \begin{tabular}{|l|l|ll|cccc|}
    \hline
       \textbf{Category} & \textbf{\#} %&
        \textbf{Approach} & \multicolumn{2}{c|}{\textbf{
%       Seeded or Generated i
Test Inputs}} & \textbf{Coverage} & \textbf{\#Exceptions} & \textbf{\#Mappings} & \textbf{Run Time} \\
        \hline
        \multirow{6}{*}{\rotatebox[origin=c]{0}{\textbf{Baselines}}} & \multirow{4}{*}{\textbf{1}} 
%        & 
        \multirow{4}{*}{Seed Inputs} &
%        \multicolumn{2}{c}{
%see \autoref{tab:seedfiles-example}} 
%\multirow{3}{*}{\parbox[l]{6cm}{
%euclid(36,20), euclid(1,40), euclid(56,19), \\ 
%euclid(5,307) euclid(92,81) euclid(1032,45), \\
%euclid(19,23) euclid(54,36)
%}}}
euclid(36,20) & euclid(1,40)
        &  \multirow{4}{*}{82\%} &  \multirow{4}{*}{0} &  \multirow{4}{*}{36 } &  \multirow{4}{*}{36} \\
        &&euclid(56,19) & euclid(5,307)  &&&&\\
        && euclid(92,81) & euclid(1032,45) &&&&\\
        && euclid(19,23) & euclid(54,36) &&&&\\ %[0.5em]
        \cline{2-8}
%        [1em]
        &  \multirow{2}{*}{\textbf{2}} \multirow{2}{*}{Random Fuzzer} &
            8nJqQWl1HX&
            BeDa.Ay!bw\&
    &  \multirow{2}{*}{0\%} &  \multirow{2}{*}{0} &  \multirow{2}{*}{0} &  \multirow{2}{*}{0} \\
    &&
            HHbA/He&
            .JVatD\%(T
                      &&&&\\
                       \hline
                     \multirow{6}{*}{\parbox[l]{1cm}{\textbf{Grammar}\\\textbf{-based}\\\textbf{Baselines}}}  &
                     \multirow{2}{*}{\textbf{3}}  
    \multirow{2}{*}{Random Fuzzer} &
            euclid(300,3000)&
            euclid(8500,100)
    & \multirow{2}{*}{64\%} & \multirow{2}{*}{0} & \multirow{2}{*}{28} & \multirow{2}{*}{28} \\
    &&
            euclid(50,3100)&
            euclid(300,150)
            &&&&\\
            &  \multirow{2}{*}{\textbf{4}}
        \multirow{2}{*}{Probabilistic Fuzzer}  & 
            euclid(586,20)&
            euclid(997,21)
    & \multirow{2}{*}{82\%} &  \multirow{2}{*}{0} &  \multirow{2}{*}{36} &  \multirow{2}{*}{36} 
    \\
    &  &
            euclid(71,92)&
            euclid(28,597)
            &&&&\\
            &  \multirow{1}{*}{\textbf{5}}
        Inverse Probabilistic &
            euclid(1,0)&
            euclid(0,2)
    & \multirow{2}{*}{36\%} & \multirow{2}{*}{3} & \multirow{2}{*}{12} & \multirow{2}{*}{16}\\
    &  \multirow{1}{*}{\quad Fuzzer} &
            euclid(0,0)&
            euclid(0,0)
            &&&&\\%
            \hline
            \multirow{8}{*}{\parbox[l]{1cm}{\textbf{\approach}\\\textbf{Single}\\\textbf{Goal}}} &
%%
%  Begin of Simulated Inputs
%%
  \multirow{2}{*}{\textbf{6}}  
        \multirow{2}{*}{Maximizing Coverage} &
            euclid(0,4149)&
            euclid(3,79)
            & \multirow{2}{*}{\textbf{100\%}} & \multirow{2}{*}{0} & \multirow{2}{*}{44} & \multirow{2}{*}{44} \\
    && 
            euclid(10,5)&
            euclid(11,7)
            &&&&\\%
            &   \multirow{2}{*}{\textbf{7}}  
        \multirow{2}{*}{Maximizing Run Time} &
            euclid(121393,75025) &
            euclid(28657,17711)
            & \multirow{2}{*}{82\%} & \multirow{2}{*}{0} & \multirow{2}{*}{36} & \multirow{2}{*}{\textbf{36}} \\
    && 
            euclid(4181,2584)&
            euclid(377,233)
            &&&&\\%
            &  \multirow{2}{*}{\textbf{8}} 
        \multirow{2}{*}{Maximizing Mappings} &
            euclid(0,221)&
            euclid(0,31)
            & \multirow{2}{*}{100\%} & \multirow{2}{*}{0} & \multirow{2}{*}{\textbf{44}} & \multirow{2}{*}{44} \\
    && 
            euclid(4,101)&
            euclid(21,6)
            &&&&\\%
            &  \multirow{2}{*}{\textbf{9}} 
        \multirow{2}{*}{Maximizing Exceptions} &
            euclid(1,0)&
            euclid(39,0)
            & \multirow{2}{*}{18\%} & \multirow{2}{*}{\textbf{4}} & \multirow{2}{*}{8 } & \multirow{2}{*}{8} \\
    && 
            euclid(12424,0)&
            euclid(2,0)
            &&&&\\%
            \hline
            \multirow{3}{*}{\parbox[l]{1cm}{\textbf{\approach}\\\textbf{Multiple}\\\textbf{Goals}}}  &  \multirow{3}{*}{\textbf{10}}  
        \multirow{3}{*}{Maximizing All} & 
            \multirow{3}{*}{\parbox[l]{1cm}{euclid(0,0)\\euclid(1,0)}} &
            \multirow{3}{*}{\parbox[l]{1cm}{euclid(75025,121393)\\ euclid(2,2)}}
            &  \multirow{3}{*}{\textbf{100\%}} &  \multirow{3}{*}{\textbf{2}} &  \multirow{3}{*}{\textbf{44}} &  \multirow{3}{*}{\textbf{44}} \\
    &&&&&&&\\%
            &&&&&&&\\%
            \hline
            \multirow{8}{*}{\parbox[l]{1cm}{\textbf{\approach}\\\textbf{Ignore}\\\textbf{Goal}}}  &  \multirow{2}{*}{\textbf{11}}  
        \multirow{2}{*}{Ignoring Coverage} &
            euclid(28657,17711)&
            euclid(722,503)
            &  \multirow{2}{*}{54\%} &  \multirow{2}{*}{\textbf{1}} &  \multirow{2}{*}{\textbf{24}} &  \multirow{2}{*}{\textbf{24}} \\
    &&
            euclid(32,0)&
            euclid(0,1)
            &&&&\\%
             &  \multirow{2}{*}{\textbf{12}}  
        \multirow{2}{*}{Ignoring Mappings} &
            euclid(0,31912)&
            euclid(421,301)
            &  \multirow{2}{*}{\textbf{54\%}} &  \multirow{2}{*}{\textbf{1}} &  \multirow{2}{*}{24} &  \multirow{2}{*}{\textbf{24}} \\
    &&
            euclid(102,0)&
            euclid(44,44)
            &&&&\\%
            &  \multirow{2}{*}{\textbf{13}} 
        \multirow{2}{*}{Ignoring Run Time}  &
            euclid(1,0)&
            euclid(0,31)
            &  \multirow{2}{*}{\textbf{54\%}} &  \multirow{2}{*}{\textbf{1}} &  \multirow{2}{*}{\textbf{24}} &  \multirow{2}{*}{24} \\
    &&
            euclid(9,7)&
            euclid(2,2)
            &&&&\\%
            & \multirow{2}{*}{\textbf{14}} 
        \multirow{2}{*}{Ignoring Exceptions}  &  
            euclid(302,302)&
            euclid(0,21)
            & \multirow{2}{*}{\textbf{64\%}} &  \multirow{2}{*}{0} &  \multirow{2}{*}{\textbf{28}} &  \multirow{2}{*}{\textbf{28}} \\
    &&
            euclid(121393,75025)&
            euclid(132,9)
            &&&&\\%
            \hline
    \end{tabular}
    \caption{\centering 
 \lrevise{   Inputs generated by \approach and baselines for different 
%    using different fitness functions as 
    testing goals: Coverage,  Exceptions,  Mappings, and Run Time (in this example, the run time equals the number of executed instructions).
Grammar-based test generators 
%are using 
use the grammar shown in \autoref{examplegrammar}.
%        The b
        \textbf{Bold} text 
%        values 
shows the values for the testing goals of \approach. }
% focused its generation towards.
        %The run time is estimated by counting the executed lines of code.
        %To generate grammar-based inputs, we used Tribble with a seed of 1234 and a tree depth of 30. The runtime is given as the number of instructions needed to execute all inputs for this synthetic example.
            %TODO -- We generated the inputs "by hand" 
    %
}\label{tab:inputstable}
\end{table*}

\begin{figure}[tbp!]
%\scriptsize
%\hspace*{1.5in}
    \centering
%\begin{flushright}
%    \advance\rightskip-4cm
%Input: Seed files S, Fitness function F
%Output: 

\lstdefinelanguage{algocode}{%
    mathescape=true,
    extendedchars=false,
    keywords={for,in,return,Input},
    keywordstyle=\bfseries,
    commentstyle=\normalfont\itfamily,
    morecomment=[l][\normalfont]{//},
    columns=fullflexible,
    numbers=left,
    stepnumber=1,
}
% Grammar Learning:
\colorlet{grammarred}{red!15!white}\newcommand{\algoHighlightG}{\makebox[0pt][l]{\color{grammarred}\rule[-4pt]{0.8\linewidth}{14pt}}}
% Input Generator:
\colorlet{inpgenblue}{blue!15!white}\newcommand{\algoHighlightI}{\makebox[0pt][l]{\color{inpgenblue}\rule[-4pt]{0.8\linewidth}{14pt}}}
% Input Selector:
\colorlet{selectoryellow}{orange!15!white}\newcommand{\algoHighlightS}{\makebox[0pt][l]{\color{selectoryellow}\rule[-4pt]{0.75\linewidth}{14pt}}}
\hfill \begin{lstlisting}[language=algocode,escapeinside={*@}{@*}]
Input: $\text{FitnessFunction} ~f$, $\text{SeedInputs}$
// Starting grammar
*@\algoHighlightG@*$G_0 := \text{learnProbabilisticGrammar}(\text{SeedInputs})$  *@\label{alg:lrnPG}@*
$I := \emptyset$
for $i$ in $0,\dots,\text{numGenerations}$:
    // Generate inputs
    *@\algoHighlightI@* $T := \text{generateInputs}(G_i)$ *@\label{alg:gen}@*
    // Mutate inputs and add mutated inputs to $t$
    *@\algoHighlightI@* $T = T \cup \text{mutateInputs}(T)$*@\label{alg:mut}@*
    $P,F = \emptyset,\emptyset$*@\label{alg:instart}@*
    for $t$ in $T$:
    	// Collect information from the parse trees
    	*@\algoHighlightS@*$P = P \cup \text{examineParseTree}(t)$ 
        // Execute subject & collect feedback
        *@\algoHighlightS@*$F = F \cup \text{executeSubjects}(I \cup \{t\})$ *@\label{alg:inend}@*
    // Select the best-performing input based on $f$
    *@\algoHighlightS@*    $s := \text{selectBestInput}(f,T,P,F)$ *@\label{alg:sel}@*
    $I = I \cup \{s\}$ // Add selected input to set of inputs
    // Learn new grammar probabilities
    *@\algoHighlightG@*    $G_{i+1} := \text{learnProbabilities}(s)$ *@\label{alg:lrn}@*
    // Mutate the grammar probabilities
    *@\algoHighlightG@*    $G_{i+1} = \text{mutateGrammar}(G_{i+1})$*@\label{alg:gmu}@*
return I
\end{lstlisting}
%\end{flushright}
    \caption{\approach algorithm}\label{algorithm}
\end{figure}

\section{Overview}
\label{sec:overview}

\subsection{Motivating Example}
%TODO Introduction
%\revise{%
\autoref{exampleprogram} shows a sample program \texttt{euclid()} that computes the greatest common divisor (GCD) of two integers \texttt{x} and \texttt{y}.  \autoref{examplegrammar} shows an excerpt of the \textit{input grammar} defining the syntax of the acceptable inputs for this program. 
%is shown in \autoref{examplegrammar}. 
A sample \textit{valid input} for this grammar \revise{is} shown in the seed inputs \autoref{tab:seedfiles-example}.  
\rev{We assume the program only accepts user inputs in the following format -- ``\texttt{euclid(x, y)}''. }

%Given this grammar,  an example of an valid input would be 
%\texttt{euclid(123,456)}.
%}

%\revise{
To demonstrate our approach (\approach), we 
%test this program by 
%first 
collect
%ing 
eight valid sample inputs, shown in \autoref{tab:seedfiles-example}.  We then 
%using this grammar 
%and 
feed each input to the \texttt{euclid()} function to collect test feedback -- code coverage,  exceptions, run time and mappings. 
The performance of the seed inputs \revise{is} shown in \autoref{tab:inputstable} (row \textit{one}). 
 We also 
learn the input distribution  of the sample inputs (i.e., the 
%nature of the 
%sample inputs by collecting the 
%    ing 
%grammar 
probabilities of input elements in the sample inputs) by parsing each input using the input grammar.   
% for
% We annotate  
 The learned grammar probabilities are 
% each expansion of the grammar with probabilities, as 
 shown in \autoref{examplegrammar} in {\textbf{\color{kulerD}green}} boxes before each expansion.
%}

\newsavebox{\cfgone}
\begin{lrbox}{\cfgone}
    \begin{minipage}[b]{.33\textwidth}
\begin{cfglisting}
start = !\annotprob{1.0}! "euclid(" integer "," integer ")"
integer = !\annotprob{0.04}!digit | !\annotprob{0.96}!nzdigit number
number = !\annotprob{0.74}!digit | !\annotprob{0.26}!digit number
digit = !\annotprob{0.09}!"0" | !\annotprob{0.91}!nzdigit
nzdigit = !\annotprob{0.12}!"1" | !\annotprob{0.14}!"2" | !\annotprob{0.11}!"3"
          !\annotprob{0.14}!"4" | !\annotprob{0.13}!"5" | !\annotprob{0.10}!"6"
          !\annotprob{0.09}!"7" | !\annotprob{0.09}!"8" | !\annotprob{0.08}!"9"
\end{cfglisting}
\end{minipage}
\end{lrbox}

\newsavebox{\cfgtwo}
\begin{lrbox}{\cfgtwo}
    \begin{minipage}[b]{.33\textwidth}
\begin{cfglisting}
start = !\annotprob{1.0}! "euclid(" integer "," integer ")"
integer = !\annotprob{0.04}!digit | !\annotprob{0.96}!nzdigit number
number = !\annotprob{0.74}!digit | !\annotprob{0.26}!digit number
!\textbf{\underline{digit}}! = !\annotprob[kulerE]{\textbf{\underline{0.5}}}! !\text{\textbf{\underline{"0"}}}!  | !\annotprob[kulerE]{\textbf{\underline{0.5}}}! !\text{\textbf{\underline{nzdigit}}}!
nzdigit = !\annotprob{0.12}!"1" | !\annotprob{0.14}!"2" | !\annotprob{0.11}!"3"
          !\annotprob{0.14}!"4" | !\annotprob{0.13}!"5" | !\annotprob{0.10}!"6"
          !\annotprob{0.09}!"7" | !\annotprob{0.09}!"8" | !\annotprob{0.08}!"9"
\end{cfglisting}
\end{minipage}
\end{lrbox}

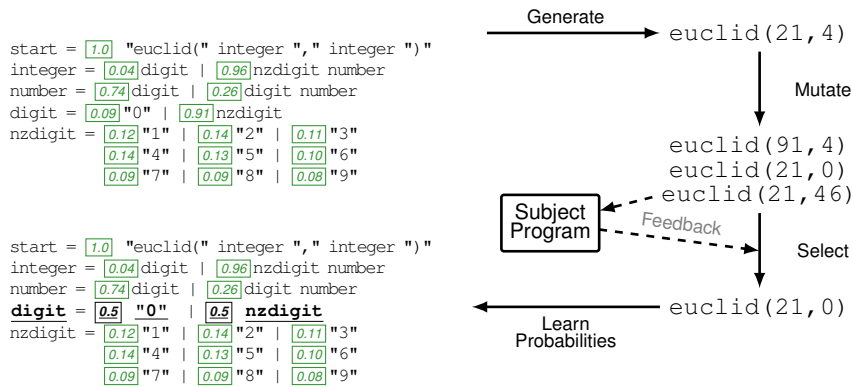
\begin{figure*}[tbp!]
    \centering
    \begin{tikzpicture}
        \node[] at (0,0) (grammar1) {\usebox\cfgone};
        \node[bdcode,right=2.5 of grammar1.north east,anchor=north west] (code1) {euclid(21,4)};
        \draw[bdarrow,shorten <=2mm] (grammar1.east |- code1) -- (code1) node[bdcaption,pos=.5] {Generate};

        \node[bdcode,below=1 of code1] (code2) {euclid(91,4)\\euclid(21,0)\\euclid(21,46)};
        \draw[bdarrow] (code1) -- (code2) node[bdcaption,pos=.5,anchor=west] {Mutate};

        \node[bdblock,below left=1.0 of code2.west,anchor=east] (program) {Subject\\Program};
        \node[bdcode,below=1 of code2] (code3) {euclid(21,0)};
        \draw[bdarrow] (code2) -- (code3) node[bdcaption,pos=.5,anchor=west] (selanch) {Select};
        \draw[bdarrow,dashed] (code2) -- (program);
        \draw[bdarrow,dashed] (program) -- (selanch.west) node[bdcaption,text=gray,pos=.5,sloped] {Feedback};

        \node[below=0 of grammar1] (grammar2) {\usebox\cfgtwo};
        \draw[bdarrow] (code3) -- (grammar2.east |- code3.center) node[bdcaption,pos=.5,below] {Learn Probabilities};
    \end{tikzpicture}

    \caption{\approach 's Input Mutation Details showing how input mutation in combination with the test feedback iteratively guides input generation towards the testing goal (failure/exception).  
In this example,  the \textit{mutated input} 
%    directing an iterative approach quickly finds an input that 
triggers an exception in the program~(\autoref{exampleprogram}).  The 
%    hypothetical subject 
program has a bug where an \texttt{division by zero} exception is thrown when \texttt{y = 0}. 
\lrevise{After the grammar learning, grammar probabilities under the digit rule are mutated (\textit{see} \textbf{\underline{bold and underlined}} \texttt{digit} rule).}
}\label{approachbetter}
\end{figure*}

\subsection{\approach Approach}
%\revise{
\rev{\autoref{algorithm} describes the \approach algorithm and \autoref{approachoverview} shows a high-level workflow of \approach, with components color-mapped to the \approach algorithm (\autoref{algorithm}). 
Specifically,  in both the algorithm (\autoref{algorithm}) and workflow diagram (\autoref{approachoverview}), the  \textit{grammar learning} steps are in {\textbf{\color{red}red}},  the \textit{input generation} steps are in {\textbf{\color{blue}blue}} and the \textit{input selection} steps are in {\textbf{\color{orange}orange}}. 
}

\rev{The key insight of \approach is to employ \textit{grammar learning} and \textit{evolutionary testing} to generate test inputs that achieve specific testing goal(s).  We  illustrate the evolution process of \approach in \autoref{approachbetter}, where expanding the grammar that we have already seen in \autoref{examplegrammar} results in generating the input ``\texttt{euclid(21,\rev{4})}''.
In the following, we illustrate \approach's test generation process by evolving a sample input for the motivating example -- \autoref{exampleprogram}. }

\rev{As shown in \autoref{algorithm}, 
\approach takes as inputs the seed inputs.  It first learns the probabilistic grammar for the seed inputs (\autoref{algorithm}: \autoref{alg:lrnPG}).  
Then, for each generation, it generates new inputs using the learned grammar (\autoref{algorithm}: \autoref{alg:gen}).  Next, the generated inputs are mutated using bit flips, changing single characters in the code (\autoref{algorithm}: \autoref{alg:mut}).  We illustrate input mutation in our motivating example (\autoref{approachbetter}).  Input mutation allows inputs to cover cases that are not covered by the grammar -- in this example,  the new input ``\texttt{euclid(21,0)}'' that contains a zero as second call argument is discovered.  Since only few of the seed inputs contained such a character, it would have been improbable to cover a case where a zero (0) is generated by expanding the grammar using the probabilistic approach.}

\rev{In an iterative step (\autoref{algorithm}: \autoref{alg:instart} to \autoref{alg:inend}),  \approach feeds all inputs obtained from mutating the generated inputs into the subject program,  and collects test feedback (e.g.,  the number of triggered exceptions for each input). It then selects (\autoref{algorithm}: \autoref{alg:sel}) which input performed best (e.g.,  triggers the the most exceptions).  In this example, an exception is triggered as soon as there is an occurrence of a zero (0)  as second argument (division by zero in line 10 of the subject program). Thus, the input generator selects this specific input out of all tested inputs.}

\rev{In the next step,  \approach learns the new probabilities for the grammar from the selected input (\autoref{algorithm}: \autoref{alg:lrn}). This allows it to evolve the grammar probabilities (and thus the inputs).  Then, the grammar probabilities are also mutated (\autoref{algorithm}: \autoref{alg:gmu}), which might also uncover more inputs that we have not seen before in the next generation.
This aforementioned steps are then repeated in the next generation,  starting from the input generation step (\autoref{algorithm}: \autoref{alg:gen}),  albeit using the learned/mutated grammars from the previous generation.  \approach stops when it achieves the specified number of generations or the testing goal. }

\subsection{Grammar-based Test Generators}\label{sec:grammar-baselines}
%\revise{
%\subsubsection{Grammar-based Test Generators}
In our evaluation,  we compare \approach to the following grammar-based fuzzing techniques: 
%\todo{ES: Random Baseline is not grammar-based, so we should give it a different name}
% 
% baselines that represent the traditional grammar-based fuzzing methods:

\smallskip
\noindent
\textbf{Random Grammar-based Fuzzer:} 
\rev{This is a random grammar-based fuzzer where the choice between productions is governed by
a uniform distribution~\cite{8952419, 9154602}.
It generates the same number of inputs as our approach (\approach), 
%selected inputs (5 inputs) per generation as our approach,  
albeit without grammar learning,  grammar probabilities,  or program feedback. 
%any feedback from the subject program.
%After each generation, no new grammar is being learned from the set of selected input files (if probabilities were learned, the next generation's purely random input generator would just ignore the probabilities).
%Only one single set of inputs is generated for each generation.
%}\par
%\revise{
%The random baseline 
In our setting,  to avoid expanding into an unbounded tree, the generator chooses the alternatives that lead to the shortest possible subtree after the grammar expansion tree of the generated input has reached a certain depth (3). \autoref{tab:inputstable} (\#3 Random Fuzzer) shows the inputs generated by the random grammar-based fuzzer when using the example grammar in \autoref{examplegrammar}.  For instance,  the generated inputs contain lots of zeroes (0) since the chance of a digit expanding to ``0'' is equal to the chance of a digit expanding to all other digits (\texttt{nzdigits} [1-9]). This illustrates the effect of a uniform distribution on the \texttt{digit} grammar rule (\autoref{examplegrammar}). }

\smallskip
\noindent
\textbf{Probabilistic 
Grammar-based Fuzzer:} 
%\revise{
%
\rev{The probabilistic grammar-based fuzzer generates inputs based on the probabilities learned 
%in the learned grammar.
while parsing seed sample inputs.  
In this fuzzer,  the choice between productions is governed by the distribution specified by the learned probabilities in the grammar. This distribution is learned from the seed inputs. 
%This way, it 
The aim is to generate inputs \emph{similar} to the inputs that were used to learn the initial probabilities of the grammar.
This baseline is similar to the approach ``\textsc{PROB}'' of \emph{Inputs From Hell}~\cite{9154602}. 
The goal is 
% that aims 
to generate inputs similar to a set of seed inputs.
%For this expansion technique, the authors were inspired by the PTC2 algorithm by Sean Luke~\cite{ptc2}, a genetic programming algorithm that allows a user to specify a probability distribution and an approximate tree size and generates a tree with the given distribution.
Unlike our approach (\approach), the seed input selection of \textsc{PROB} is random, and it does not employ a feedback loop. 
% just as the random baseline selection.
\autoref{tab:inputstable} (\#4 Probabilistic Fuzzer)  shows the inputs generated by the probabilistic baseline.  In contrast to the random fuzzer (row \#3) or inverse probabilistic fuzzer (row \#5), these inputs are similar to the seed inputs that were used to learn the grammar probabilities (row \#1). }
%In particular, these inputs contain similar digits, i.e., all digits except eight (8). }

\smallskip
\noindent
\textbf{Inverted Probabilistic Fuzzer:}
\rev{This baseline generates inputs that are \textit{dissimilar} to a set of seed inputs. 
In this approach,  the
production choice is governed by the distribution obtained by a probability inversion process, where 
we invert the probabilities in the  probabilistic grammar learned from parsing the seed inputs. 
This is similar to the (``\textsc{INV}'') approach in \emph{Inputs From Hell}~\cite{9154602}.
%from the probabilistic baseline to generate 
%inputs probabilistically.
%Thus, it aims to generate 
The aim is to generate inputs that are \emph{dissimilar} to the inputs used to the seed inputs used in learning the grammar probabilities, in order to 
%For some programs, this might 
trigger unexpected behaviors. 
\approach is different from the inverted probabilistic baseline since \approach iteratively selects and refines seed inputs via a feedback loop.  The inputs shown in \autoref{tab:inputstable} (row \#5) were generated by the inverse probabilistic fuzzer and they are \textit{dissimilar} to the seed inputs (row \#1). that the probabilities were learned from:
Since with inverted probabilities, the chance of a digit expanding to ``0'' is very high, most generated numbers are zeroes (0).  We note that this is due to the lack of single-digit zero (0) in the seed inputs (row \#1). }

%\begin{table*}[!tbp]\centering
%\caption{\centering Effectiveness of \approach in achieving a single goal in comparison to seed files 
% This table contains the results for all subjects, i.e. "JSONJava", "MinimalJson", "Genson", "Pojo", "Argo", 'Gson', "Jackson", "json-simple", "JsonToJava", "FastJson", "json-flattener", "json2flat", "json-simple-cliftonlabs" "Rhino", "rhino-sandbox" "cssValidator", "closure-stylesheets", "cssparser", "jstyleparser", "sac"
% => Number of files: 5         *              50     *      20          = 5000
%                     (selected files per gen) (generations) (Subjects)
%\begin{tabular}{|l | l | l | l | l | l | l |}
%\scriptsize
%\hline
%\textbf{Approach} & \textbf{Coverage} & \textbf{Mapping} & \textbf{\#Exc.} & \textbf{Uniq\_Exc.Exc.} & \textbf{Run time} & \textbf{Size[kB]} \\
%\hline
%\textbf{Seed files}    & 56.66 & 317182433& 1069 & 15 & 54547.53 & 64625 \\
%\hline
%TODO Mention in the text: "Single" means "Dampening Strategy"
%\textbf{\approach (Single)} & 59.88 & 37548316 & 2020 & 40 & 6218.11 & 697 \\
%\textbf{\approach (Multiple)}  & 60.49 & 39411942 & 1876 & 36 & 5640.10 & 700 \\
%\hline
%\textbf{Impr. (Single)}        & 5.68\%& -88.16\% & 88.96\%& 166.67\% & -88.60\% &\\
%\textbf{Impr. (Multiple)}      & 6.76\%& -87.57\% & 75.49\%& 140.00\% & -89.66\% &\\
%\hline
%}
%\end{tabular}
%\label{tab:seeds}
%\vspace{-0.5cm}
%\end{table*}

\subsection{\approach versus Grammar-based Test Generators}

%\revise{
\autoref{tab:inputstable} shows the test inputs generated by our approach for \texttt{euclid()} function and the testing goals achieved by the resulting test inputs, in comparison to the baselines. % and \approach 's ablation.  

Notably,  the baselines were ineffective in targeting specific testing goal(s): As expected,    random fuzzer (row two) produced \textit{invalid inputs} and did not achieve any of the four testing goals.
Meanwhile,
%human-written seed inputs (row one) and 
the random grammar-based fuzzer (using Tribble~\cite{8952419}), produced \textit{syntactically valid inputs} (row three).
However,  these test inputs have a low performance on the four testing goals. 
\revise{Likewise,  probabilistic fuzzer~\cite{9154602} (row four) could not target most of the testing goals beyond those achieved by the seed inputs (row one).
%However,  
Similarly,  the inverse probabilistic fuzzer~\cite{9154602} (row five) performs worse than the seed inputs (row one) for most (three out of four) goals.  However,  it triggers three \texttt{division by zero} exceptions missed by the seed inputs since it generates inputs different from the seed inputs (\textit{less of the same}).  }%
%\footnote{Except for the inverse probabilistic fuzzer which triggers three \texttt{division by zero} exceptions.}   
As an example,  our input corpus had many multiple digit numbers  and 
fewer single digit numbers.  This  results in a low probability \annotprob{0.04} of \texttt{integer} being expanded to \texttt{digit}.  As shown in \autoref{tab:inputstable},  running the probabilistic fuzzer generates inputs that also have multiple digits (row four) and its inverse probabilistic fuzzer generates single digit numbers (row five).  However, these baselines
were unable to effectively achieve the testing goal(s).  
%}

On the other hand,  
\autoref{tab:inputstable}  (rows six to nine) illustrates that the test inputs generated by \approach achieve each testing goal better than the baselines.  
For instance,  \approach achieves the maximum code coverage (row six) and the highest number of mappings (row eight),  and  exceptions (row nine). 
% and runtime.  
%We attribute the performance of 
This is due to the evolutionary nature of \approach. 
%, i.e., it is able to 
It systematically evolves input generation towards the target testing goal using the test feedbacks.  Finally,  \approach 's ignore goal mode (last four rows, rows 11-14) further shows that \approach can be tuned to ignore a specific testing goal. 
%indeed targets the intended goal. 
Notably, its performance decreases for each ignored goal,  i.e.,  not targeting a specific testing goal (\textit{see} \autoref{tab:inputstable}).  For instance, the code coverage achieved by \approach when it ignores code coverage is almost half (54\%) of that achieved when it targets code coverage (100\%).
%-at-hand. 
%}

%\revise{
%In comparison, 
%% to the baselines, 
%we observed that .... 
%%In o
%
%% \recheck{Since there are few single digiti  
%    }
%%    \par
%    
%    \revise{
%    Using the shown grammar without any probabilities allows us to generate the following six example inputs, using Tribble with the 30-random-6 mode and seed 1234:\\
%    %\begin{center}
%    %\end{center}
%    We use the grammar with the learned probabilities to generate new inputs to test \texttt{euclid()} with and collect code coverage, run time and mappings.
%}
%
%    \todo{ES: explain mappings or reference section?}\todo{ES: ...continue}
%

%\subsection{Main Idea}

%\subsection{Motivating Example}

%\section{Approach}\label{approach}

%%%%%%%%%%%%%%%%%%%%%%%%%%%%%%%%%%%%%%%

\newlength\arrowsep\setlength\arrowsep{2mm}
\newsavebox\annotgr
\begin{lrbox}{\annotgr}
    \resizebox{!}{.75\boxh}{
    \begin{tikzpicture}[microtree]
        \node[]{5}
            child{node[] {3}
                child{node[] {4}}
                child{node[] {1}}
            }
            child{node[] {2}
                child{node[] {3}}
                child{node[] {2}}
            };
    \end{tikzpicture}}
\end{lrbox}
\newsavebox\fuzzer
\begin{lrbox}{\fuzzer}
    \resizebox{\boxw}{!}{
    \begin{tikzpicture}
        \node {Generator};
    \end{tikzpicture}
    }
\end{lrbox}
\newsavebox\mappingb
\begin{lrbox}{\mappingb}
    \begin{tikzpicture}
        \node[align=center,font=\ttfamily\footnotesize] {1,3\\3,4,5};
    \end{tikzpicture}
\end{lrbox}
\newsavebox\feedback
\begin{lrbox}{\feedback}
    \resizebox{\boxw}{!}{
    \begin{tikzpicture}
        \node {Program};
    \end{tikzpicture}
}
\end{lrbox}

\begin{figure*}[tbp!]\centering
\begin{tikzpicture}[%
    files/.pic={%
        \node[bdblock,draw=none,text=black] (-centernode) {#1};
        \draw[bdlinewidth] (-centernode.north west) -- (-centernode.south west) -- (-centernode.south east) -- ([yshift=-2mm]-centernode.north east) -- node[shape=coordinate,pos=0] (an1) {} node[shape=coordinate,pos=1] (an2) {} ([xshift=-2mm]-centernode.north east) -- cycle;
        \draw[bdlinewidth] (an1) -- (an1 -| an2) -- (an2);
        }%
]

    \node[bdblock] (gr1) {\usebox\annotgr};
    \bdcaptionbot{gr1}{Probabilistic Grammar}
    \multiconnectnodeeast{gr1}

    \node[bdblock,above=1.5\nsep of gr1] (parser1) {Parser};

    \pic[above=\nsep of parser1] (crawledfiles) {files={Crawled\\Inputs}};

    \node[bdblock,right=\nsep of gr1] (tribble) {\usebox\fuzzer};
    \bdcaptionbot{tribble}{Input\\Generator};
    \multiconnectnodeeast{tribble}

    \node[bdblock,right=\nsep of tribble-east21,anchor=south west] (mut) {Input\\Mutator};
    \multiconnectnodeeast{mut}

    \node[shape=coordinate,right=\nsep of mut] (ifanchor) {};
    \pic[above=.5\nsep of ifanchor,anchor=south west] (in1) {files={\tiny{Mutated}\\[-.3em]\footnotesize{Inputs}}};
    \pic[below=.5\nsep of ifanchor,anchor=north west] (in2) {files={Inputs}};
    \multiconnectnodeeast{in1-centernode}
    \multiconnectnodeeast{in2-centernode}

    \node[bdblock,right=\nsep of in1-centernode] (parser2) {Parser};
    \multiconnectnodewest{parser2}

    \node[bdblock,right=\nsep of in2-centernode] (program) {\usebox\feedback};
    \multiconnectnodewest{program}

    % Ugly dynamically drawn huge node with other nodes inside of it:
    \node[bdblock,draw=none,right=\nsep of parser2] (phantom1) {};
    \node[bdblock,draw=none,right=\nsep of program] (phantom2) {};
    \node[bdblock,draw=none,anchor=west,right=\nsep of phantom1.west] (phantom3) {};
    \node[bdblock,draw=none,anchor=west,right=\nsep of phantom2.west] (phantom4) {};
    \draw[bdline] (phantom1.north west) -- (phantom2.south west) -- (phantom4.south east) -- (phantom3.north east) -- cycle;
    \node[bdslimblock,anchor=west,right=.5\nsep of phantom1.west] (features) {Features};
    \node[bdslimblock,anchor=west,right=.5\nsep of phantom2.west] (coverage) {Coverage};
    \draw[draw=none] (features.south) -- node[bdslimblock,pos=.5] (fitness) {Fitness} (coverage.north);
    \node[shape=coordinate] at (coverage.center -| phantom2.west) (cov) {};
    \node[shape=coordinate] at (features.center -| phantom1.west) (feat) {};
    \draw[draw=none] (fitness.east) -- node[shape=coordinate,pos=.5] (fitmid) {} (fitness.center -| phantom4.east);
    \node[shape=coordinate] at (phantom3.north -| fitmid) (fit) {};
    \draw[bdarrow] (cov) -- (coverage);
    \draw[bdarrow] (feat) -- (features);
    \draw[bdarrow] (coverage) -- (fitness);
    \draw[bdarrow] (features) -- (fitness);
    \draw[bdarrow,rounded corners=3pt] (fitness.east) -- (fitmid) -- (fit);

    \node[bdblock,above=1.5\nsep of phantom1] (select) {Select};

    \pic[left=\nsep of select] (selected) {files={Inputs}};

    \node[bdblock,left=\nsep of selected-centernode] (parser3) {Parser};

    \node[bdblock,left=\nsep of parser3] (gr2) {\usebox\annotgr};
    \bdcaptionbot{gr2}{Learned Grammar}

    \node[bdblock,above=0.6\nsep of tribble] (gr3) {\usebox\annotgr};

    \node[bdblock,left=1.0\nsep of gr2] (grmut) {Grammar Mutator};
    %\bdcaption{gr3}{Mutated Grammar}

    %Arrows
    \draw[bdarrow] (crawledfiles-centernode) -- (parser1);
    \draw[bdarrow] (parser1) -- (gr1);
    \draw[bdarrow] (gr1) -- (tribble);
    \draw[bdarrow] (tribble-east21) to[out=0,in=180] (mut);
    \draw[bdarrow] (tribble-east22) to[out=0,in=180] (in2-centernode);
    \draw[bdarrow] (mut) to[out=0,in=180] (in1-centernode);
    \draw[bdarrow] (in1-centernode-east21) to[out=0,in=180] (parser2-west21);
    \draw[bdarrow] (in1-centernode-east22) to[out=0,in=180] (program-west21);
    \draw[bdarrow] (in2-centernode-east21) to[out=0,in=180] (parser2-west22);
    \draw[bdarrow] (in2-centernode-east22) to[out=0,in=180] (program-west22);
    \draw[bdarrow] (parser2) to[out=0,in=180] (feat);
    \draw[bdarrow] (program) to[out=0,in=180] (cov);
    \draw[bddashedarrow,-,densely dotted] (in1-centernode) -- (in2-centernode);
    \draw[bddashedarrow,densely dotted] (in1-centernode) to[out=90,in=270,looseness=.5] (select);
    \draw[bdarrow] (fit) to[out=270,in=0] (select);
    \draw[bdarrow] (select) -- (selected-centernode);
    \draw[bdarrow] (selected-centernode) -- (parser3);
    \draw[bdarrow] (parser3) -- (gr2);
    \draw[bdarrow] (gr2) to[] (grmut);
    \draw[bdarrow] (grmut) -- (gr3);
    \draw[bdarrow] (gr3) -- (tribble);

    %----- Dashed Boxes
    \begin{pgfonlayer}{background}
    \draw[bddashedbox={red}] %
        ($(crawledfiles-centernode.north west) + (-\dmarg,\dmarg)$) -- %
        ($(gr1.west |- program.south) + (-\dmarg,-\dmarg)$) -- %;
        ($(gr1.east |- program.south) + (\dmarg,-\dmarg)$) -- %;
        ($(parser1.north east |- gr2-caption.south) + (\dmarg,-\dmarg)$) --
        ($(parser3.east |- gr2-caption.south) + (\dmarg,-\dmarg)$) -- %;
        ($(parser3.east |- crawledfiles-centernode.north) + (\dmarg,\dmarg)$) -- node[pos=0,bddashedcaption={red},anchor=north east] {Grammar Learning} %;
        cycle;
    \draw[bddashedbox={blue}] %
        ($(in2-centernode.south -| tribble.west) + (-\dmarg,-\dmarg)$) -- %;
        ($(tribble.west |- in1-centernode.north) + (-\dmarg,\dmarg)$) -- %;
        ($(in1-centernode.north east) + (\dmarg,\dmarg)$) --
        ($(in2-centernode.south east) + (\dmarg,-\dmarg)$) -- node[pos=.5,bddashedcaption={blue},anchor=south] {Input Generator}
        cycle;
    \draw[bddashedbox={orange}] %
        ($(program.west |- crawledfiles-centernode.north) + (-\dmarg,\dmarg)$) -- %;
        ($(program.south west) + (-\dmarg,-\dmarg)$) -- %;
        ($(phantom4.south east) + (\dmarg,-\dmarg)$) --
        ($(phantom4.east |- crawledfiles-centernode.north) + (\dmarg,\dmarg)$) node[pos=1,bddashedcaption={orange},anchor=north east] {Input Selector} --
        cycle;
    \end{pgfonlayer}
\end{tikzpicture}
\caption{Overview of our approach (\approach)}\label{approachoverview}
\end{figure*}
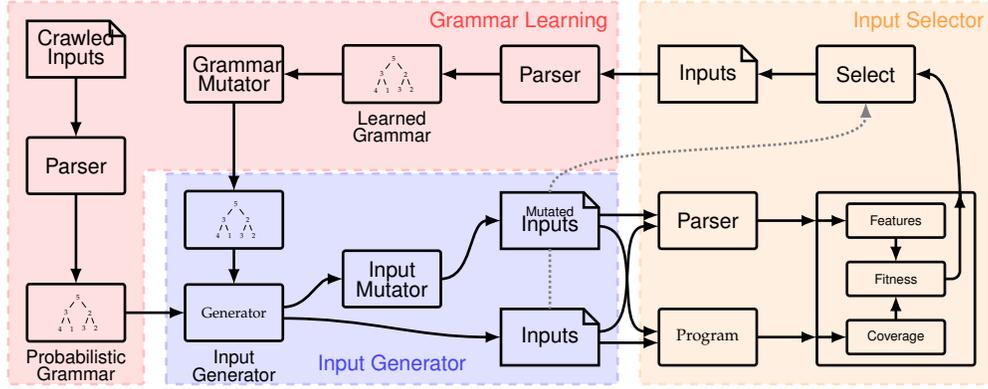

\section{\approach Approach}
\label{sec:approach}

\subsection{Approach workflow}\label{infrastructure}

In \autoref{approachoverview}, we give a high-level overview of the \revise{workflow} of our approach.
The evaluation starts with crawled real-world seed inputs (top left) that are used to create the probabilities for the first generation's starting grammar (bottom left).
An input generator then uses the grammar to generate one set of inputs.
This set is then duplicated and one of the two resulting sets is mutated (bottom middle).
Both sets of inputs are processed by a parser and fed into the subject program, which gives the necessary feedback that the algorithm needs to select the best-performing inputs, i.e., the code coverage, the number of features, the subject run time (bottom right).
An input selector then selects the best-performing inputs and feeds those back into a grammar learning tool (top middle) and a grammar mutator which mutates the grammar probabilities.
The resulting grammar is used in the next generation and the whole process continues with the input generation step.
\par
Generally, our toolset can be split into three main groups of tools that work together to evolve inputs, as seen in \autoref{approachoverview} in three different shade colors:
\begin{enumerate}
    \item A grammar learning tool and a grammar mutator (the {\color{red}red} shaded area in \autoref{approachoverview}).
        This toolset has the task of learning probabilities from the distributions of grammar expansions that are present in a set of inputs and mutating the learned probabilities in the generated grammar.
    \item An input generator and mutator (the {\color{blue}blue} shaded area).
        The input generator generates $n$ inputs using a fuzzer that makes use of the probabilities that were learned in the grammar learning tool.
        Then, the generated set of inputs is duplicated and the duplicated set is mutated using parse tree swaps and bit flips.
        Both sets of inputs are then further processed in the input selector step, i.e. there are now $2n$ inputs in total.
    \item An input selector (the {\color{orange}yellow} shaded area).
        All the $2n$ inputs are fed into a parser and into the subject program.
        From those tools, the input selector gets the feedback it needs about the fitness of each run, i.e. input features, program status, run time, coverage, \revise{etc.}
        The fitness of each input is then calculated and the best-performing input is selected to generate the grammar for the next generation.
\end{enumerate}
To generate and evolve inputs, the whole approach can be run for an arbitrary number of generations, until a goal is reached, or for a certain number of generations.
%\todo{ES: Better change the colors in a way that can be easily read on a black/white print?}

\subsection{Mapping (Input-to-code-complexity)}\label{mapping}

%\todo{Reviewer 1:An example is given of input-to-code complexity based on some tree of features, but the general form is not described or formalized. nzdigit appears undefined}
%
To evaluate the complexity of our inputs, we propose a mapping from features that are present in the input set to executed methods in the subject program.
This mapping serves as an enhanced test coverage metric
%asurement 
that takes into account which 
%accounts for the 
input feature triggers which function/method.
%program element(s) it exercises.
% feature of the input triggered which program element.
\rev{We map rule types to executed methods, both individually and in combination, to capture more context for each mapped grammar rule in the mapping.
This way, we do not only capture which rule triggers which executed part of the program, but also the context in which each rule was selected in the grammar.}

%\todo{R3: I suggest providing justification for selecting the grammar rule types as the feature. Additionally, it is unclear why these rule types are mapped to the executed lines both individually and in combination.}

\subsubsection{Formalization}
%
%\todo{consider fomalization (see revise in blue) before example}
%
\lrevise{
%Let us assume 
Consider that 
%we have a set of executed 
a set of functions $E$ were executed 
%during the program execution of the 
for a test input with 
% that were analyzed to generate the 
feature set $F$.
The mapping $M$ is then defined as the cross product $M := F \times E$, i.e. it is a set of 2-tuples $(f,e) \in (F \times E)$, where each input feature $f$ is mapped to a function $e$.
That is,  all subsets of the parse tree of a test input are mapped to the functions that were executed when running the input on the subject program.
}

\subsubsection{Example}
\lrevise{Let us illustrate our mapping with 
% approach with 
an example.  
Consider 
%we assume 
that an example set of LOCs shown in \autoref{mapexploc} is executed. For simplicity, we employ the executed line numbers $\mathit{2}$ and $\mathit{3}$ in this example, however, we note that \approach maps functions instead.
\approach uses ANTLR~\cite{antlr} to extract features from the test input using the \texttt{getRuleIndex()} method.  
For each node of the resulting parse tree, 
%This method
%that 
\approach extracts the index of the parser rule that was applied to the token stream of the input.
\autoref{mapexpmap} presents a subset of the mapping that would result from the examples in \autoref{fig:complexfeatureset}. 
%can be seen in .
}\par\lrevise{
Assume that the sample \texttt{euclid()} input seen in \autoref{fig:examplejson} is processed by \approach using the program from \autoref{exampleprogram}.
First, the feature set, i.e. the information about all subtrees of up to a depth of $d$, is generated.
% (in our experiments we use $d=3$ to avoid memory problems).
\autoref{fig:ruleindices} shows the index for each grammar feature (non-terminals). 
%parse rule. 
The parse tree (\autoref{mapexptree}) shows that the start rule \texttt{start} (index 1) has one child \texttt{integer} (index 2) which has two children \texttt{digit} (index 4) with one of them expanding to \texttt{nzdigit} (index 5).
Therefore, the resulting executed lines of code and features with a maximum tree depth of $d=3$ are}
\begin{align*}
    E = \{&2,3\}\\
    F = \{&\{1\},\{2\},\{4\},\{5\},\{1,2\},\{1,2,4\},\\
      &\{1,2,4,5\},\{2,4\},\{2,4,5\},\{4,5\}\}
\end{align*}
\lrevise{%
This results in the mappings $M = F \times E$ shown in \autoref{mapexpmap}.%
}
\par

\forestset{parsetree/.style={for tree={parent anchor=south,child anchor=north,l sep-=2mm,l=0}}}
\newsavebox\jsontree\savebox\jsontree{\small{\begin{forest}
            parsetree
            [start
                [integer
                    [digit
                        [nzdigit]
                    ]
                    [digit]
                ]
            ]
\end{forest}}}
\newsavebox\jsontreemut\savebox\jsontreemut{\small{\begin{forest}
            parsetree
            [start
                [integer
                    [digit]
                    [digit
                        [nzdigit]
                    ]
                ]
            ]
\end{forest}}}

\begin{figure}[tbp!]
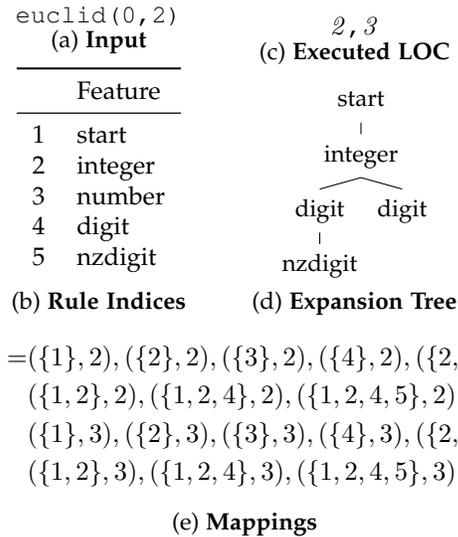

    \begin{minipage}[b]{3.4cm}%
        \begin{minipage}[b]{\textwidth}%
            \centering \texttt{euclid(0,2)}\\[-0.5em]%
            \subcaption{\textbf{Input}}\label{fig:examplejson}%
        \end{minipage}%
        \vspace{.5em}%

        \begin{minipage}[b]{\textwidth}%
        \centering
            \begin{tabular}{ll}%
                \toprule ~&Feature\\%
                \midrule 1&start\\2&integer\\3&number\\4&digit\\5&nzdigit\\%
            \end{tabular}\\%
            \subcaption{\textbf{Rule Indices}}\label{fig:ruleindices}%
        \end{minipage}
    \end{minipage}%
    %\hspace{3em}%
    \begin{minipage}[b]{3.4cm}%
        \begin{minipage}[b]{\textwidth}%
            \centering%
            \texttt{$\mathit{2}$,$\mathit{3}$}\\[-.5em]%
            \subcaption{\textbf{Executed LOC}}\label{mapexploc}%
        \end{minipage}%
        \\[.5em]%
        \begin{minipage}[b]{\textwidth}%
            \centering%
            \usebox\jsontree\\[-.5em]%
            \subcaption{\textbf{Expansion Tree}}\label{mapexptree}%
        \end{minipage}
    \end{minipage}\\%
    %\hfill%code coc
    %\begin{minipage}[b]{1cm}$\to$\\[4em]~\end{minipage}%
    %\begin{minipage}[b]{2.5cm}\centering\begin{tabular}{>{\ttfamily}l}\{1\}$\to\mathit{2}$\\\{1\}$\to\mathit{3}$\\\{1,2,4,5\}$\to\mathit{2}$\\\{1,2,4,5\}$\to\mathit{3}$\\\{1,2,4\}$\to\mathit{2}$\\~~~~$\vdots$\end{tabular}
    \begin{minipage}[b]{0.4\textwidth}%
    \centering
\begin{align*}
    M = & (\{1\},2),  (\{2\},2),  (\{3\},2),  (\{4\},2), (\{2,4\},2),\\
        & (\{1,2\},2),  (\{1,2,4\},2),  (\{1,2,4,5\},2),(\{4,5\},2),\\
        & (\{1\},3),  (\{2\},3),  (\{3\},3),  (\{4\},3), (\{2,4\},3),\\
        & (\{1,2\},3),  (\{1,2,4\},3),  (\{1,2,4,5\},3),(\{4,5\},3)\\
\end{align*}
\\[-2em]\subcaption{\textbf{Mappings}}\label{mapexpmap}\end{minipage}
    \caption{
    \lrevise{The resulting mappings from evaluating the example input from \autoref{fig:examplejson}.
        For simplicity,  this example illustrates mappings by showing executed line numbers.  However, we note that \approach uses executed functions.}
    }\label{fig:complexfeatureset}
\end{figure}

\subsection{Fitness Functions}\label{fitness}\label{fitnessfunction}
In each step of the input evolution, the program selects the best-performing inputs based on a fitness function.
Our fitness functions aim to maximize the following feedbacks:
\begin{enumerate}
    \item Code \textit{coverage} (\rev{functions covered})
    \item The number of \revise{triggered \textit{exceptions}}  
    \item \revise{The number of \textit{unique exceptions}}
%    (failures/crashes)
    \item Long execution time (\textit{run time})
%    Worst-case execution time
    \item The number of unique \textit{mappings} (\textit{see} Section \ref{mapping})
\end{enumerate}
\newcommand\fitfcc{\textit{Code Coverage}\xspace}
\newcommand\fitfnfc{\textit{Functions Covered}\xspace}
\newcommand\fitfnemm{\textit{Number of Mappings}\xspace}
\newcommand\fitfute{\textit{Number of triggered unique Exceptions}\xspace}
\newcommand\fitfrt{\textit{Program Run-Time}\xspace}
\newcommand\fitfte{\textit{Number of triggered Exceptions}\xspace}
By maximizing code coverage without any mappings, we are able to compare our work against most other work done in the topic of input generation, since most use code coverage as performance metric.
Additionally, the mappings are an important metric in this work, because they combine code coverage and input features in a way that newly covered functions can be easily assigned to input features that are responsible for triggering code coverage.
This way, we can not only increase code coverage, but also discover different features of the input that are responsible for triggering the same line of code.
\par
\revise{
To target \textit{new} exceptions in the subject program,  we count both the number of exceptions thrown by the subject program and the number of unique exceptions.}
%.  We aim to 
%Since we want to focus 
%on triggering new exceptions.}
% the total numbers are weighed (i.e. multiplied by) 10% for the number of triggered exceptions and 90% for the number of unique exceptions, respectively
Additionally, we collect and show all triggered unique exceptions in a table in the result reports of the evaluation.
To generate inputs that cause worst-case execution time to increase in the subject program, we measure the program execution time.

%\todo{R1: motivate the choice of scalarization vs. many-objective optimization: At the top of my head,  scalaraization is a better approach, which is tunable to achieve our three different testing strategies, generation modes....  (i.e.,  single,  multiple and ignore ) \\ ``authors choice of scalarization should be motivated in light of such works. It should be discussed if there are specific reasons for the choice.''
%}

\subsubsection{Weighted Sum}

\rev{
To support multiple testing goals and different testing strategies,  \approach's fitness function employs \textit{weighted sum scalarization} as its many-objective optimization algorithm.  This method combines multiple test feedbacks into a single scalar value.  
This is particularly useful in our goal-directed testing scenario where the developer has a preference for targeted testing goal(s), i.e,.  wants to focus on (or ignore) specific testing goal(s).  In this setting,  scalarization allows for finding a single optimal solution that balances the developer's goal preferences.  In particular, scalarization allows \approach to be tunable by developers to support the three different testing strategies, i.e.,  \textit{single} goal mode,  \textit{multiple} goal mode,  and \textit{ignore} goal mode (\textit{see} Section \ref{strategies}).  
For instance,  \approach supports \textit{single goal mode} by setting a high weight value (10) for the relevant test feedback (e.g., exception) and a low weight value (one (1)) for all other test feedback (e.g.,  coverage and runtime).  It supports \textit{multiple goal mode} by setting all (four) test feedback weights to the same value (one (1)), and supports \textit{ignore goal mode} by setting the feedback weight of the ignored goal to zero (0) and all other feedback to one (1). 
}

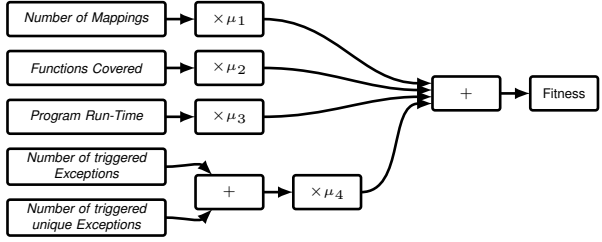
\begin{figure}[tbp!]
    \centering
    \newlength\wssep\setlength\wssep{2mm}
    \newcounter{varcnt}\setcounter{varcnt}{1}\newcommand\mulpli[1]{\node[bdslimblockxm,right=2\wssep of #1] (#1{}mulpli) {$\times \mu_\arabic{varcnt}$};\stepcounter{varcnt}}
    \begin{tikzpicture}[bdslimblockx/.style={bdslimblock,minimum width=2cm,text width=2cm,inner sep=.5mm,outer sep=0},bdslimblockxm/.style={bdslimblockx,minimum width=1mm,text width=8mm},node distance=\wssep]
        \node[bdslimblockx] (m1) {\fitfnemm};
        \node[bdslimblockx,below= of m1] (m2) {\fitfnfc};
        \node[bdslimblockx,below= of m2] (m3) {\fitfrt};
        \node[bdslimblockx,below= of m3] (m9) {\fitfte};
        \node[bdslimblockx,below= of m9] (m5) {\fitfute};
        \node[bdslimblockxm,anchor=west] at ($(m9.south east)!0.5!(m5.north east) + (2\wssep,0)$) (m4) {$+$};
        \mulpli{m1}
        \mulpli{m2}
        \mulpli{m3}
        \mulpli{m4}
        \draw[draw=none] (m1.north east) -- (m9.south east) node[pos=.5,bdslimblockxm,xshift=20\wssep] (addnode) {$+$};
        \multiconnectnodewest{addnode}
        \node[bdslimblockxm,right=2\wssep of addnode] (result) {Fitness};
        \foreach \i in {1,2,3,4}{
            \draw[bdarrow] (m\i) to[out=0,in=180] (m\i{}mulpli);
            \draw[bdarrow] (m\i{}mulpli) to[out=0,in=180] (addnode-west4\i);
        }
        \draw[bdarrow] (m5) to[out=0,in=225] (m4);
        \draw[bdarrow] (m9) to[out=0,in=135] (m4);
        \draw[bdarrow] (addnode) to[out=0,in=180] (result);
    \end{tikzpicture}
    \caption{The weighted sum fitness function of \approach. Each number is multiplied by a weight $\mu_{1\dots 4}$. One or multiple metrics may be focused by increasing their weight.}\label{weightedsumfitf}
\end{figure}

\smallskip
\noindent
\textbf{Fitness Function Computation:} The fitness function that the feedback loop maximizes is calculated as follows:
Each feedback $x_n$ is multiplied by a factor $\mu_n$.
All the results are then added up into a fitness $F$ that describes the performance of an input which will be maximized by the feedback loop.
A flowchart of the fitness function can be seen in \autoref{weightedsumfitf}.
\par

To be able to assign weights by changing the multipliers $\mu_1,\dots,\mu_4$, all of the metrics need to be normalized to the interval $[0,1]$ in the following way:
\begin{itemize}
\item \fitfnemm: 
%\todo{Under the weighted sum section and number of mappings, the \textbf{meaning of x} and the \textbf{choice of sigmoid function needs justification}. How the \textbf{pre-determined maximum} is obtained is not described.}
    We normalize this metric to $[0,1]$ using the sigmoid function $\frac{x}{1+\left|x\right|}$\lrevise{, with $x$ being the number of mappings}.
    \lrevise{We employ the sigmoid function since our preliminary  expreriments show that the sigmoid function performs best,  resulting in the highest improvements.} 
%     in our experiments
%    was experimentally determined to 
%result in the greatest improvements. 
% in our experiments.}
    The numbers returned by the non-normalized metric can become huge for subjects with a large grammar size.
    %For instance, in previous experiments, there were cases with more than $600\,000$ mappings even for a subject using a grammar as small as JSON.
    Therefore, we make up for the precision problems that result in feeding such huge numbers into the sigmoid function by dividing the number with a quotient determined by hand.
    %We found $24\,000$ to be a reasonable upper bound for this quotient, which is the number we experimentally determined before the evaluation and which produced reasonable results.
    Since we still use a sigmoid function, we only need to determine an approximate upper bound here.
    If the number of mappings increases to a much higher number than the determined quotient, the calculations will still work, but have less precision.
    Our sigmoid function, making use of that upper bound, returns an almost linear result in the interval $[0,max]$ that is inside the range $[0,\frac{2}{3}]$ and in case our metric returns a higher number than the pre-determined maximum, we still get a number in the interval $(\frac{2}{3},1)$.
    \lrevise{The maximum value was determined by running the subject programs on a random set of inputs and inspecting the number of mappings that were produced during those runs.}
%    If we are completely off calculating the maximum, we might run into precision problems, though.
\item \fitfnfc: To normalize the number of new functions covered, the number of new functions covered is simply divided by the total number of functions.
    Since there will never be more new functions covered than there are functions in total, this will never be greater than one (1).
\item \fitfrt: Likewise, we normalize this fitness function by dividing its result by the timeout we set for each subject program run.
    The program should never run longer than the timeout, and therefore if a timeout occurs, the normalized fitness function should return $1.0$.
\item \fitfte: \lrevise{Since each subject program run can only throw one exception at a time, we divide the number of exceptions by the number of program runs (test executions).  
We note that only one 
% are at most one 
%This is a sensible divisor because there can never be more than 
%one 
exception can be thrown in one test execution.} %
    %\newrevise{(which is the number of test files per run).\footnote{We note that ...  \\
%For instance ... } }
%\todo{For the fitness functions, the number of test files does not seem to be a sensible divisor to normalize by when accounting for the number of triggered exceptions}
    To take account of unique exceptions, we add both the total and unique number of exceptions before normalization.
    \revise{Since we want to focus on triggering \textit{new} exceptions, the total numbers are weighed (i.e. multiplied by) 10\% for the number of triggered exceptions and 90\% for the number of unique exceptions, respectively.
    The rationale is that rewarding unique exceptions more increases the likelihood of triggering \textit{new} exceptions. }
\end{itemize}

Therefore, the fitness will be calculated as follows:
\begin{center}
\begin{tabular}{cccc}
    $x_1 = \frac{\frac{a}{a_{max}}}{1+\left|\frac{a}{a_{max}}\right|}$&$x_2 = \frac{b}{b_{tot}}$&$x_3 = \frac{c}{\text{timeout}}$&$x_4 = \frac{0.1d + 0.9e}{\text{inputs}}$\\[1em]
\multicolumn{4}{c}{$\text{Fitness } F = \mu_1x_1 + \mu_2x_2 + \mu_3x_3 + \mu_4x_4$}\\
\end{tabular}
\end{center}
with $a$ being the number of mappings, $a_{max}$ the maximum number of mappings that has been determined before the experiment, $b$ the number of covered functions, $b_{tot}$ the total number of functions in the program, $c$ the program run-time, $timeout$ the timeout of each subject program run, $d$ the number of exceptions, $e$ the number of unique exceptions and $inputs$ the number of inputs in each test run.

\subsection{Gene Representation}

\subsubsection{Grammar Probability Representation}
%\todo{R1: The gene representation used by FDLoop is unspecified.}

\noindent
\textbf{Formalization:}
\lrevise{\approach improves the 
%target achieved 
test suite in each generation by mutating the grammar probabilities. 
%such that the resulting input performs better in the next generation.  
Specifically,  
%of the resulting inputs in each generation by mutating 
we mutate the probabilities of the probabilistic grammar in each generation of \approach without changing the grammar's tree structure.  To this end,  we employ the following two-dimensional real-valued array as the genetic representation for the probabilistic grammar:
% can be defined as a two-dimensional real-valued array 
$$[[p_{0,0},p_{0,1},\dots],[p_{1,0},p_{1,1},\dots],\dots]$$ 
with $p_{i,j}$ being the probability of the $j$-th grammar alternative  under the $i$-th grammar expansion rule. 
% and $J$  is the total number of alternatives.
}

%When mutating a grammar, it randomly selects a given percentage of grammar expansion tree nodes and changes the probabilities of all children to a uniform distribution.
\lrevise{
\approach uses grammar probability mutations that achieve a uniform distribution.  
The grammar mutator only mutates the probabilities of the grammar and does not change the grammar expansions themselves to preserve all grammar features.
Let $J$  be the total number of alternatives for a specific rule. 
%That is,  w
When mutating the grammar, an expansion rule with the index $i_0$ is randomly selected and all probabilities $p_{i_0,j}$ are changed to a constant value $c=1/J$, such that 
$$p_{i_0,0} = p_{i_0,1} = \dots = p_{i_0,J-1}  = c$$ 
%i.e.,  select a constant $c$ such that 
%is equal to 
%$$c=1/J$$, 
and sum of all possible alternatives (or terminals) is one (1):
%, i.e., 
$$\textstyle \sum_0^{j}(p_{i_0,j}) = 1$$
%.
}

%\par
%\skip
\noindent
\textbf{Example:}
\lrevise{%
    For example, the learned grammar in \autoref{examplegrammar} is genetically represented by the following real-valued array
%     ( representation) 
    \\
%    \texttt{
 $$[[1.0], [0.04 | 0.96], [0.74 | 0.26], [0.09 | 0.91], \dots]$$
%    }.
    We show an example grammar mutation in \autoref{approachbetter}, where the grammar mutator randomly selects the \texttt{digit}  rule three ($3$) with probabilities 
%    \texttt{
$$[0.09 | 0.91]$$
and changes the distribution of the rule to a uniform distribution, i.e.,  the probability of both alternatives is \texttt{0.5}. 
}\par

\lrevise{
%Formally, 
 \autoref{approachbetter} shows the resulting probabilities.   
%Formally,  i
In this example,  
%For instance,  in the mutation shown in \autoref{approachbetter},  
$i$ is the index of expansion rule \texttt{digit} (index three (3)), $J=2$,  and $c=0.5$.  
%In this example.,  
The probability of the \texttt{digit} rule before mutation ($p$) is $p_{3,0}= 0.09$ and  $p_{3,1}= 0.91$. 
After mutation ($p^\prime$),  
\approach makes the probability distribution of the 
\texttt{digit} rule uniform,  such that  $p^\prime_{3,0}= 0.5$ and $p^\prime_{3,1}= 0.5$.  
%Intuitively,  this probability makes a skewed distirbution for the \texttt{digit} rule uniform after mutation. 
}

\subsubsection{Input Representation}
\noindent
\textbf{Formalization:}
\lrevise{The generated inputs are represented as bit arrays (i.e., text encoded in UTF-8).
\approach uses mutation operators that work on the generated inputs like bit-flips and parse tree mutations.
In a bit-flip, a random bit from the bit array is selected and then filpped from 0 to 1 or vice-versa.
}

\begin{figure}
{\footnotesize
    \begin{tikzpicture}[remember picture,node distance=0.4,every node/.style={inner sep=0,outer sep=0,font=\ttfamily},explain/.style={font=\sffamily\tiny,anchor=east, outer sep=2mm},mutarrow/.style={-latex,shorten >=1mm,shorten <=1mm},mutmark/.style={fill=lightgray,draw=lightgray,line width=1.5mm}]
        \node (inpnode) {euclid(21,\subnode{unmutSix}{4})};
        \node[below=of inpnode] (ascii) {[101 117 99 108 105 100 40 50 49 44 \subnode{lparenAscii}{52} 41]};
        \node[below=of ascii] (bitarray) {[$\dots$ 110001 101100 \subnode{lparenBinary}{110\underline{\textbf{1}}00} 101001]};
        \node[below=of bitarray] (flippedbitarray) {[$\dots$ 110001 101100 \subnode{lparenFlipped}{110\underline{\textbf{0}}00} 101001]};
        \node[below=of flippedbitarray] (flippedascii) {[101 117 99 108 105 100 40 50 49 44 \subnode{lparenMut}{48} 41]};
        \node[below=of flippedascii] (minpnode) {euclid(21,\subnode{mutSix}{\underline{\textbf{0}}})};

        % Explain:
        \node[explain,align=right] at (inpnode -| ascii.west) {\textbf{Original}\\\textbf{Input}};
        \node[explain,align=right] at (ascii -| ascii.west) {\textbf{Original}\\\textbf{Bytes}};
        \node[explain] at (bitarray -| ascii.west) {\textbf{Bits}};
        \node[explain,align=right] at (flippedbitarray -| ascii.west) {\textbf{Flipped}\\\textbf{Bits}};
        \node[explain,align=right] at (flippedascii -| ascii.west) {\textbf{Mutated}\\\textbf{Bytes}};
        \node[explain,align=right] at (minpnode -| ascii.west) {\textbf{Mutated}\\\textbf{Input}};

        % Arrows:
    \begin{pgfonlayer}{background}
        \draw[mutmark] (unmutSix.north west) rectangle (unmutSix.south east);
        \draw[mutmark] (lparenAscii.north west) rectangle (lparenAscii.south east);
        \draw[mutarrow] (unmutSix) -- (lparenAscii.north west);
        \draw[mutmark] (lparenBinary.north west) rectangle (lparenBinary.south east);
        \draw[mutarrow] (lparenAscii.south west) -- (lparenBinary);
        \draw[mutmark] (lparenFlipped.north west) rectangle (lparenFlipped.south east);
        \draw[mutarrow] (lparenBinary) -- (lparenFlipped);
        \draw[mutmark] (lparenMut.north west) rectangle (lparenMut.south east);
        \draw[mutarrow] (lparenFlipped) -- (lparenMut.north west);
        \draw[mutmark] (mutSix.north west) rectangle (mutSix.south east);
        \draw[mutarrow] (lparenMut.south west) -- (mutSix);
    \end{pgfonlayer}
    \end{tikzpicture}
    }
    \caption{\lrevise{Mutating a generated example from \autoref{approachbetter} using a random bit-flip.}
    }
    \label{fig:bitflip}
\end{figure}
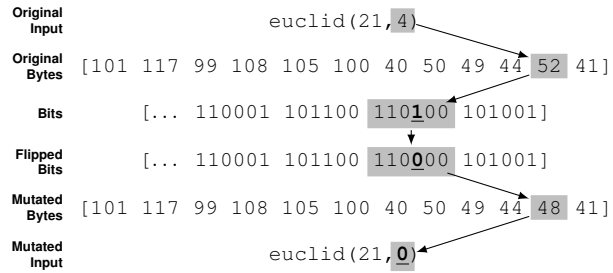

\noindent
\textbf{Example:}
\lrevise{
\autoref{fig:bitflip} shows an example where the generated input from \autoref{approachbetter} is mutated using a random bit-flip.}

%\todo{formalize the gen representation \\
%- discuss how the input files are represented as binary arrays (hence, bit flips) and \\
%- how they are represented as probabilistic parse trees  \\
%	* a parse tree structure (but never mutated, we want onlyvalid inputs) \\
%	* probabilities of input (tree) features , represented as a real-valued array (hence, mutation of probabilities) \\
%}

\subsection{Mutators}

%\todo{R1: Parse tree swap mode needs to be further explained, justified, and may need an example. The authors should consider providing a formal specification of parse tree swap mode}

%\todo{specify given percentages for swaps}

%\todo{formalization}
%
%\todo{ES Parse Tree Swap = similar to ``Introduce Choice'' mode of 
%\url{https://dl.acm.org/doi/abs/10.1145/3605157.3605170}
%?
%}
\rev{
We employ bit-flip as an input mutation operation.  This is a 
%As mutation operators, we chose Bit-Flip as input mutator operator - a 
commonly used mutation operator in state-of-the-art fuzzers~\cite{DBLP:journals/corr/abs-1812-00140}.  For grammar mutations, we chose a parse tree swap because it allows us to easily mutate valid input parse trees in a way that the mutated parse tree still produces a valid input~\cite{10.1145/3708517, fuzzingbook2024}.
This is due to the fact that we only swap parse trees with the same grammar rule index as the root node.  For instance,  our parse tree swap mutation is similar to the ``Introduce Choice'' mode in previous works~\cite{10.1145/3708517}. 
}

\subsubsection{Formalization}

%\todo{consider fomalization (see revise in blue) before example}

The mutators we use in the input generation are split into a \emph{grammar mutator} and an \emph{input mutator}.
The input mutator first parses each input into a parse tree.
Then, a mutation mode is selected for each input - the mutator supports \emph{bit flip} and \emph{parse tree swap}.
In the \emph{bit flip} mode, numbers and strings are randomly changed in a given percentage of leaf nodes in the parse tree of the input \lrevise{(cf. \autoref{fig:bitflip})}.
\lrevise{The \emph{parse tree swap} mode swaps 
%a given percentage of 
subtrees in the parse tree of the input \textit{before} the input's parse tree is flattened (pretty-printed).
Swapping two parse subtrees for a grammar expansion $e$ means randomly selecting two distinct parse subtrees that have $e$ as root node and swapping them inside the parse tree of the generated input. 
This ensures that the resulting mutated input is syntactically valid. }
\par

\begin{figure}[t]
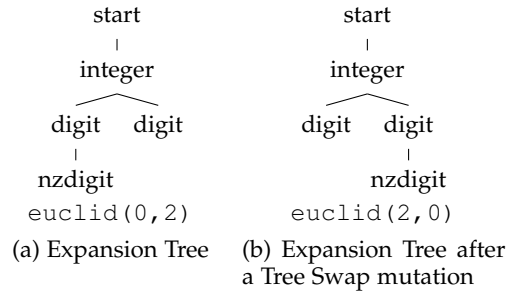

        \begin{minipage}[b]{3.5cm}%
            \centering%
            \usebox\jsontree\\[-.0em]\texttt{euclid(0,2)}%
            \subcaption{Expansion Tree\\\strut}\label{mapexptreeUnmut}%
        \end{minipage}%
        %\hfill%
        \begin{minipage}[b]{3.5cm}%
            \centering%
            \usebox\jsontreemut\\[-.0em]\texttt{euclid(2,0)}%
            \subcaption{Expansion Tree after a Tree Swap mutation\strut}\label{mapexptreeMut}%
        \end{minipage}%
        \caption{\lrevise{Applying a parse tree swap mutation swaps two subtrees with the same root node inside the parse tree.}}\label{fig:parsetreeswap}
\end{figure}

\subsubsection{Examples}
\textbf{Bit-Flip Example:} \lrevise{We show an example of mutating the input \texttt{euclid(21,4)}
    %from \autoref{approachbetter} 
    using bit flips in \autoref{fig:bitflip}.
In this example, the digit \texttt{4}, which is \texttt{52} in ASCII (\texttt{110100} in binary), is mutated using a bit flip, i.e. \texttt{110\textbf{\underline{1}}00} becomes \texttt{110\textbf{\underline{0}}00}, which translates back to the digit \texttt{0}.}\par

\noindent
\textbf{Parse-tree Swap Example:} \lrevise{
A parse tree swap is shown in \autoref{fig:parsetreeswap}, where the input \texttt{euclid(0,2)} is first parsed into the derivation tree shown in \autoref{mapexptreeUnmut}.
Then, two random nodes with the same root rule are selected (\texttt{digit}) and swapped, resulting in the parse tree shown in \autoref{mapexptreeMut}.}

\subsection{Algorithmic Complexity}

%%%%%%%%%%%%%%%%%%%%%%%%%%%%%%%%%%%%
The complexity of the \approach's
test input generation 
%(PTC2) 
is 
%effectively 
linear under reasonable constraints, being in $\mathcal{O}(r_{mean} + b_{max})$ with the average tree size $r_{mean}$ and the maximum arguments to a nonterminal $b_{max}$~\cite{ptc2}.
%~\footnote{Since \approach is built on 
%%We use 
%\emph{Tribble},  its input generator has the same time complexity as \emph{Tribble} ~\cite{8952419}. } 
% as input generator, 
%These generated files are then mutated (\autoref{alg:mut}), which doubles the number of generated files in $T$.
ANTLR - the parser generator \approach uses for parsing inputs - has a worst-case complexity of $\mathcal{O}(n^2)$~\cite{llstar}.
Selecting and mutating parse tree nodes in the next step is done in $\mathcal{O}(k * \log(n))$ with the number of mutations $k$.
Finally, the parse tree needs to be un-parsed, i.e., transformed back into an input, which is in linear time $\mathcal{O}(n)$ (each parse tree node needs to be visited exactly once).
%\par
After input generation,  each input is parsed in ANTLR, ran through the subject program and evaluated through a fitness function.    % fed into the subject program (\autoref{alg:instart}-\autoref{alg:inend}).
%This parser again uses ANTLR with a complexity of $\mathcal{O}(n^2)$~\cite{llstar}.
%Based on the results that are collected from the subject program and the parse tree, the best input is selected using a fitness function that takes those results as input(\autoref{alg:sel}).
The fitness function has a complexity of $\mathcal{O}(1)$, since it only multiplies the existing results with given factors (\textit{see}  Section \ref{fitnessfunction}).%}\par\revise{
%The selected input is added to the set of inputs $I$ and new grammar probabilities are learned and stored in the grammar for the next iteration (\autoref{alg:lrn}).
%The learning step involves parsing all input files with a total length of $n$ with ANTLR ($\mathcal{O}(n^2)$) and flattening the resulting parse tree with the annotated probabilities which is in $\mathcal{O}(\log(n) * b_{avg})$ with $b_{avg}$ as the average number of expansions for each grammar rule that is annotated with a probability.
%This grammar learning approach is similar to the approach by Eberlein et al.~\cite{eberlein2020evolutionary}.
%As grammar learning tool, we modified the ANTLR counting grammars presented in \emph{Inputs from Hell} by Soremekun et al.~\cite{9154602}}\par\revise{
%In the last step, the new grammar is mutated (\autoref{alg:gmu}), before being used to generate new input files in the next iteration - this is in $\mathcal{O}(n^2 + (k+1)\log(n))$ since the grammar is first parsed with ANTLR, then $k$ random probabilities in the parse tree are mutated and the parse tree is flattened to a file.
%After a fixed number of iterations,the set of generated inputs $I$ is returned.

\section{Experimental Setup}\label{evaluationsetup}

In this section, we discuss the experimental setup for the evaluation of % the effectiveness of 
our approach (\approach).

\subsection{Research Questions}\label{rqs}
We pose the following research questions: % are the following:

\begin{enumerate} %[label=\textbf{RQ\arabic{enumi}}]
\item \rqone
\item \rqtwo
\item \rqthree
\item \rqfour
%\item \rqfive
\end{enumerate}

%A detailled description of the research questions and overview of our goals can be found in \autoref{rqdefinitions}.

%\todo{R1: motivate the choice of scalarization vs. many-objective optimization: briefly justify why scalarization supports our three testing modes}

\subsection{Test Generation Modes}\label{strategies}

%\revise{%In this experiments, 
\approach operates in %is evaluated using 
three testing modes, namely {\em single goal mode}, {\em multiple goals mode} and {\em ignore goal mode}. 
%To achieve a specific testing mode, e
Each mode employs a different weight in the described fitness function (\autoref{weightedsumfitf}) to focus or ignore one or more specific test feedback(s)/goal(s). 
%In this work, a
All our experiments (RQ1 to RQ4) are performed in both the {\em single goal mode} and {\em multiple goals mode}. The  {\em ignore-goal mode} is \textit{only} employed for the ablation study (RQ3) to determine the contribution of the ignored goal to the performance of the approach. 
%Throughout the experiments, we use four strategies that each employ different weights in the fitness function:
%}
The following describes each testing mode supported by \approach: 

\begin{enumerate}
    \item \textbf{Single Goal Mode}.
        To focus on one single testing goal, \approach employs a \textit{goal-specific} fitness function across generations. 
        Specifically, \approach sets the weight of the test feedback corresponding to the goal-at-hand to a high value 
%        one of the fitness function's weights to 
        ($10$) while setting the weights of the other (three) test feedbacks  to a low value (one). 
%        This way, 
        The goal is to make \approach %the fitness score focuses 
        focus on the feedback that should be maximized. 
%        and still improves the other test feedbacks when the focused one stays the same between two generations.
%        \par
        For example, given that the single testing goal-at-hand is code coverage, 
%        focusing on code coverage using the dampening strategy, 
        \approach sets a high weight for code coverage ($\mu_2 = 10$) and the weights for all other test feedbacks (e.g., exception) is set to $\mu_1 = \mu_3 = \mu_4 = 1$. In this work, we consider this testing mode for all experiments (RQ1 to RQ4). 
        %The exact definition of the fitness function and its weights is stated in \autoref{fitness}.
        
%    \item \textbf{Single weighing strategy}.
%        For this strategy, the fitness function is only focused on one certain feedback, i.e., the focused feedback gets a weight of one while all other feedbacks are set to a weight of zero.
%        This way, we only focus on the feedback that is to be improved.\par
%        Setting all weights except for one to zero effectively transforms the fitness function into the maximized feedback alone.
%        For example, focusing on coverage using the single weighing strategy means setting $\mu_2=1$ and $\mu_1=\mu_3=\mu_4=0$.
%        Looking at the definition of the fitness function, we get $F = 0x_1 + 1x_2 + 0x_3 + 0x_4$ which can be simplified to $F = x_2 = \frac{b}{b_{tot}}$, inserting the definition from \autoref{fitness}.

    \item \textbf{Multiple Goals Mode}.
    To simultaneously target multiple testing goals, 
    \approach employs an \emph{equally weighted} fitness function which sets equal weights for all test feedbacks
%    .             The fourth strategy uses the \emph{equally weighted} fitness function where all weights are set to 
        (e.g., all set to $1$).
        This strategy does not target one specific testing goal, instead 
%         certain goal, but equally focuses on all four feedbacks.
%        For 
        it enables a developer to generate a test suite that targets/achieves numerous goals simultaneously.  
%        , this might be the most convenient option for generating inputs.
%        Equally targeting all feedbacks means that the developer ends up with 
%        a single test suite to target most goals, instead of one test suite for each goal.
        For the fitness function, this means setting all weights to one ($\mu_1 = \mu_2 = \mu_3 = \mu_4 = 1$) which results in a 
%         resulting in a 
fitness function $F = x_1 + x_2 + x_3 + x_4$. This mode is evaluated for all experiments (RQ1 to RQ4).

    \item \lrevise{ \textbf{Ignore Goal Mode}. 
    To ignore a specific testing goal means the goal is \emph{not a testing goal-at-hand}, i.e., a developer is not interested in generating tests to achieve that goal. This could be because a developer is interested in other goals (e.g., a  subset of goals), or there already exists many tests achieving the ignored goal. 
%    To achieve this, 
    In this mode, \approach employs a 
%    When \emph{ignoring} a certain feedback in the 
fitness function that does not consider the test feedback associated with the goal. 
In particular, \approach sets the weight of the ignored feedback to zero (0) while setting all other weights to one (1).  
In this work, we employ this goal to determine the contribution of a test feedback to the effectiveness of \approach  (ablation study, RQ3).  
%impact 
%This way, we can exclude a certain fitness from the fitness function to see its impact on the end result in an experiment which is important especially for research question 3.
As an example, if code coverage is \emph{not a testing goal-at-hand}, all weights $\mu_1,\mu_3,\mu_4$ are set to one except for the weight of the test feedback corresponding to coverage (i.e., $\mu_2$) which is set to zero (0).
This results in a fitness function $F = x_1 + x_3 + x_4$ where the code coverage is completely ignored.}

\end{enumerate}

%\todo{R1: consider moving gramma-based baselines before Section 2.3}

%\todo{R1: add examples for each gramma-based baselines using Table 1 and Figure 1. ``I believe a brief recap would make the presentation self-
%contained without forcing the reader to refer to the referenced work.''}

\begin{table}[tb!]
    \centering
    \begin{tabular}{>{}l>{}p{.18\textwidth}|lrr}
        \toprule
        ~ & \normalfont{Subject} & Version & \# Methods & \# LOC\\
        \midrule
        \multirow{13}{*}{\rotatebox[origin=c]{90}{JSON}} & JSONJava & 20180130 & 202 & 3\,742 \\
                                                         & MinimalJson & 0.9.5 & 224 & 6350 \\
                                                         & Genson & 1.4 & 1\,182 & 18\,780\\
                                                         & Pojo & 0.5.1 & 445 & 18\,492 \\
                                                         & Argo & 5.4 & 523 & 8\,265 \\
                                                         & Gson & 2.8.5 & 793 & 25\,172 \\
                                                         & Jackson & 2.9.0 & 5\,378 & 117\,108 \\
                                                         & json-simple & a8b94b7 & 63 & 2\,432 \\
                                                         & JsonToJava & 1880978 & 294 & 5\,131 \\
                                                         & FastJson & 1.2.51 & 2\,294 & 166\,761 \\
                                                         & json-flattener & 0.6.0 & 138 & 1\,522 \\
                                                         & json2flat & 1.0.3 & 37 & 659 \\
                                                         & json-simple-cliftonlabs & 3.0.2 & 183 & 2\,668\\
        \midrule
        \multirow{5}{*}{\rotatebox[origin=c]{90}{CSS}} & cssValidator & 1.0.4 & 7\,774 & 120\,838 \\
                                                       & closure-stylesheets & 0.9.27 & 3\,029 & 35\,401 \\
                                                       & cssparser & 0.9.27 & 2\,014 & 18\,465 \\
                                                       & jstyleparser & 3.2 & 2\,589 & 26\,287 \\
                                                       & sac & 1.3 & 368 & 8\,250\\
        \midrule
        \multirow{2}{*}{\rotatebox[origin=c]{90}{JS}} & Rhino & 1.7.7 & 4\,873 & 100\,234 \\
                                                      & rhino-sandbox & 0.0.10 & 49 & 529\\
        \bottomrule
    \end{tabular}
     \caption{Details of Subject Programs used in the Evaluation
%\todo{Why are subject programs in \texttt{this format} here. ... inconsistent with other tables }     
     }\label{subjectstable}
\end{table}

\begin{table*}[!tbp]\centering
    \caption{\centering Effectiveness of \approach (\emph{Single Goal}) in comparison to seed inputs. 
 \textbf{Bold} text indicates statistically significant 
% Mann-Whitney U 
 test results (p-value$\leq$0.05) or cases where \approach outperforms ($>=$) seed inputs, or vice versa.
%  are in \textbf{bold} text. 
``\#'' = ``Number of''. }
% This table contains the results for all subjects, i.e. "JSONJava", "MinimalJson", "Genson", "Pojo", "Argo", 'Gson', "Jackson", "json-simple", "JsonToJava", "FastJson", "json-flattener", "json2flat", "json-simple-cliftonlabs" "Rhino", "rhino-sandbox" "cssValidator", "closure-stylesheets", "cssparser", "jstyleparser", "sac"
% => Number of files: 5         *              50     *      20          = 5000
%                     (selected files per gen) (generations) (Subjects)
\begin{tabular}{l | l | l | l | l | l | l | l | l | l | l | l |}
    &&\multicolumn{2}{c|}{Code Coverage}&\multicolumn{2}{c|}{\#Exceptions}&\multicolumn{2}{c|}{Mappings}&\multicolumn{2}{c|}{Run Time (sec.)}&\multicolumn{2}{c|}{\#Unique Exceptions}\\
	                     & Subjects  & Seeds      & \approach     & Seeds   & \approach  & Seeds     & \approach   & Seeds     & \approach   & Seeds   & \approach     \\
	\hline\multirow{13}{*}{\rotatebox{90}{JSON}}
        &JSONJava                & 24.75   & \textbf{25.74} & \textbf{0}       & \textbf{0}       & 5250      & \textbf{5355}     & 0.467     & \textbf{18.266}   & \textbf{0}       & \textbf{0}          \\
        &MinimalJson             & 44.64   & \textbf{45.54} & \textbf{0}       & \textbf{0}       & 10500     & \textbf{10710}    & 0.235     & \textbf{18.039}   & \textbf{0}       & \textbf{0}          \\
        &Genson                  & 12.94   & \textbf{13.45} & \textbf{499}     & 204     & 16065     & \textbf{16800}    & 0.357     & \textbf{24.958}   & 1       & \textbf{2}          \\
        &Pojo                    & \textbf{36.40}   & \textbf{36.40} & 0       & \textbf{154}     & \textbf{17010}     & \textbf{17010}    & 1.397     & \textbf{27.528}   & 0       & \textbf{2}          \\
        &Argo                    & 40.73   & \textbf{41.68} & 2       & \textbf{125}     & 22365     & \textbf{23520}    & 0.367     & \textbf{18.704}   & \textbf{1}       & \textbf{1}          \\
        &Gson                    & 22.32   & \textbf{22.57} & 0       & \textbf{245}     & 18585     & \textbf{18795}    & 0.362     & \textbf{23.592}   & 0       & \textbf{3}          \\
        &Jackson                 & 15.79   & \textbf{16.31} & 0       & \textbf{1}       & 89145     & \textbf{92400}    & 0.809     & \textbf{41.573}   & 0       & \textbf{1}          \\
        &json-simple             & 34.92   & \textbf{36.51} & \textbf{0}       & \textbf{0}       & 2310      & \textbf{2415}     & 0.281     & \textbf{16.92}    & \textbf{0}       & \textbf{0}          \\
        &JsonToJava              & 24.83   & \textbf{25.17} & 0       & \textbf{246}     & 7665      & \textbf{7770}     & 0.733     & \textbf{20.551}   & 0       & \textbf{4}          \\
        &FastJson                & 87.33   & \textbf{100.00}  & 1       & \textbf{39}      & 264810    & \textbf{272475}   & \textbf{347.366}   & 311.724  & 1       & \textbf{2}          \\
        &json-flattener          & 40.06   & \textbf{93.26} & 5       & \textbf{181}     & 122955    & \textbf{287490}   & \textbf{41852.989} & 366.808  & 3       & \textbf{5}          \\
        &json2flat               & \textbf{90.67}   & 88.64 & 2       & \textbf{249}     & \textbf{420525}    & 419265   & \textbf{764.324}   & 518.084  & 1       & \textbf{6}          \\
        &json-simple-cliftonlabs & 86.51   & \textbf{91.45} & 1       & \textbf{51}      & 245070    & \textbf{256095}   & \textbf{282.401}   & 250.405  & 1       & \textbf{2}          \\
    \hline\multirow{3}{*}{\rotatebox{90}{CSS}}
    &cssValidator            & \textbf{93.78}   & 92.21 & \textbf{0}       & \textbf{0}       & \textbf{27364024}  & 7036688  & \textbf{3230.706}  & 2612.491 & \textbf{0}       & \textbf{0}          \\
    &closure-stylesheets     & \textbf{94.22}   & 93.82 & \textbf{0}       & \textbf{0}       & \textbf{17817992}  & 4166764  & \textbf{3599.661}  & 360.713  & \textbf{0}       & \textbf{0}          \\
    &cssparser               & 91.99   & \textbf{93.89} & \textbf{0}       & \textbf{0}       & \textbf{11842214}  & 3011424  & \textbf{522.735}   & 291.714  & \textbf{0}       & \textbf{0}          \\
    \hline\multirow{4}{*}{\rotatebox{90}{\footnotesize JavaScipt}}
    &jstyleparser            & \textbf{91.95}   & 90.82 & 14      & \textbf{31}      & \textbf{16234910}  & 3854676  & \textbf{3118.831}  & 502.653  & \textbf{2}       & 1          \\
    &sac                     & 89.45   & \textbf{89.94} & 47      & \textbf{240}     & \textbf{11205178}  & 2279596  & \textbf{376.372}   & 260.149  & 1       & \textbf{4}          \\
    &Rhino                   & \textbf{23.17}   & 20.62 & 0       & \textbf{7}       & \textbf{54581505}  & 4126584  & 0.896     & \textbf{186.958}  & 0       & \textbf{2}          \\
    &rhino-sandbox           & \textbf{86.73}   & 79.59 & \textbf{498}     & 247     & \textbf{176894355} & 11642484 & \textbf{446.239}   & 346.276  & 4       & \textbf{5}          \\
	\hline
\multicolumn{2}{c|}{Total }                  & 56.66   & 59.88 & 1069    & 2020    & 317182433 & 37548316 & 54547.528 & 6218.106 & 15      & 40         \\
        	\hline
\multicolumn{2}{c|}{\approach Improvement}             & \multicolumn{2}{c|}{5.38\%}             & \multicolumn{2}{c|}{47.08\%}         & \multicolumn{2}{c|}{-744.73\%}          & \multicolumn{2}{c|}{-777.24\%}        & \multicolumn{2}{c|}{62.50\%}    \\

\multicolumn{2}{c|}{\# Cases \approach $\geq$ Seeds} & \multicolumn{2}{c|}{14} & \multicolumn{2}{c|}{18} & \multicolumn{2}{c|}{12} & \multicolumn{2}{c|}{10} & \multicolumn{2}{c|}{19} \\
\hline 
\multicolumn{2}{c|}{\rev{Odds Ratio (p-value)}} & \multicolumn{2}{c|}{\rev{0 (0.3522)}} &  \multicolumn{2}{c|}{\rev{\textbf{0 (0.0001)}}}   & \multicolumn{2}{c|}{\rev{\textbf{0 (0.0001)}}}   & \multicolumn{2}{c|}{\rev{0 (0.7031)}}  & \multicolumn{2}{c|}{\rev{\textbf{0 (0.0001)}}}  \\
	\hline 
\multicolumn{2}{c|}{Mann-Whitney U (p-value)} & \multicolumn{2}{c|}{-0.79 (0.432)} & \multicolumn{2}{c|}{\textbf{-2.62 (0.009)}} & \multicolumn{2}{c|}{-0.58 (0.564)} & \multicolumn{2}{c|}{-0.73 (0.463)} & \multicolumn{2}{c|}{\textbf{-2.79 (0.005)}} \\
\end{tabular}
\label{tab:seeds-single}
%\vspace{-0.5cm}
\end{table*}

\subsection{Evolutionary Test Generators}\label{sec:evo-baselines}

\rev{
We  compare \approach to two evolutionary test generators,  namely \evogfuzz~\cite{eberlein2020evolutionary}  and \dynamosa~\cite{panichella2017automated}.  
In total we employ five baselines,  including the three (3) grammar-based baselines (Section \ref{sec:grammar-baselines}).  In the following, we describe each evolutionary baseline and its experimental setting.  
}

\smallskip
\noindent
\textbf{\evogfuzz:}
This approach by Eberlein et al.~\cite{eberlein2020evolutionary} combines grammar-based test generation with an evolutionary approach.
By mutating probabilities stored inside the input grammar, inputs can be evolved to trigger certain program behavior, similar to \approach.
\evogfuzz  is the closest related work to \approach. It has been shown to effectively find defects (exceptions).  Unlike \approach, 
\evogfuzz 
%However, it 
is focused on one test feedback (exceptions) and it does not leverage input mutation.  We compare the effectiveness of \approach versus EvoGFuzz in our evaluation (\textit{see} \textbf{RQ2}).

%\todo{R1: explain the modification performed when using DynaMOSA as a baseline, it should be described in the paper}

%\todo{R1: in paper and response, why we compared to Dynamosa?  non-grammar-based baseline, and also multi-objective approach similar to \approach's (multiple goals). }

\smallskip
\noindent
\textbf{\dynamosa:}
\revise{
Dynamic Many-Objective Sorting Algorithm (\dynamosa)~\cite{panichella2017automated} is a many-objective search algorithm for structural test case generation. The goal of \dynamosa is to 
%address the problem of covering 
cover multiple test targets (e.g., statements or branches) via dynamic selection 
%of the coverage targets using 
%based on 
and control dependency hierarchy.  
%The focus of \dynamosa is 
%coverage testing, i.e.,  
%achieving multiple test coverage goals (e.g., statements or branches).  
It has been shown to be effective in generating test suites that improve code coverage.
%~\cite{panichella2017automated}.  
%However, 
\rev{\dynamosa is similar to \approach due to its many-objective search method, 
%Unlike \dynamosa, 
albeit \approach employs weighted sum scalarization as its many-objective optimizer.  In addition, \approach is designed for grammar-based system-level testing,  and it targets (three) more testing goals, beyond coverage. }
%\todo{ES: Why DynaMOSA? $\to$ Describe here or in response letter?}
\rev{In particular,  we compare against \dynamosa as a non-grammar-based baseline.}  
%since a reviewer requested us to.}
%}
%
%\revise{
%Besides,  \approachwhich ensures that generated inputs are 
%syntactically valid.
In our experiments (\textbf{RQ2}), we compare the performance of \approach to \dynamosa using \evosuite~\cite{fraser2011evosuite}.  
%Indeed,  \dynamosa has been integrated into \evosuite as its \textit{default} search strategy. 
\evosuite\footnote{https://www.evosuite.org/} is a popular, state-of-the-art search-based unit test generator for Java programs. %~\cite{fraser2011evosuite}.
We note that \dynamosa is the \textit{default} search strategy of \evosuite.
%\footnote{
%\todo{
%We note that t
%The authors of \dynamosa recommended using \evosuite (with \dynamosa as default search strategy) for comparison when we contacted them about comparing to their tool. } %}
 We also set the parameters of \evosuite to values that are comparable to our experimental setting.  
%In particular, 
%Similar to 
To match the settings of \approach and \evogfuzz (in \textbf{RQ2}),
we set the 
%stopping condition (\texttt{stopping\_condition})
% and time taken (\texttt{MAXTIME}) 
%and number of generations (\todo{MAXGENERATIONS}) 
%of \evosuite to 
%``\texttt{MAXTIME}'' and 
%``600'' seconds per subject program,  respectively.  
%Specifically, 
%Specifically,  we set the 
the search budget (\texttt{-Dsearch\_budget}) to 48 and 
the stopping condition (\texttt{-Dstopping\_condition}) to ``\texttt{MAXGENERATIONS}''. 
%%we set the the search budget (-Dsearch_budget) to 48 and the stopping condition (-Dstopping_condition) 
%to “MAXGENERATIONS”.
}\rev{Since \dynamosa generates unit test cases and \approach targets the system level,  we manually extract  test inputs from \dynamosa's  generated unit test cases and convert them into system-level  tests, before comparing against \approach.
This allows the execution of the unit tests as system test inputs.}

\subsection{Research Protocol}

%\todo{R2: I suggest to report in 4.4 how the subsets of applications were chosen, e.g., random sampling.}

%\todo{This appears to be related to RQ3 and RQ4 where we have fewer subject programs}

For each research question, we ran a different set of experiments.
Since all inputs are directly dependent on the feedback of each subject program, each of the evaluations had to be done for each subject program and maximized feedback independently.
In total we had to run four (4) evaluations each for \emph{focusing on a single goal},  
%four (4)evaluations 
%\emph{only focusing on a single goal alone},
%four evaluations 
and \emph{ignoring a single goal}. 
Then,  for multiple goal mode of \approach, we had one evaluation with equal weights for each of the 20 subjects  (\textit{see} \autoref{subjectstable}).  
\revise{Similar to previous work~\cite{9154602},  we employ these set of 20 subject programs and three (3) input formats. We also note that our subject programs subsume the set used in the closest baseline (EvoGFuzz)~\cite{eberlein2020evolutionary}.
In each experiment, we generate five inputs per generation for 50 total generations.
This makes a total of 100 experiments 
%which had to be run
 to evaluate our approach on all subjects for \textbf{RQ1}.
 }

For the comparison to baselines (\textbf{RQ2}), 
%In addition, 
we ran one evaluation for each grammar-based baseline and format. This is sufficient since the baselines are not subject-specific and do not have any other parameters like weights.
%For the comparison against \evogfuzz, we 
%conducted an 
\revise{In this experiment,  we employ 
%experiment 
%ran an additional evaluation 
%using 
all 10 subject programs reported in the \evogfuzz, work including eight (8) subject programs for JSON,  and one subject program \textit{each} for CSS and JavaScript~\cite{eberlein2020evolutionary}. 
%We note that we execute \evogfuzz for each subject program 
%Similar to \approach 's experimental setting,  w
%We run \evogfuzz to 
For all approaches, we generate five inputs per generation, for 48 total generations.}
%\todo{In particular,   describle exact \evogfuzz setting  and how it matches \approach setting}

%from 
%were both available in 
%the \evogfuzz work~\cite{eberlein2020evolutionary}.  
%and \approach.

In addition,  we conducted an ablation study (\textbf{RQ3}) where we examined the impact of certain components in \approach by executing five (5) runs of \approach \revise{
(a) without grammar mutations,  and (b) without an input mutator. We further inspect the contribution of each test feedback to \approach 's performance.  Our ablation study involved the use of \revise{four (4)} 
\lrevise{randomly selected} 
subject programs, namely \revise{MinimalJSON, Argo, Genson, and Pojo}.}
%(similar to \textit{RQ1)}).

%\todo{Update the sensitivity analysis settings to account for newer subjects} \\
%Finally,  o
Our sensitivity analysis (\textbf{RQ4}) inspects the stability of \approach using different parameter values. \revise{The sensitivity analysis employs the default settings of \approach similar to (\textbf{RQ1}), i.e,  we generate five inputs per generation, for 50 total generations.  However, 
we employ only 
%but involves using
 \lrevise{four (4) randomly selected}
subject programs  (similar to \textbf{RQ3})
 -- MinimalJSON, Argo, Genson, and Pojo. 
This is because of the high computational cost and the large number of experimental settings for this analysis. 
%}
%
%that consume JSON inputs, namely  
%We note that the experiments in RQ4 involves only four subject programs, namely  for 
%\textsc{MinimalJSON},  \textsc{Argo},  \textsc{Genson},   and \textsc{Pojo}.}  
We examined the sensitivity of \approach to varying seed input size using five different number of initial seed inputs (between five to 1000 inputs, – \{5, 50, 200, 500, 1000\}).  We execute each seed input setting for both single goal and multiple goals mode of \approach, resulting 
%in 10 total settings (\todo{25}  
25 experimental runs.  
In this work, an \textit{experimental run} refers to a specific fitness function setting, such that multiple goal mode of \approach has \textit{only} one setting, and single goal mode as one setting per testing goal. 
%Meanwhile, an experimental setting means either the single goal or multiple goal mode of \approach. \todo{...}. 
As an example, we conducted 25 experimental runs in total for seed input size sensitivity. 
For multiple goal mode,  we executed one run for each configuration of seed inputs in the multiple goal mode, i.e., total five runs. 
On the other hand,  we performed one run for each (of the four) testing goals and each of the five seed inputs setting in the single goal mode of \approach resulting in a total of 20 runs for the single goal mode.}

\revise{
We also experimented with four different numbers or generated inputs between one and 25 – \{1, 5, 10, 25\} for both single goal and multiple goals, resulting in twenty (20) experimental runs in total.
%\revise{
We perform the sensitivity of \approach to different input generation sizes using six different generation sizes (\{5, 10, 25, 50, 100, 200\}), resulting in five (5) total runs since we only execute for 200 generations but report intermediate results until generation $N$.
We then conclude with the five varying random seed values (\{1, 2, 3, 4, 5\}) for both single goal and multiple goals, resulting in 25 experimental runs.
% runs in total. 
%using the multiple goal mode.
}

%\todo{How many subject programs are involved in our ablation study? and which programs?}

\revise{
%Overall,  we note that our research protocol involves over \revise{526} experimental runs and \revise{166} CPU days.  
Due to time limit and the huge computational cost of both experiments, we have employed a \rev{randomly selected subset of subject programs in} 
the comparison to the baselines (\textbf{RQ2} - 10 subjects), 
our ablation study (\textbf{RQ3} - \revise{four} subjects) and sensitivity analysis (\textbf{RQ4} - four subjects).  For instance,  
%In particular, 
our sensitivity analysis (\textbf{RQ4}) with \revise{four} subject programs
%, being the the most computationally intensive experiment, 
involved a total of  \revise{300} experimental runs and \revise{72} CPU days.
%despite involving only a subset of  
%\todo{Remember to stress the total number of experiments (526+) and time taken (166 CPU days) required to do all experiments, especially sensitivity analysis  This is important to rebut Reviewer 3's sensitivity analysis request}
}

\subsection{Statistical Analysis}

In our experiments, 
we conducted statistical analysis to determine statistical significance and correlation in our study (\textbf{RQ1, RQ2} and \textbf{RQ4}).
We check whether there is a 
statistically significant 
%statistical 
difference between the performance of \approach in comparison to the best-performing baseline 
(\evogfuzz) and seed inputs.  
\rev{We employ the Mann-Whitney U-test~\cite{10.1214/aoms/1177730491} to measure the differences in the stochastic ranking of \approach and  baselines (seed inputs or \evogfuzz). We also report effect sizes using Odds ratio.  This setting is in line with the recommendations for statistical analysis in software testing~\cite{arcuri2014hitchhiker}.  The Mann-Whitney U-test compares the stochastic ranking of the distribution of the performance of \approach versus the distribution of the performance of baselines (seed inputs or \evogfuzz).  
It checks whether observations in \approach are more likely to be larger than the observations in the seed inputs (or \evogfuzz).  It allows us to 
%In particular, 
%we c
\rev{compare} the distribution of the performance of \approach to each baseline 
%(e.g., seed inputs or \evogfuzz) 
per subject. 
Meanwhile, the odds ratio quantifies the strength of the association between \approach and the baselines.  }
%-level. 
%In our analysis,  w
% for different experimental 
%\todo{describe setup for statistical tests (Mann-Whitney U test). }
\rev{The last rows of \autoref{tab:seeds-single} and \autoref{tab:seeds-multiple} show the Mann-Whitney U-test results comparing \approach to the seed inputs,  and the second to last row reports the odds ratio results. 
For instance,  consider the Mann-Whitney U test results for exceptions (\#exceptions) in \autoref{tab:seeds-single}, 
%Consider 
comparing the per-subject distribution of \approach (column 6) to seed inputs  (column 5).  
%The values in each subject row are reported per subject, and the Mann Whitney U-test is performed on these per-subject results to compare \approach against the seed inputs.
%\todo{what are the sample distributions compare in the Mann Whitney test }
%For example, 
%even though the seed inputs solely triggered more exceptions for Genson in single-goal mode (499 vs 204) in \autoref{tab:seeds-single}, the total result shows that \approach triggered more exceptions (2020 vs 1069).
%Furthermore,  
The Mann-Whitney U-test on the distribution of triggered exceptions shows that the result is statistically significant (p-value of $0.009$,  which is $\le$ the $0.05$ threshold).
Likewise, the odds ratio results show that the observations for seed inputs and \approach are  significantly uncorrelated (p-value of $0.0001$). }
We also report the results for the Mann-Whitney U-test (statistic and p-values) and odds ratio when comparing the performance of \approach versus seed inputs (\textbf{RQ1},   \autoref{tab:seeds-single} and \autoref{tab:seeds-multiple}) and versus baselines (\textbf{RQ2}, and \autoref{tab:comparison}). Lastly, we conduct a correlation analysis in our sensitivity study (\textbf{RQ4}) using the Spearman’s correlation coefficient.  \autoref{tab:sensitivity-num-seed-inputs} to \autoref{tab:sensitivity-rand-seeds} reports our correlation analysis.  \rev{Our statistical analysis computation and results are provided in our artifact~\cite{artifact}.}

\begin{table*}[!tbp]\centering
    \caption{\centering Effectiveness of \approach (\emph{Multiple Goals}) in comparison to seed inputs.
 \textbf{Bold} text indicates statistically significant 
% Mann-Whitney U 
 test results (p-value$\leq$0.05) or cases where \approach outperforms ($>=$) seed inputs, or vice versa.
%  are in \textbf{bold} text. 
``\#'' = ``Number of''. } 
%        (Cases where subjects outperforms ($>=$) seed inputs are in \textbf{bolden} text.}
% This table contains the results for all subjects, i.e. "JSONJava", "MinimalJson", "Genson", "Pojo", "Argo", 'Gson', "Jackson", "json-simple", "JsonToJava", "FastJson", "json-flattener", "json2flat", "json-simple-cliftonlabs" "Rhino", "rhino-sandbox" "cssValidator", "closure-stylesheets", "cssparser", "jstyleparser", "sac"
% => Number of files: 5         *              50     *      20          = 5000
%                     (selected files per gen) (generations) (Subjects)
\begin{tabular}{ l | l | l | l | l | l | l | l | l | l | l | l |}
%   \hline\multirow{2}{*}{\rotatebox{90}{Format}}  
   & &\multicolumn{2}{c|}{Code Coverage}&\multicolumn{2}{c|}{\#Exceptions}&\multicolumn{2}{c|}{Mappings}&\multicolumn{2}{c|}{Run Time (sec.)}&\multicolumn{2}{c|}{\#Unique Exceptions}\\
	                  &   Subjects   & Seeds      & \approach    & Seeds   & \approach & Seeds     & \approach  & Seeds     & \approach  & Seeds   & \approach    \\
    \hline\multirow{13}{*}{\rotatebox{90}{JSON}}
    &JSONJava                & 24.75 & \textbf{25.74} & \textbf{0    }   & \textbf{0    }   & 5250      & \textbf{5460}     & 0.467     & \textbf{18.161}   & \textbf{0}       & \textbf{0}          \\
    &MinimalJson             & 44.64 & \textbf{45.54} & \textbf{0    }   & \textbf{0    }   & 10500     & \textbf{10710}    & 0.235     & \textbf{18.06}    & \textbf{0}       & \textbf{0}          \\
    &Genson                  & 12.94 & \textbf{13.54} & \textbf{499  }   & 208     & 16065     & \textbf{16800}    & 0.357     & \textbf{33.066}   & 1       & \textbf{2}          \\
    &Pojo                    & \textbf{36.40} & \textbf{36.40} & 0       & \textbf{113  }   & \textbf{17010}     & \textbf{17010}    & 1.397     & \textbf{26.531}   & 0       & \textbf{3}          \\
    &Argo                    & 40.73 & \textbf{43.21} & 2       & \textbf{94   }   & 22365     & \textbf{23730}    & 0.367     & \textbf{18.959}   & \textbf{1}       & \textbf{1}          \\
    &Gson                    & 22.32 & \textbf{22.57} & 0       & \textbf{239  }   & 18585     & \textbf{18795}    & 0.362     & \textbf{23.091}   & 0       & \textbf{2}          \\
    &Jackson                 & 15.79 & \textbf{16.14} & \textbf{0    }   & \textbf{0    }   & 89145     & \textbf{91140}    & 0.809     & \textbf{41.077}   & \textbf{0}       & \textbf{0}          \\
    &json-simple             & 34.92 & \textbf{36.51} & \textbf{0    }   & \textbf{0    }   & 2310      & \textbf{2415}     & 0.281     & \textbf{17.038}   & \textbf{0}       & \textbf{0}          \\
    &JsonToJava              & 24.83 & \textbf{25.17} & 0       & \textbf{247  }   & 7665      & \textbf{7770}     & 0.733     & \textbf{20.353}   & 0       & \textbf{3}          \\
    &FastJson                & 87.33 & \textbf{100.00}& 1       & \textbf{44   }   & 264810    & \textbf{270480}   & \textbf{347.366}   & 298.682  & \textbf{1}       & \textbf{1}          \\
    &json-flattener          & 40.06 & \textbf{93.22} & 5       & \textbf{141  }   & 122955    & \textbf{283185}   & \textbf{41852.989} & 343.858  & 3       & \textbf{5}          \\
    &json2flat               & 90.67 & \textbf{90.70} & 2       & \textbf{248  }   & \textbf{420525}    & 408450   & \textbf{764.324}   & 590.966  & 1       & \textbf{6}          \\
    &json-simple-cliftonlabs & 86.51 & \textbf{93.83} & 1       & \textbf{20   }   & 245070    & \textbf{255465}   & \textbf{282.401}   & 254.994  & 1       & \textbf{2}          \\
    \hline\multirow{3}{*}{\rotatebox{90}{CSS}}
    &cssValidator            & 93.78 & \textbf{93.99} & \textbf{0    }   & \textbf{0    }   & \textbf{27364024}  & 7433342  & \textbf{3230.706}  & 2075.303 & \textbf{0}       & \textbf{0}          \\
    &closure-stylesheets     & \textbf{94.22} & 94.02 & \textbf{0    }   & \textbf{0    }   & \textbf{17817992}  & 4312920  & \textbf{3599.661}  & 353.607  & \textbf{0}       & \textbf{0}          \\
    &cssparser               & 91.99 & \textbf{94.21} & \textbf{0    }   & \textbf{0    }   & \textbf{11842214}  & 3335080  & \textbf{522.735}   & 283.606  & \textbf{0}       & \textbf{0}          \\
    \hline\multirow{4}{*}{\rotatebox{90}{\footnotesize JavaScript}}
    &jstyleparser            & 91.95 & \textbf{93.01} & 14      & \textbf{31   }   & \textbf{16234910}  & 3701486  & \textbf{3118.831}  & 512.947  & \textbf{2}       & 1          \\
    &sac                     & 89.45 & \textbf{90.44} & 47      & \textbf{237  }   & \textbf{11205178}  & 1752275  & \textbf{376.372}   & 251.89   & 1       & \textbf{4}          \\
    &Rhino                   & \textbf{23.17} & 20.71 & 0       & \textbf{7    }   & \textbf{54581505}  & 3427573  & 0.896     & \textbf{64.075}   & 0       & \textbf{2}          \\
    &rhino-sandbox           & \textbf{86.73} & 80.80 & \textbf{498  }   & 247     & \textbf{176894355} & 14037856 & \textbf{446.239}   & 393.835  & \textbf{4}       & \textbf{4}          \\
	\hline
        \multicolumn{2}{c|}{Total}                   & 56.66 & 60.49 & 1069    & 1876    & 317182433 & 39411942 & 54547.528 & 5640.099 & 15      & 36         \\
        	\hline
\multicolumn{2}{c|}{\approach Improvement}              & \multicolumn{2}{c|}{6.33\%}            & \multicolumn{2}{c|}{43.02\%}         & \multicolumn{2}{c|}{-704.79\%}         & \multicolumn{2}{c|}{-867.14\%}        & \multicolumn{2}{c|}{58.33\%}         \\
%	\hline
        \multicolumn{2}{c|}{\# Cases \approach $\geq$ Seeds} & \multicolumn{2}{c|}{17} & \multicolumn{2}{c|}{18} & \multicolumn{2}{c|}{12} & \multicolumn{2}{c|}{10} & \multicolumn{2}{c|}{19} \\
        \hline
        \multicolumn{2}{c|}{\rev{Odds Ratio (p-value)}} & \multicolumn{2}{c|}{\rev{0 (0.341)}} &  \multicolumn{2}{c|}{\rev{\textbf{0 (0.0001)}}}   & \multicolumn{2}{c|}{\rev{\textbf{0 (0.0001)}}}   & \multicolumn{2}{c|}{\rev{\textbf{0 (0.0001)}}}  & \multicolumn{2}{c|}{\rev{0 (0.7003)}}  \\
        	\hline
        \multicolumn{2}{c|}{Mann-Whitney U (p-value)} & \multicolumn{2}{c|}{-1.06 (0.287)} & \multicolumn{2}{c|}{\textbf{-2.36 (0.018)}} & \multicolumn{2}{c|}{-0.54 (0.588)} & \multicolumn{2}{c|}{-0.80 (0.422)} & \multicolumn{2}{c|}{\textbf{-2.32 (0.020)}} \\
\end{tabular}
\label{tab:seeds-multiple}
%\vspace{-0.5cm}
\end{table*}

%\todo{R1: explain how mann whiteny test is done for a specifci example, e.g., sac: ``From the description given, it is not clear to me how the statistical tests were performed. I was not able to interpret the results (e.g.,
%Table 3) with respect to statistical significance. Let's take the subject "sac" in Table 3, and the "code coverage" criterion. We see the
%values 89.45 and 89.94 are reported, 89.94 in bold face (i.e., statistically significant). So what do these median values represent? The
%median coverage induced by the respective tools in each generation? I believe it would be useful if authors clearly describe how they
%performed the tests.''}

\subsection{Implementation and Platform}
Our tool was implemented in 5.3K lines of Python code and 3.9K lines of Java code.
\approach was built on the Tribble grammar-based fuzzer~\cite{8952419} since it supports several grammar-based baselines.
%  including EvoGFuzz.}
The parsers were implemented using ANTLR~\cite{antlr}, and several libraries were used for the Python implementation (lz4~\cite{lz4}, matplotlib~\cite{matplotlib}, numpy~\cite{numpy}, pygraphviz~\cite{pygraphviz}, matplotlib\_venn~\cite{matplotlibvenn}, tqdm~\cite{tqdm}, sortedcontainers~\cite{sortedcontainers} and pytest~\cite{pytest}).
The experiments were conducted inside a Docker~\cite{docker} container running Debian Bullseye.
The host system was an ASRock X470D4U with a Ryzen 5600X (6x3.70GHz) and 32GB of ECC memory. \approach 's implementation and experimental data are available online~\cite{artifact}: 
\begin{center}
%\url{https://tinyurl.com/DirectedGTG}
\url{https://tinyurl.com/FDLoop-V3}
\end{center}

%\todo{link Github and update readme}

%\todo{new link for artifact -- FDLOOP}

\subsection{Seed Inputs}

We crawled a large corpus of $300\,000$ JSON, $5\,500$ CSS and $200\,000$ JavaScript files from GitHub using the GitHub Crawling API~\cite{crawlingapi} and randomly selected 1000 files as seed inputs for the evaluation.
%\todo{ES}
In all experiments, we used 500 randomly selected seed inputs selected from the 1000 seed inputs, and for the experiments involving a different number of seed inputs (RQ4, see Section \ref{rq4}), we selected smaller sets of 5, 50, and 200 seed inputs from the same set.

\section{Evaluation Results}\label{evaluation}

This section discusses our evaluation results
% (section \ref{results}), 
and the implications of our findings: 
%% and future outlook 
%(section \ref{discussion}).  
%
%\subsection{Results}\label{results}
%
%This section reports 
We report the effectiveness of our approach (\textbf{RQ1}) and how it compares to the state-of-the-art baselines (\textbf{RQ2}). In \textbf{RQ3}, we conduct an \textit{ablation study} to examine the contribution of our
design choices (i.e., mutators and test feedbacks)
%our test feedbacks and design choices 
to the 
%each test feedback (e.g., code coverage or triggered exceptions) and design choice (i.e., input mutation and grammar mutation) to the 
effectiveness of \approach. Finally, \textbf{RQ4} reports the sensitivity of \approach to several 
%factors and 
parameter settings (e.g., number of initial seed inputs).

\subsection{RQ1 Effectiveness}\label{rq1} 
We examine if our approach effectively targets the 
four aforementioned testing goals, namely \textit{code coverage}, \textit{number of (unique) exceptions}, \textit{input complexity} and \textit{
%worst case 
execution time} (runtime).  
In this experiment, we compare the performance of \approach to that of the 500 real-world \textit{initial seed inputs}, which can be considered as an existing test suite. For each subject configuration, 
%our experiments involved the generation of 
\approach generates five inputs over 50 generations, resulting in 250 generated inputs in total.
For each of our four testing goals, we investigate the effectiveness of \approach when targeting a \textit{single testing goal} (e.g., \textit{only} code coverage). %, 
%We evaluate 
Specifically, we examine if \approach effectively achieves a specific testing goal when it 
%focuses solely on the goal and its corresponding (program or input) feedback. 
%\revise{
predominantly focuses on that goal. This is achieved by the \textit{single goal strategy} which sets a high weight (e.g., \revise{10}) for the targeted goal (e.g., exception), while 
% (e.g., exception) (0.7) and 
the weight of the other goals \revise{(e.g., mappings, etc.)} is set to one-tenth (e.g., \revise{1}) of the focused goal's weight. %}
%, i.e., focusing on the goal's corresponding program/input feedback. 
In addition, we investigate if \approach is effective  
%and 
%when it 
in simultaneously targeting \textit{multiple testing goals}, (i.e., all four testing goals). In this setting, \approach focuses on all four testing goals and aims to achieve all goals equally. 
%\revise{
This is achieved by the \textit{single goal strategy} which sets the weight for all goals to a constant value (e.g., 0.25). 
%}
%This experiment involved all 
% when the feedback are equally weighted. 
%In this experiment, we execute \approach using the research protocol discussed in \checknumber{X}.  Specifically, given \checknumber{x} initial benign seed inputs, our approach learns a probabilistic grammar and evolves  the initial input and learned grammar over \checknumber{X} generations to produce a test suite that targets the goal-at-hand. The initial seed input utilized in this experiments are randomly selected \revise{and benign}. In this experiment we evaluated on all three input formats and \todo{X} subject programs (see figure \checknumber{X}). 
%\checknumber{ and figure Y} 
\autoref{tab:seeds-single} and \autoref{tab:seeds-multiple} report the effectiveness of \approach %results %obtained by our approach 
%for 
in targeting a single goal and multiple goals, %for all input formats and subject programs, 
respectively. 
%We report the results for multiple testing goals in \checknumber{Table x and figure Y}. 

\smallskip
\noindent %\revise{ 
\textbf{\textit{Single Testing Goal:} } 
Our evaluation results show that our approach \textit{is highly effective in achieving a single testing goal}. 
\approach effectively achieves all four testing goals across our subject programs. In addition, we found that \textit{\approach outperforms the initial seed inputs (aka existing test suite) in achieving a targeted testing goal.} % and configurations.}  
\autoref{tab:seeds-single} shows that \textit{\approach outperforms %or is at least simi\textit{better than or similarly to} ($\ge$) 
real-world seed inputs in most (73 out of 100) cases} (\textit{see} ``$\# \ge$ Seeds'' in \autoref{tab:seeds-single}). %, for all testing goals. % .
%  \autoref{tab:seeds-single} further shows that \approach \textit{strictly} outperforms the initial seed inputs for all four tested goals in most (58 out of 100) cases. 
%Generally, o 
\approach is up to 63\% more effective than the initial seed inputs in achieving a single testing goal. This is especially 
evident for %two testing goals -- 
code coverage and triggered (unique) exceptions. 
For instance, \approach found 47\% more exceptions and 63\% more unique exceptions than real-world seed inputs across all subject programs. We also observed that %Even though 
\approach outperforms the seed inputs in targeting 
input-to-code complexity (aka mappings) and long execution time (aka Run Time) for most (up to 12) subject programs. However, \approach is less effective than the initial seed inputs in achieving 
mappings and long executions for certain subjects (e.g., rhino-sandbox). 
%\revise{x
%\todo{
%[Done]
%Fix the following text, this is not necessarily true, instead, the concern is that our approach still misses the input semantics, context, orderins/precedence and constraints that is missed by grammars, e.g.,  reuse of variables in different context, common in human-written  CSS and JS files}
\revise{
On inspection, 
we observed that human-written seed inputs
% especially 
for JavaScript 
are richer and more expressive than the inputs generated by \approach because   
%In particular, s
%Seed inputs 
they capture complex input semantics. 
%better 
%at a higher level 
%than \approach.
%is not yet achievable by \approach, e.g., 
As an example,  
%unlike \approach, 
seed inputs often contain repeated/nested input features (e.g., variable names) or
%often 
%encapsulate the 
%contain 
specific input ordering
% or other constraints among input features 
 (e.g., variable definition before use). }
%}
%In addition,  
%we attribute this to the sheer large size/complexity of these programs in comparison to the 
%small number/size of generated inputs in this experiments.  
%\revise{
%Further experiments clearly 
%%with larger number of generated input files 
%show that \approach effectively achieves these testing goals % to these complex programs 
%when given a larger number of input generations %time to generate larger number of input files 
%(see RQ4 - Sensitivity Analysis).\footnote{We note that the experiments in RQ4 involves only four subject programs, namely  for \textsc{MinimalJSON},\textsc{Argo},\textsc{Genson}, and \textsc{Pojo}. 
%}
%
\revise{This implies that 
%there is room for improving \approach in comparison to human-written seed inputs, i.e.,  
effectively capturing input semantics will improve the performance of \approach. 
Overall, these results suggest that \approach is effective 
%in aiding developers 
in targeting a testing goal.
}
 %-at-hand. %targeting a specific single testing goal.  
%}

\begin{result}
%\revise{
\approach is effective in achieving %a targeting 
a single testing goal. It outperforms the 
%improves over the 
initial 
seed inputs in most (73\% of) cases. 
% by \checknumber{up to 63}\%. 
% to achieve the  testing goal-at-hand. %, on average.
%}
\end{result}

\begin{table*}[!tbp]\centering
    \caption{\centering \lrevise{Effectiveness of \approach in comparison to the baselines.  
    The best results are highlighted in \textbf{bold text}, and the result for the best performing baseline (excluding seed inputs) is highlighted in 
    \underline{underlined text}.  \rev{``MW'' =  Mann-Whitney U (p-value) and ``OR'' = Odds Ratio (p-value). 
    \textbf{Bold} MW and OR values are statistically significant (p-value $\leq$0.05).}
%     or cases where FDLOOP outperforms (>=) seed inputs (or vice versa) (“#” = “Number of”).
%    \todo{Runtime is reported in seconds}
%    (``Uniq\_Exc.'' = unique exceptions)
}}
% This table contains the results for EvoGFuzz, i.e. only the subjects: "Argo", "Genson", "Gson", "JSONJava", "JsonSimple", "MinimalJson", "Pojo", "Rhino",
% => Number of files: 5         *              48     *      8          = 1920
%                     (selected files per gen) (generations) (Subjects)
\begin{tabular}{l | l | l | l | l | l | l}
%\hline
%\textbf{Approach}        
      & \textbf{Coverage} & \textbf{Mappings} & \textbf{\#Exceptions} & \textbf{\#Unique Exceptions} & \textbf{Run Time} (sec.)  & \textbf{Size (kB)} \\
\hline
\textbf{Seed Inputs}  &  35.85	&	82045279	& 501		&		15 	& 3235.80	&	 19537.62 	\\	
\hline
\textbf{Random Baseline}        &   21.73	&	23838	& 	328	&	 	3	& \textbf{\underline{2871.63}}	&	4.34	\\	
\textbf{Probabilistic Baseline}       & \underline{32.30} 	&	5454939	& 	449	&		\underline{9}	& 		2833.06 &	1203.40 	\\	
\textbf{Inverse Prob.  Baseline}        & 	 14.56	&  406734  	& 	480	&		2	& 	2866.93	&	 187.13 	\\	
\hline
\textbf{\evogfuzz }   & 32.15 	&	\textbf{\underline{13173177}} 	& 	 \underline{575} 	&	\underline{7}	& 	7.11 	& 2414.77 \\	
%\hline
\textbf{\dynamosa (\evosuite)} 	& 	11.21  & 145722	& 	6	&	5		& 	1810.70	&	0.24	\\	
\hline 
\textbf{\approach (Single)}    &  	35.89 &	11147735	&  \textbf{937}		&	\textbf{14}		& 		2847.10 &	179.84	\\	
\textbf{\approach (Multiple)}  & \textbf{36.25}  &	10901060	& 	865	&		13	& 	2210.51	&	192.55	\\	
\hline
\textbf{Impr.  (Single) vs.  \underline{Best Baseline}}   & 11.11\% 	&	-15.38\% 	& 	62.97\%	&	100\% &	-0.85\% 		&		\\	
\textbf{Impr.  (Multi.) vs.  \underline{Best Baseline}}   & 12.23\% 	&	-17.25\%	& 	50.43\% &	85.71\%	& 	-23.02\%	&		\\	
\hline
\textbf{\# \approach  (Single) $>$ Baselines} & All (5/5)	& 	4/5	& 	All (5/5)	& 	All (5/5)	& 	3/5	& 		\\
\textbf{\# \approach  (Multi.) $>$ Baselines} & 	All (5/5)	& 	4/5	& 	All (5/5)	& 	All (5/5)	& 	2/5 	& 		\\
\hline
\rev{\textbf{OR \approach (Single) vs.  \evogfuzz}} & \rev{1 (1.0)} & \rev{\textbf{0 (0.0001)}}& \rev{\textbf{0 (0.0001)}} & \rev{\textbf{0 (0.0001)}} & \rev{ 0 (0.7577)} &   \\
\rev{\textbf{OR \approach (Multi) vs.  \evogfuzz}} & \rev{1 (1.0)} & \rev{\textbf{0 (0.0001)}}& \rev{\textbf{0 (0.0001)}}& \rev{\textbf{0 (0.0002)}} & \rev{0.015 (0.8921)} &   \\
\hline
\textbf{MW \approach (Single) 
%$>$ 
vs.  \evogfuzz} 
%Mann Whitney U (p-value)
& -0.63 (0.526)	&   -0.49 (0.626)	& -1.17 (0.242)	& -1.37 (0.172)	& \textbf{-3.42 (0.001)}	& 		\\
\textbf{MW \approach (Multi.) 
%$>$ 
vs. \evogfuzz} 
& -0.68 (0.494)	&   -0.49 (0.626)	& -1.07 (0.283)	& -1.32 (0.188)	& \textbf{-3.42 (0.001)}	& 		\\
\end{tabular}
\label{tab:comparison}
%\vspace{-0.5cm}
\end{table*}

\noindent %\revise{  
\textbf{\textit{Multiple Testing Goals:} }
We found that \approach  \textit{is effective in simultaneously achieving multiple testing goals}. \autoref{tab:seeds-multiple} shows that 
%\approach achieves all four testing goals simultaneously for most programs. 
%In particular, 
\approach outperforms the initial seed inputs in most (76 out of 100) cases when it equally targets all testing goals. 
\approach is up to 58\% more effective than the seed inputs for goals such as code coverage and the number of triggered (unique) exceptions. 
We also observed that \approach mostly outperforms the initial seed inputs for the other testing goals (mapping and long execution time), e.g., 12 out of 20 cases for mappings. %However, the s
\revise{
Seed inputs perform better for certain subject programs and input formats. 
On one hand, we observed that this is
due to %Again, we attribute this performance to a lack of enough 
insufficient number of %fewer 
generations to reach the targeted goals for JSON subject programs. 
In our sensitivity analysis (\textbf{RQ4}),  we found that
\approach outperforms the JSON seed
inputs 
%s, for JSON subjects, 
when given enough generations (e.g.,  \textsc{json2flat}).
%  \approach outperforms these subjects for all testing goals. 
On the other hand, 
%we observed that 
JavaScript and CSS seed inputs are often richer, more diverse and larger than 
\approach's generated inputs. 
% for J 
We attribute the better performance of seed inputs for these formats to the lack of input constraints and semantics in \approach. 
}
%Seed inputs perform better for certain subject programs (e.g. , \textsc{json2flat}) 
%%\revise{
%due to %Again, we attribute this performance to a lack of enough 
%insufficient number of %fewer 
%generations to reach the targeted goals for these programs. %Again, o
%Our sensitivity analysis (\textbf{RQ4}) further shows that given enough generations, \approach outperforms these subjects for all testing goals. %} 

Comparing the results for single goal versus multiple goals (\textit{cf.} \autoref{tab:seeds-single}  vs. \autoref{tab:seeds-multiple}), we observed that \approach performs similarly well in both settings. Although \approach (in multiple goals mode) appears to be more effective for code coverage and unique exceptions than in the single goal mode.  We attribute the better performance of the multiple goal mode to the interaction among program and input feedbacks captured by equally targeting all goals. 

\begin{result}
%\revise{
Our approach (\approach) 
%effectively achieves multiple testing goals. It 
simultaneously achieves all testing goals better than the initial seed inputs in most (76\% of) % out of 100) 
cases.  
%}
\end{result}

\noindent  
\textbf{\textit{Statistical Analysis:} }
Our analysis shows that \textit{there is a statistically significant difference in the 
performance of the inputs generated by \approach versus seed inputs in 40\% of the tested configurations.}  For two out of the five testing goals (namely  unique exceptions and number of triggered exceptions),   the distribution of the behavior triggered by \approach's generated input and seed inputs are non-identical.  This result holds for both the single and multiple goal modes of \approach. 
\rev{The Mann-Whitney U-test results (\autoref{tab:seeds-single} and \autoref{tab:seeds-multiple}) show that the difference between the distributions of seed inputs and exception-inducing inputs generated by \approach is statistically significant (p-value [0.0005, 0.02] $\leq$ 0.05 ($\alpha$)). For the other three testing goals (coverage,  mappings and runtime), 
despite differences in the performance of the seed inputs versus \approach, we observed that the difference in the distributions are not statistically significant.  This suggests that 
the exception-inducing inputs generated by \approach are non-identical to the initial seed inputs and emphasizes why  \approach triggers exceptions and erroneous behaviors better than the initial seed inputs. }

\begin{result}
\rev{Exception-inducing inputs generated by \approach are
significantly different from seed inputs.}
\end{result}

\rev{We also observed that \textit{there is no statistical association between the performance of \approach and seed inputs}.   \autoref{tab:seeds-single} and \autoref{tab:seeds-multiple} reports our odds ratio results analysing  the association between \approach and seed inputs.  Both tables show that the performance of seed inputs and \approach performances are uncorrelated -- all odds ratio values are zero (0) ($<$ one (1)) for both single and multiple goal modes.  We observed that the odds ratio results are statistically significant in most (six out of 10) settings.  This is statistically significant  for both exceptions and mappings across single and multiple goal modes.  Overall, these results suggest that there is no association between the performance of \approach and Seed Inputs. We attribute this difference to the grammar learning and evolutionary testing nature of \approach.}

\begin{result}
\rev{There is no statistical association (odds ratio $<$ 1) between the 
performance of Seed Inputs and \approach:  
There is a statistically significant  lack of association 
%(p-value $\le$ 0.05) 
in 60\% of settings. }
\end{result}

\subsection{RQ2 Comparison to the State-of-the-art}\label{rq2}

%\todo{update RQ2 discussion w.r.t.  to the changes in (unique exceptions) \autoref{tab:comparison}}

%\todo{Change the number of subject programs to the new number:  is it 9 or 10? I expected 10 but results has 9 subjects } \\
%\todo{LK: how come we have nine subjects in the latest results, including cssvalidator and Rhino? But we had eight before including CSS and JS } \\
%\todo{update \autoref{tab:comparison} }\\
\revise{Let us examine how our approach (\approach) compares to the state-of-the-art 
%grammar-based 
baselines in effectively targeting specific testing goals.  In this experiment, we compare \approach to 
five baselines including three grammar-based test generators and 
two evolutionary approaches (\evogfuzz~\cite{eberlein2020evolutionary} and \dynamosa ~\cite{panichella2017automated}).
%\dynamosa is a multi-objective evolutionary search method for unit test generation which is built on \evosuite ~\cite{fraser2011evosuite}. 
The grammar-based test generators are namely \textit{random generator}, \textit{probabilistic generator}, and \textit{inverted-probabilistic generator}~\cite{8952419,9154602}.  
% and an \textit{evolutionary grammar-based test generator} (EvoGFuzz).  
%\revise{
%We employ 
This experiment involves 10 subject 
programs that are common to the 
\evogfuzz evaluation and the 
grammar-based baselines.  It includes 
%encompassing 
%using, 
%i.e.,
all of the subject programs 
used in 
\evogfuzz~\cite{eberlein2020evolutionary}, 
namely 
%  These includes 
  eight programs that process JSON inputs,  one program for CSS (cssValidator) and a JavaScript engine (Rhino).   
%since these %set of 
%programs have been previously 
%We employ these 10 subject porgrams since they evaluated for all approaches (including EvoGFuzz). }
%Similar to previous 
In this experiment, we set \approach to generate five inputs per generation for 48 generations.  
Similarly,  we executed \evosuite using the default settings with \dynamosa as the search algorithm for 48 generations.  
%\footnote{Specifically,  we set the 
%the search budget (\texttt{-Dsearch\_budget}) to 48 and the stopping condition (\texttt{-Dstopping\_condition}) to ``\texttt{MAXGENERATIONS}''.}.
% of %our baseline comparison 
%this experiment.  
%how our approach compares to the baselines. 
}

%\todo{consider discussing results along input format and subject programs}. 
\noindent \revise{  
\textbf{\textit{\approach vs.  All Baselines:}}
%\revise{
%Results show that Overall,  
We found that \approach outperforms the baselines in 86\% (43/50) of all tested configurations.  \autoref{tab:comparison} highlights our results.
The single goal mode of \approach (Single) outperforms the baselines in 92\% (23/25) settings. 
This is despite the fact it generates smaller inputs than most of the baselines (except the random baseline and \dynamosa). The multiple goal mode of \approach (Multi.) outperforms the baselines in 88\% (22/25) settings.    
%\autoref{fig:effectiveness} shows that 
\approach is the most effective in achieving 60\% (three out of five) testing goals  when compared 
%in comparison 
to the 
baselines.   It is up to twice as 
%up to 100\% more 
effective as the baselines in triggering (unique) exceptions and coverage. 
}

%\todo{\evogfuzz vs \approach : I still think there may be a bug here, it is weird that we have less mappings but more coverage}
\noindent \lrevise{  
\textbf{\textit{Best Baseline Per Goal:}}
%\revise{
We observed that\textit{ \evogfuzz is the most effective baseline.} It outperforms other baselines in maximizing mappings and (unique) exceptions.  Meanwhile, the probabilistic and random baselines are the best baselines in maximizing coverage and runtime, respectively.  Notably,  
\autoref{tab:comparison} 
%and  \autoref{fig:effectiveness}  
illustrates that 
\approach (Single) is comparable to the random baseline in maximizing run time,  but it is outperformed by \evogfuzz in maximizing the number of mappings.  
On inspection,  we observed that the better performance of  \evogfuzz holds for \textit{only} one program (CSSValidator/CSS) out of ten programs.  Indeed,  \approach outperforms \evogfuzz for all other programs.  Further inspection showed that the CSS inputs generated by \evogfuzz are larger than the inputs generated by \approach. These larger inputs generated by \evogfuzz trigger more mappings than \approach within the 13th and 19th generations.  
% which we leave for the future. 
%\todo{XXX} subjects and \todo{XXX} input formats, namely ...... This is because .... 
%\todo{but why, is this only for a specific subject or format? This was not the case when we had only JSON subjects}.  
%\todo{why results for \dynamosa is poor}  
%}
%
%\revise{
We further found that \dynamosa is outperformed by most baselines and \approach.  
We attribute the performance of \dynamosa 
%(\evosuite) 
to the lack of input grammar 
%These results demonstrate the importance of leveraging input grammars during input generations., in particular, we attribute the performance of \dynamosa to the lack of grammar 
which is required to generate structured inputs necessary 
%that 
to deeply assess the program logic of 
%consumed by 
our subject programs.  Inspecting the inputs generated by \dynamosa, we observed that 
%\todo{
they are mostly syntactically \textit{invalid} inputs which can not be generated by a valid input grammar.
For instance,  \dynamosa generated an empty JSON input which triggered the exception StringIndexOutOfBoundsException in Jackson. This exception was not triggered by the other baselines since they are all  
grammar-based approaches.
%other approaches using the input grammar. 
Such inputs demonstrate the importance of also generating syntactically \textit{invalid} inputs
during testing.  %}
% in triggering some erroneous behavior.}  
Overall,  evaluation results demonstrate that \approach outperforms the (best) 
%in comparison to the 
baselines in most settings.   
%The performance of \evogfuzz (for CSSValidator) also suggests that 
%%there is room for improving \approach
%%when targeting mappings.  It implies that 
%incorporating additional test feedbacks (e.g.,  large file sizes) and leveraging correlations among test feedbacks 
%%, and that targeting larger file sizes 
%may improve mappings.  We leave this investigation 
%%of the correlation among test feedbacks and additional test feedbacks 
%for
%%The effect of file size on our testing goals require 
%future work. 
% formats used in this work. 
%Finally, we observed that human-written seed files outperforms all baselines in terms of mappings, runtime and file size.  These suggests that there is room for improving the baselines, e.g., 
%%for these testing goal, 
%by 
%%We posit that 
%capturing input semantics. 
% will alleviate this concern. 
% contribution of input semantics 
}

%\todo{Discuss comparison to seeds,  }

\begin{result}
\lrevise{
\approach outperforms the baselines in 86\% (43/50) of all tested settings. It is (up to 100\%) better than the best baseline in maximizing (unique) exceptions and coverage.  
}
\end{result}

\noindent \lrevise{  
\textbf{\textit{Statistical Analysis:} }
\textit{We observe a statistically significant difference between the inputs generated by \evogfuzz
and the inputs generated by \approach when targeting long runtime.}
%For runtime,  t
The performance of the inputs generated by \approach (for runtime) versus \evogfuzz 's inputs are statistically non-identical.  This result holds for both the single and multiple goal modes of \approach. 
\autoref{tab:comparison} 
%reports the results of our Mann-Whitney U-test 
%and 
show that the difference between the medians of the distributions of \evogfuzz 's inputs and the inputs generated by \approach (for runtime) is statistically significant (p-value 0.001 $\leq$ $\alpha=0.05$).  
%We observe statistically significant difference between \approach  and \evogfuzz for only runtime.  
Even though there are differences in the performance of \approach versus \evogfuzz for the other four testing goals,  we could not reject the null hypothesis since Mann-whitney U test (p-value is greater than $\alpha=0.05$).   
%Overall, t
All in all,  the statistical difference between \approach and \evogfuzz 
explains the better performance of \approach 
%for runtime. 
when targeting runtime. 
%se results suggest that 
}

\begin{result}
\lrevise{
The 
%runtime-targeting 
inputs generated by 
\approach (for runtime)   are 
significantly different 
from  the inputs generated by \evogfuzz.
}
\end{result}

\rev{
We observed that \textit{there is no statistical association between the performance of \approach and \evogfuzz in 80\% of the settings}.  \autoref{tab:comparison} (``OR \approach vs.  \evogfuzz'') reports the odds ratio results for \approach versus \evogfuzz. We also observed that this result is statistically significant (p-values $\le$ 0.05) in six out of ten (60\%) settings.  This result further explains the difference int he performance of both approaches.  However, we observed that there is an association between the coverage performance of both approaches (OR = 1).  We attribute this to the grammar-based nature of both approaches. Overall,  there is no association between the approaches for all tested goals, except coverage.  
}

\begin{result}
\rev{
In most (80\%) settings, there is no statistical association between \approach and \evogfuzz. The 
lack of association is statistically significant  in six out of ten (60\%) settings.
}
\end{result}

\begin{figure}[!tbp]
\centering
\caption{
\lrevise{Venn diagram showing the number of unique exceptions triggered by \approach in comparison to the baselines}
}
\includegraphics[scale=0.95]{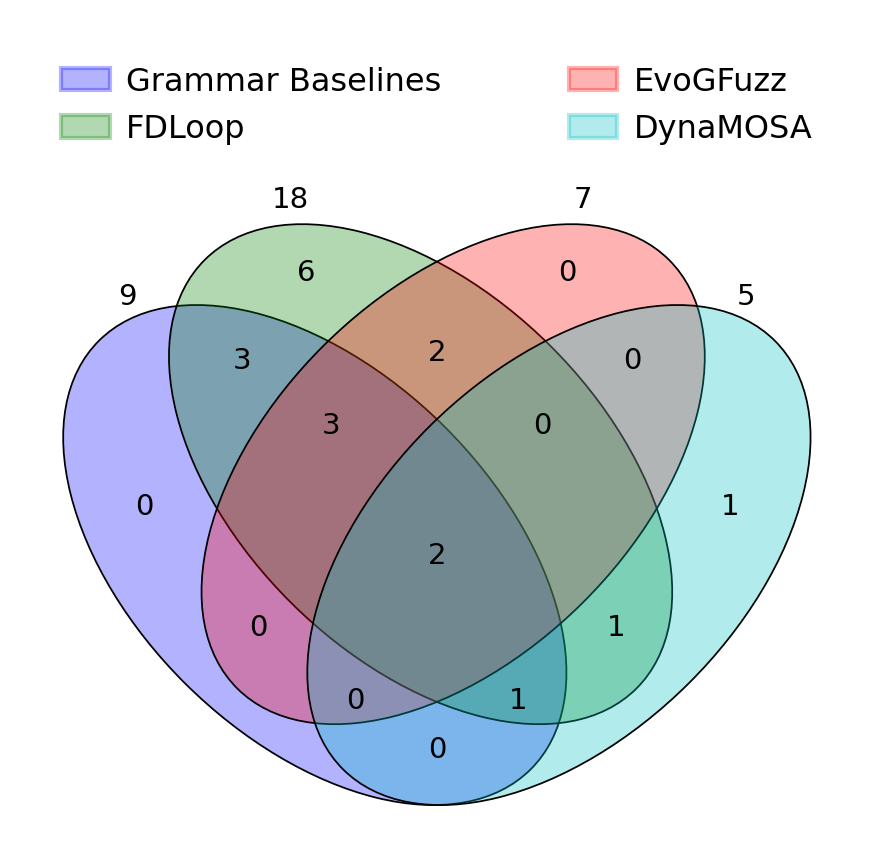}
\vspace{-0.4cm}
\label{fig:venn-diagram-1}
\vspace{-0.4cm}
\end{figure}

%\todo{what is special about the six exceptions only triggered by \approach}

%\todo{What is the input that induced the exception that is unique to \dynamosa and why could other approaches induce this exception?}

\noindent \lrevise{  
\textbf{\textit{Unique Exceptions:} }
\textit{\approach is twice (2X) as effective as the best 
%outperforms the 
baseline 
%by more than two times 
in exposing  
unique exceptions. } 
\autoref{fig:venn-diagram-1} shows the distribution of unique exceptions among all approaches.  
We found that \approach triggers the most (18) unique exceptions,  six of these exceptions were triggered by \textit{only} \approach. In comparison,   the best baseline (\evogfuzz) induce seven (7) unique exceptions.  Meanwhile,  \dynamosa trigger five (5) unique exceptions
%. 
%only one baseline (\dynamosa) induced 
%In addition, 
and grammar-based baselines (Random, PROB and INV) induce nine (9) unique exceptions in total.  
\autoref{fig:venn-diagram-1} shows that \approach induced almost all exceptions in this experiment (except one).  
%However,  
All exceptions induced by the grammar baselines were also induced by at least one other approach. 
Most exceptions were triggered by more than one approach,  but seven exceptions were unique to a single approach. 
%except for the 
Six (6) exceptions were triggered by \textit{only} \approach and one exception (StringIndexOutOfBoundsException in Jackson) was triggered by \textit{only} \dynamosa.  
%  were induced by .    
%\todo{
On inspection, we  observed that \dynamosa triggered this exception because it generated an empty input that can not be generated by a valid JSON grammar. 
%} 
%Overall,  t
In summary, these results demonstrate that 
%the superior ability of 
\approach outperforms the baselines in inducing error-prone behaviors (exceptions). 
% and its complementarity to . 
}

\begin{result}
\lrevise{
\approach is twice (2X) as effective as the baselines in inducing unique exceptions.  
%\approach 
It triggered 18 exceptions,  six of which were triggered by only \approach. 
}
\end{result}

\begin{table*}[!tbp]\centering
\caption{\centering 
\lrevise{Ablation Study showing the contribution of input mutation and grammar mutation to the effectiveness of \approach, for each testing goal.}
% (``w/o'' means ``without'', ``Mut.'' means ``Mutator', ``Contr.'' means ``Contribution of'')
%Effect of the features of our approach
}
\begin{tabular}{|l|l | l | l | l | l | l |}
\hline
\textbf{Mode} & \textbf{\approach Setting} & \textbf{Cov.} & \textbf{Map.} & \textbf{\#Exc.} & \textbf{\#Uniq\_Exc.} & \textbf{Run time} \\
\hline
\multirow{5}{*}{\rotatebox{90}{\textbf{Single}}} 
& \textbf{\approach (Single)}         & 34.27 & 68040 & 483 & 5 & 89.23 \\
%\textbf{\approach (Multiple)}       & 34.67 & 68250 & 415 & 6 & 96.62 \\
\cline{2-7}
& \textbf{Without Grammar Mutator}      & 31.41 & 61980 & 576 & 5 & 262.58 \\
& \textbf{Without Input Mutator}      & 34.55 & 66885 & 480 & 6 & 103.18 \\
%\textbf{Multiple w/o Gram. Mut.}    & 31.60 & 60467 & 400 & 5 & 193.34 \\
%\textbf{Multiple w/o Input Mut.}    & 34.60 & 68145 & 413 & 6 & 218.82 \\
\cline{2-7}
& \textbf{Contribution of Grammar Mutator} & 8.34\%&8.91\%&-19.25\%& 0\%&-194.27\%\\
& \textbf{Contribution of Input Mutator}   & -0.82\%&1.70\% &0.62\%&-20\%&-15.63\%\\
%\textbf{Multi Contr. Grammar Mut.}  & 8.85\%&11.40\%&3.61\%& 20\%&-100.10\%\\
%\textbf{Multi Contr. Input Mut.}    & 0.20\%&0.15\% &0.48\%&0\%&-126.47\%\\
\hline
\multirow{5}{*}{\rotatebox{90}{\textbf{Multiple}}} 
&  \textbf{\approach (Multiple)}       & 34.67 & 68250 & 415 & 6 & 96.62 \\
\cline{2-7}
& \textbf{Without Grammar Mutator}    & 31.60 & 60467 & 400 & 5 & 193.34 \\
& \textbf{Without Input Mutator}    & 34.60 & 68145 & 413 & 6 & 218.82 \\
\cline{2-7}
& \textbf{Contribution of Grammar Mutator}  & 8.85\%&11.40\%&3.61\%& 20\%&-100.10\%\\
& \textbf{Contribution of Input Mutator}    & 0.20\%&0.15\% &0.48\%&0\%&-126.47\%\\
\hline 

\end{tabular}
\label{tab:ablation}
%\vspace{-0.5cm}
\end{table*}

\subsection{RQ3 Ablation Study}\label{rq3}

In this experiment, we examine the contribution of our design choices (i.e., mutators and test feedbacks) to the effectiveness of \approach. Firstly, we examine the contribution of \approach 's mutator components to its effectiveness. % of \approach. 
Specifically, we compare the performance of \textit{default} \approach to \approach variants without the \textit{mutator components}, i.e.,  \approach \textit{without an input mutator} and \approach \textit{without a grammar mutator}. 
%The goal is to evaluate the contribution of each mutator component to the effectiveness of \approach. 
\autoref{tab:ablation} reports the effectiveness of \approach with and without each mutator component. 
Secondly, we examine the contribution of each test feedback to the effectiveness of \approach in the multiple goal model. The goal is to determine the contribution of a test feedback to the performance of \textit{\approach in the multiple goal mode}. To determine the contribution of a specific test feedback, we compare the performance of 
%\textit{default multiple mode of} 
\approach (when it uses all test feedbacks)  versus when it ignores a specific test feedback. 
\lrevise{Due to the computational cost, this experiment was conducted using four (4) randomly selected subject programs, namely MinimalJSON, Argo, Genson, and Pojo}.  
% specific  on the goal
\autoref{tab:feedback-contribution} reports the results for this experiment. 
%versus without the components. 

\smallskip
\noindent 
\textbf{\textit{Contribution of Mutators:}}
Our evaluation results show that \textit{grammar mutation contributes more than input mutation to the effectiveness of \approach across all testing goals}. 
In particular, we observed that grammar mutation contributes (up to 20\%) in achieving almost all goals (7 out of 10) than input mutation. Grammar mutation contributes (up to 11.4\%) to the effectiveness of \approach when targeting coverage and mappings goals. It also contributes to triggering unique exceptions especially in the multiple goal mode. 
For instance, \autoref{tab:ablation} illustrates that grammar mutation contributed 20\% to the number of triggered unique exceptions in the multiple goal mode of \approach, than input mutation (0\%). Meanwhile, input mutation contributed (at best) 1.7\% to the effectiveness of \approach in discovering new input-to-code complexity (\textit{see} mappings (``Map.'') \autoref{tab:ablation}). 
The better performance of grammar mutation versus input mutations holds for both the single goal and multiple goals modes of \approach.  
We attribute the better performance of grammar mutation vs.  input mutation to the larger effect size 
%and effect of 
%its coarse grained granularity,  i.e..,  the larger size and impact of 
of mutating grammar nodes,  e.g., non-terminals vs.  input characters. 
We also observed that both mutation components contribute to almost all testing goals, except for the long execution time (\textit{see} ``Run Time'' in  \autoref{tab:ablation}). 
This suggests that our mutation operators do not aid in targeting long execution (runtime).  We believe that effectively inducing long execution time may require employing better code-based features that are strongly coupled with runtime, e.g., number of (executed) loops, etc.

\begin{result}
Grammar mutation contributes the most (up to 20\%) to the effectiveness of \approach in targeting most (70\% of) goals. 
\end{result}

\begin{table}[!tbp]\centering
\caption{\centering 
\lrevise{Contribution (Contr.) of test feedbacks (``Fb.'') to \approach's effectiveness in achieving 
%mul on 
    multiple testing goals (``Uniq\_Exc.'' = unique exceptions)}
%means ``Test Feedback'')
}
\begin{tabular}{| l | l | l | l | l | l |}
\hline
\textbf{Approach} & \textbf{Cov.} & \textbf{Map.} & \textbf{Exc.} & \textbf{\#Uniq\_Exc.} & \textbf{Run t.} \\
\hline
 \textbf{Multiple}        & 60.49   & 39411942  & 1876 & 36& 5640.10 \\
\hline
 \textbf{Ignore Fb.}& 60.07   & 23573943   & 1015 & 31& 5619.93 \\
 %\textbf{No Code Coverage}& 60.07   & X         & X    & X & X \\
 %\textbf{No Mappings}     & X       & 23573943  & X    & X & X \\
 %\textbf{No Exceptions}   & X       & X         & 1015 & 31& X \\
 %\textbf{No Run Time}     & X       & X         & X    & X & 5619.93 \\
\hline
 \textbf{Contr.}    &  0.69\% & 40.19\%   &45.89\%&13.89\%& 0.36\%\\
\hline
\end{tabular}
\label{tab:feedback-contribution}
%\vspace{-0.5cm}
\end{table}

\noindent \lrevise{
\textbf{\textit{Contribution of Test Feedbacks:}} To determine the contribution of each test feedback (``Fb.''), 
%\approach ignores the test feedback 
%in this expciment 
we compare the effectiveness of \approach in multiple goals mode (``Multiple') versus 
%\approach 
when it 
ignores a specific test feedback (``Ignore Fb.''). 
\autoref{tab:feedback-contribution} reports the results of this experiment. 
}

\lrevise{
We observed that \textit{two major test feedbacks (the number of triggered exceptions and input-to-code-complexity mappings) contribute the \textit{most} (86\%) to the effectiveness of \approach in simultaneously achieving multiple testing goals}. \autoref{tab:feedback-contribution} shows that input-to-code-complexity (``map.'') and number of exceptions (``Exc.'') contributed 40\% and 46\% to the effectiveness of \approach, respectively. % in the multiple goals mode. 
This is followed by the number of unique exceptions which contributes about 14\% to the performance of \approach. 
Meanwhile, we observed that % tracking the 
program run time and code coverage test feedbacks contribute the \textit{least} to the effectiveness of \approach in targeting multiple testing goals. 
These results suggest that %these two test feedbacks (i.e., 
exceptions and input-to-code-complexity %) 
are important to simultaneously target multiple testing goals.
It also suggests that simultaneously targeting these two test feedbacks indirectly achieves the other two testing goals (coverage and runtime).
%may correlate with achieving other goals.  }
%Overall, we 
We attribute the better contribution 
%of these two test fe
of the two test feedbacks (i.e., mappings and exceptions) 
to the fact that they capture program behavior at a coarse level, which yields better results
than code coverage and program runtime.
%t In particular, we observed that these goals interact the 
% ost witht he other goals. 
}

\begin{result}
\lrevise{
Input-to-code-complexity %(mappings) 
and the number of triggered exceptions %contributes the most of the 
contribute the most (86\%) to the effectiveness of \approach in simultaneously achieving multiple testing goals. 
}
\end{result}

\subsection{RQ4 Sensitivity Analysis}\label{rq4}
%\todo{add experimental setting and results for varying random seed values (single \& Multiple)} \\
This experiment investigates the sensitivity of \approach to four main parameter settings, namely the number of generated inputs, the number of input generations,  the number of initial seed inputs and 
different random seed values. 
\revise{
The goal of this experiment is to examine if 
\approach is 
%\textit{still} effective 
\textit{stable} 
%under different settings 
or \textit{sensitive} to the parameter settings. Thus, we experiment with varying values for each parameter to show the effectiveness of \approach in different configurations. 
Due to the computational cost of sensitivity analysis, 
all experiments were conducted with four (4) \lrevise{randomly selected} subject programs that process JSON as input format, namely \textsc{MinimalJSON},  \textsc{Argo},   \textsc{Genson},  and \textsc{Pojo}.  This experiment involved running both single and multiple goal modes of \approach 
%in \revise{120} settings. Overall,  this accounted for 
for \revise{300} experimental runs
%total configurations 
taking a total time of \revise{72} CPU days. }

%The results of these experiments are reported in \autoref{tab:sensitivity-num-seed-inputs}, \autoref{tab:sensitivity-num-gen-inputs}, and \autoref{tab:sensitivity-num-gens}. 

\begin{table}[!tbp]\centering
    \caption{\centering Sensitivity of \approach to the \textit{number of initial seed inputs (seed)} 
%    for 50 input generations 
    (``Uniq\_Exc.'' = \# unique exceptions,  
``Cov.'' =  Code coverage, 
``Map.'' = mappings, 
   ``Exc'' = \# exceptions,  
``$\rho$'' = Spearman's correlation coefficient)
%Effect of the features of our approach
}
\begin{tabular}{| c |l | l | l | l | l | l |}
\hline
\null & \textbf{Seeds} & \textbf{Cov.} & \textbf{Map.} & \textbf{\#Exc.} & \textbf{\#Uniq\_Exc.} & \textbf{Run t.} \\
\hline
{{\multirow{4}{*}{\rotatebox[origin=c]{90}{\textbf{Single}}}}}
&\textbf{5 seeds}   & 34.55   & 67410     & 427 & 7       & 77.69 \\
&\textbf{50 seeds}  & 34.55   & 67935     & 482 & 7       & 98.07 \\
&\textbf{200 seeds} & 34.55   & 67935     & 497 & 6       & 90.94 \\
&\textbf{500 seeds} & 34.58   & 67305     & 494 & 6       & 85.60 \\
&\textbf{1000 seeds}& 34.55   & 68040     & 465 & 7       & 81.03 \\
\cline{2-7}
&$\rho$             & 0.35    & 0.29      & 0.3 & -0.29   & 0     \\
\hline
{{\multirow{4}{*}{\rotatebox[origin=c]{90}{\textbf{Multiple}}}}}
&\textbf{5 seeds}   & 34.55   & 67935     & 460 & 6       & 102.55\\
&\textbf{50 seeds}  & 34.55   & 67935     & 447 & 6       & 70.62\\
&\textbf{200 seeds} & 34.58   & 68040     & 407 & 6       & 64.59\\
&\textbf{500 seeds} & 34.27   & 67305     & 471 & 7       & 76.76\\
&\textbf{1000 seeds}& 34.29   & 67410     & 456 & 6       & 69.41\\
\cline{2-7}
&$\rho$             & -0.67   & -0.67     & 0.1 & 0.35    & -0.5 \\
\hline
\end{tabular}
\label{tab:sensitivity-num-seed-inputs}
%\vspace{-0.5cm}
\end{table}

\smallskip
\noindent
\textbf{\textit{Number of Initial Seed Inputs:}}
%In this experiment, we 
This experiment investigates the effectiveness of \approach using five (5) sizes of initial seed inputs between five (5) and 1000 
%This resulted in the following settings 
(including the default 500 seeds in previous experiments) -- $\{5, 50, 200, 500, 1000\}$, 
while employing the default settings for other parameters,
% of \approach, 
i.e., five generated inputs and 50 input generations.   

\autoref{tab:sensitivity-num-seed-inputs} shows 
%the results of this experiment.  show 
that \textit{the number of initial seed inputs has little or no effect on the performance of \approach}. All tested properties remain mostly stable as the number of initial seed inputs increases for both testing modes of our approach.  
For instance,  the number of achieved coverage (34.27 - 34.58)  and mappings (67305 - 68040) were relatively stable across all configurations. 
Besides,  
the effectiveness of most (80\%) settings  have a moderate to no 
correlation to the initial number of seed inputs. 
Spearman correlation coefficient shows that the effectiveness of 80\% of the settings
%most settings 
have a moderate to no correlation ($\rho=[0.5, -0.5]$) to the number of initial seed inputs. 
%to 
%about one in two tested settings have a weak correlation or no correlation ($\rho=[0.3, -0.3]$). 
We also observed a strong negative correlation ($\rho$=-0.67) for code coverage and mappings (both in the multiple goals mode), and  
%Furthermore,  pearson correlation coefficient shows that 
the  highest positive correlation ($\rho$=0.35) was \textit{moderate} 
%and only observed 
for code coverage (single goal) and unique exceptions (multiple goals).  
%Overall, t
These results suggests that the number of initial seed inputs has a low effect on the  performance of \approach. This suggests that \approach 
is mostly insensitive to the initial number of seed inputs.  
%Hence, we recommend the use of a minimal set of initial (five) inputs. 

\begin{result}
The effectiveness of \approach is stable as the initial number of seed inputs increases: Most (80\%) settings  have a moderate to no 
correlation to the initial number of seed inputs. 
\end{result}

\begin{table}[!tbp]\centering
    \caption{\centering Sensitivity of \approach to the \textit{number of generated inputs (inp.)}
%     for 50 input generations (``Uniq\_Exc.'' = unique exceptions)
%Effect of the features of our approach
 (``Uniq\_Exc.'' = \# unique exceptions,  
``Cov.'' =  Code coverage, 
``Map.'' = mappings, 
   ``Exc'' = \# exceptions,  
``$\rho$'' = Spearman's correlation coefficient)
}
\begin{tabular}{| c |l | l | l | l | l | l |}
\hline
\null & \textbf{Approach} & \textbf{Cov.} & \textbf{Map.} & \textbf{Exc.} & \textbf{\#Uniq\_Exc.} & \textbf{Run t.} \\
\hline
{{\multirow{4}{*}{\rotatebox[origin=c]{90}{\textbf{Single}}}}}
&\textbf{1 inp.}  & 34.03   & 66990     & 135  & 4 & 349.06\\
&\textbf{5 inp.}  & 30.38   & 68040     & 483  & 5 & 89.23\\
&\textbf{10 inp.} & 34.55   & 68145     & 870  & 7 & 324.34\\
&\textbf{25 inp.} & 34.55   & 68250     & 1229 & 7 & 111.04\\
\cline{2-7}
&$\rho$           & 0.95    & 1.00      & 1.00 & 0.95 & -0.4 \\
\hline
{{\multirow{4}{*}{\rotatebox[origin=c]{90}{\textbf{Multiple}}}}}
&\textbf{1 inp.}  & 34.10   & 66990     & 75   & 5 & 368.50 \\
&\textbf{5 inp.}  & 34.67   & 68250     & 415  & 6 & 96.62 \\
&\textbf{10 inp.} & 34.58   & 68040     & 626  & 6 & 356.20 \\
&\textbf{25 inp.} & 34.58   & 68040     & 1315 & 7 & 219.63 \\
\cline{2-7}
&$\rho$           & 0.51    & 0.51      & 1.00 & 0.92 & -0.4 \\
\hline
\end{tabular}
\label{tab:sensitivity-num-gen-inputs}
%\vspace{-0.5cm}
\end{table}

\noindent
\textbf{\textit{Number of Generated Inputs:}} Using the default settings of \approach (i.e., 500 seed inputs and 50 generations), this experiment 
%was conducted with 
examines the effect of the size of generated inputs by varying the size of generated inputs 
%four sizes 
between one and 25 
%Overall, we investigated the
%effectiveness of \approach 
%albeit the number of generated inputs is also varied 
%for four sizes of generated inputs (
(including the default size of five) 
-- $\{1, 5, 10, 25\}$.

We observed that there exists \textit{a positive relationship between the number of generated inputs and two testing goals, namely mappings and exceptions}. 
Specifically, \autoref{tab:sensitivity-num-gen-inputs} shows that 
as the number of generated inputs increases, the number of achieved input-to-code complexity mappings and (unique) exceptions also increases. 
%In addition,  we observed that 
Spearman's correlation coefficient ($\rho=[0.51,1.0]$) shows that \textit{there is a strong positive correlation between the the number of generated inputs and the effectiveness of our approach} for almost all (90\%) settings (except runtime).  This is particularly strong in the single goal mode for all testing goals, where we observed (almost) perfect positive correlation (0.95 - 1.00) for four out of five testing goals. 
%  except runtime.  
Meanwhile, we found that 
% the achieved code coverage is stable as the number of generated inputs increases. We also found that the 
program run-time \revise{has} a moderate negative correlation 
%does not seem to correlate 
with the number of generated inputs ($\rho=-0.4$). These results hold regardless of the testing mode, i.e., single or multiple goal(s) modes. 
Overall, these results suggest that there is a strong correlation between the number of inputs generated by \approach and its effectiveness in achieving most testing goals. 
% feedbacks
%, especiall
%the two test feedbacks -- input-to-code complexity mappings and (unique) exceptions. 

%\todo{ES: we should report correlation coefficients, this can be a good statistical correlation to report for all of RQ3 \\ basically, we should consider reporting the steepness of the slope of increase or decrease, is it slightly, moderately or significant, positive or negative. }

\begin{result}
There is a strong correlation between the number of generated inputs and 
\approach's effectiveness:
% in  
%achieving the testing goals.: \approach's 
Its effectiveness in 
achieving all testing goals (except runtime) 
% is sensitive to . 
%For %
%for two testing goals. % -- 
%mappings and exceptions, 
%T: T
%t
%The number of mappings and exceptions 
increases as the number of generated inputs increases.  

\end{result}

\begin{table}[!tbp]\centering
    \caption{\centering Sensitivity of \approach to the \textit{number of input generation (gen.)} 
%    using five (5) initial seed inputs (``Uniq\_Exc.'' = unique exceptions)
%Effect of the features of our approach
 (``Uniq\_Exc.'' = \# unique exceptions,  
``Cov.'' =  Code coverage, 
``Map.'' = mappings, 
   ``Exc'' = \# exceptions,  
``$\rho$'' = Spearman's correlation coefficient),
reported time in minutes.
}
\begin{tabular}{| c |l | l | l | l | l | l | l |}
\hline
\null & \textbf{Approach} & \textbf{Cov.} & \textbf{Map.} & \textbf{\#Exc.} & \textbf{\#Uniq\_Exc.} & \textbf{Time} \\
\hline
%TODO Is it necessary to include the generations in the table, or sufficient to show a line plot? The table gets very long and the line chart may show the improvement over time better than the table
% Subjects: MinimalJson, Genson, Argo, Pojo
% Number of files: 5 * 4 * {number of generations}
{{\multirow{6}{*}{\rotatebox[origin=c]{90}{\textbf{Single}}}}}
&\textbf{5 gen.}    & 32.96 & 64050 & 30   & 4 & 8.90 \\
&\textbf{10 gen.}   & 34.36 & 67305 & 89   & 4 & 32.78 \\
&\textbf{25 gen.}   & 34.55 & 67305 & 192  & 6 & 97.04 \\
&\textbf{50 gen.}   & 34.55 & 67305 & 461  & 7 & 158.57 \\
&\textbf{100 gen.}  & 34.55 & 67935 & 990  & 7 & 372.34 \\
&\textbf{200 gen.}  & 34.55 & 68040 & 1958 & 7 & 672.90 \\
% SAMPLE SIZE FOR RHO-TEST: 200 !!!
\cline{2-7}
&$\rho$             & 0.38  & 0.93  & 1.00 & 0.71 & 1.00 \\
\hline
{{\multirow{6}{*}{\rotatebox[origin=c]{90}{\textbf{Multiple}}}}}
&\textbf{5 gen.}    & 33.02 & 62480 & 25   & 5 & 14.98\\
&\textbf{10 gen.}   & 34.17 & 65642 & 62   & 5 & 35.50 \\
&\textbf{25 gen.}   & 34.44 & 67725 & 206  & 5 & 125.74 \\
&\textbf{50 gen.}   & 34.55 & 67935 & 418  & 7 & 222.31 \\
&\textbf{100 gen.}  & 34.55 & 67935 & 890  & 7 & 467.99 \\
&\textbf{200 gen.}  & 34.55 & 67935 & 1953 & 7 & 913.909 \\
% SAMPLE SIZE FOR RHO-TEST: 200 !!!
\cline{2-7}
&$\rho$             & 0.62  & 0.62  & 1.00 & 0.67 & 1.00 \\
\hline
\end{tabular}
%\caption{Effect of the features of our approach}
\label{tab:sensitivity-num-gens}
%\vspace{-0.5cm}
\end{table}

%
%We evaluate the sensitivity of \approach to several factors and parameters, namely the choice of the initial seed, the number of initial seed inputs, the number of generations, and \hecknumber{x}. \hecknumber{ Figure X and \autoref{tab:sensitivity}} highlights our findings. 

\noindent
\textbf{\textit{Number of Input Generations:}} 
%In this experiment, we employed the default parameter settings of \approach (i.e.,  five seed inputs and five generated inputs) 
This experiment was conducted with 
%and 50 generations). In addition, we examined the effectiveness of \approach with 
%five additional number of 
six generation sizes 
%between five and 1000 
(including the default 50 generations)
%.
%In total, we employed the following settings for the number of input generations 
-- $\{5, 10, 25, 50, 100, 200\}$.  
%For other parameters, 
%All experiments
We used the default settings of \approach for other parameters (i.e., 500 seed inputs and five generated inputs).

We found that 
\textit{there is a strong positive correlation between the number of input generations and the effectiveness of \approach in achieving in the targeted testing goal} (\textit{see} \autoref{tab:sensitivity-num-gens}).  This finding holds for both the single goal and multiple goals mode of \approach. \autoref{tab:sensitivity-num-gens} demonstrates that 
\textit{the effectiveness of \approach increases as the number of input generations increases, for all testing goals and both testing modes of \approach}. 
This is evident by the strong spearman's correlation coefficient ([0.61, 1.00]) across most (90\%) of our settings. 
This is most evident for two main testing goals, namely run time and triggered exceptions: 
%For these testing goals, 
For instance, consider the number of triggered exceptions by \approach in the multiple testing goals mode. 
\autoref{tab:sensitivity-num-gens} shows that the effectiveness of 
\approach %effectiveness 
in achieving the testing goal improves substantially (by up to 77X) as the number of input generation increases (by 40X).
These results suggest that \approach  is more effective as the number of input generation increases
and it effectively achieves the targeted testing goal(s) given enough number of generations.
% More importantly, \approach. 
We attribute this improvement  in the effectiveness of \approach as the number of input generation increases to the evolutionary nature of \approach. 

\begin{result}
%Our approach is insensitive to the number of input generations. 
There is a strong positive relationship between the effectiveness of \approach and the number of input generations.
For all tested goals, \approach's effectiveness improves as the  number of input generations increases. 
%\approach is more effective at achieving a targeted goal as the number of input generation increases.  
\end{result}

\begin{table}[!tbp]\centering
    \caption{\centering \revise{Sensitivity of \approach to different \textit{random seed values (seed)} 
 (``Uniq\_Exc.'' = \# unique exceptions,  
``Cov.'' =  Code coverage, 
``Map.'' = mappings, 
   ``Exc'' = \# exceptions,  
``$\rho$'' = Spearman's correlation coefficient),
reported time in minutes.
}}
\begin{tabular}{| c |l | l | l | l | l | l | l |}
\hline
\null & \textbf{Approach} & \textbf{Cov.} & \textbf{Map.} & \textbf{\#Exc.} & \textbf{\#Uniq\_Exc.} & \textbf{Time} \\
\hline
{{\multirow{6}{*}{\rotatebox[origin=c]{90}{\textbf{Single}}}}}
%&\textbf{seed 0}    & 34.27  & 68040  & 483   & 5 &89.23  \\
&\textbf{seed 1}    & 30.54 &  57330 & 388   &7  & 34.17 \\
&\textbf{seed 2}    &  30.51 & 56700  &  533  &6  & 34.44  \\
&\textbf{seed 3}    & 30.92 & 56700 &  460  & 6 & 33.93 \\
&\textbf{seed 4}    & 30.92 &  57225 &  444  & 7 &34.45  \\
&\textbf{seed 5}    & 31.05  & 56700  &  455  & 7 & 34.46 \\
% SAMPLE SIZE FOR RHO-TEST: 200 !!!
\cline{2-7}
& $\rho$               &0.87 & -0.53 & 0.10 & 0.29 & 0.70 \\

\hline
{{\multirow{6}{*}{\rotatebox[origin=c]{90}{\textbf{Multiple}}}}}
%&\textbf{seed 0}    & 34.67  & 68250  &  415  & 6 & 96.62  \\
&\textbf{seed 1}    & 34.29 & 67410  &  465  & 7 & 45.03  \\
&\textbf{seed 2}    & 34.56 & 67935  &  399  & 7 & 46.81  \\
&\textbf{seed 3}    & 34.67 & 68250  &  450  & 6 &44.71  \\
&\textbf{seed 4}    &  34.56 & 67935  &  400  & 7 & 44.61 \\
&\textbf{seed 5}    & 34.56 & 67935 & 476   & 6 & 44.81 \\
% SAMPLE SIZE FOR RHO-TEST: 200 !!!
\cline{2-7}
& $\rho$              &  0.62 & 0.62 & 0.30 & -0.58 & -0.60  \\
\hline
\end{tabular}
\label{tab:sensitivity-rand-seeds}
%\vspace{-0.5cm}
\end{table}

%\todo{Update RQ4 "Varying Random Seed Values" based on changes to Table 11}

\noindent
\textbf{\textit{Varying Random Seed Values:}} \revise{
This experiment examines the sensitivity of \approach to varying random seed (\texttt{--random-seed}) values.  The 
%goal of the
seed parameter enables the 
reproducibility of input generation.
% is reproducible. 
To examine the sensitivity of our findings to this parameter,  we experimented with five random seed values between one (1) and five (5) -- \{1, 2, 3,  4, 5\}
%.  Similar to previous settings,  we used 
while employing the default settings of \approach (500 seed inputs and five generated inputs). 
}

\revise{
We observed that \textit{\approach's effectiveness is mostly stable across different random seed values}.  
In 60\% (six out of 10) settings,  
%
%In particular,  
%For instance,  
%there was a 
there is a very low difference ($\approx \pm$2)
%  in the results 
across varying seed values (\autoref{tab:sensitivity-rand-seeds}).  
\approach's effectiveness remains stable across all settings for three out of five testing goals (coverage,  unique exceptions,  and runtime).   
%However, the results for mapping ([56700 - 68250]) and number of exceptions ([388 - 533]) 
%were less stable.  
%}
%\revise{
%
However,  
%we found 
%that 
\textit{there is a positive correlation between the varying random seed values and the effectiveness of \approach for most (70\%) of the tested settings. } In particular, 
\autoref{tab:sensitivity-rand-seeds} shows that 
there is a strong positive correlation between varying random seed value and \approach's effectiveness in 40\% (four out of 10) cases.  
%This is evident in 
Spearman's correlation coefficient ($\rho$) was strongly positive ([0.62, 0.87])
in (four) settings, 
e.g., the effectiveness of \approach for runtime and coverage (in single goal mode).
% and 
%\approach (multi. ) 
%for mapping and coverage (in the multiple goal mode).    
In addition, we observed a weak positive correlation ([0.10, 0.30]) in three cases, 
%namely \approach's effectiveness, 
e.g.,  exceptions and unique exceptions (in single goal mode). 
Finally,  there is a moderate negative correlation ([-0.53, -0.60]) between \approach's effectiveness and 
random seed values 
%testing goals 
in three settings, 
%.  As an example,  
this is evident 
%We observed this 
%for mappings (in the single goal mode),  and 
for unique exceptions and runtime (multiple goal mode). 
% and \approach's effectiveness and 
%In summary, 
%Overall, 
%t
%These r
Results suggests that 
%even though 
\approach's effectiveness is mostly stable across random seed values,  but 
%however,  
varying seed values positively correlates with \approach's effectiveness for specific goals (e.g., coverage). 
% cases.
% in most settings. 
%Thus, we recommend developers experiment with varyong seed values especially
%This implies that experimenting 
}

\begin{result}
\revise{
In 60\% of settings, 
%.  The  of 
\approach's effectiveness is stable ($\approx \pm$2) as the random seed value varies.
However,  
there is a weak to strong positive correlation between 
\approach's effectiveness and varying random seed value in most (70\%) settings. 
%The effectiveness of most (70\%) \approach settings 
%The effectiveness of \approach has 
%have moderate to no correlation to the random seed value. 
%in most (70\%) settings.  
%For all testing goals (except runtime),  the effectiveness of \approach is stable as the random seed value varies.  
}
\end{result}

\begin{figure}[tbh!]
\centering
%\vspace{-0.4cm}
\caption{
\lrevise{Venn diagram showing the number of unique exceptions triggered by \approach 
%in comparison to 
vs.  All baselines 
%and the 
vs.  Seed Inputs}}
\includegraphics[scale=0.95]{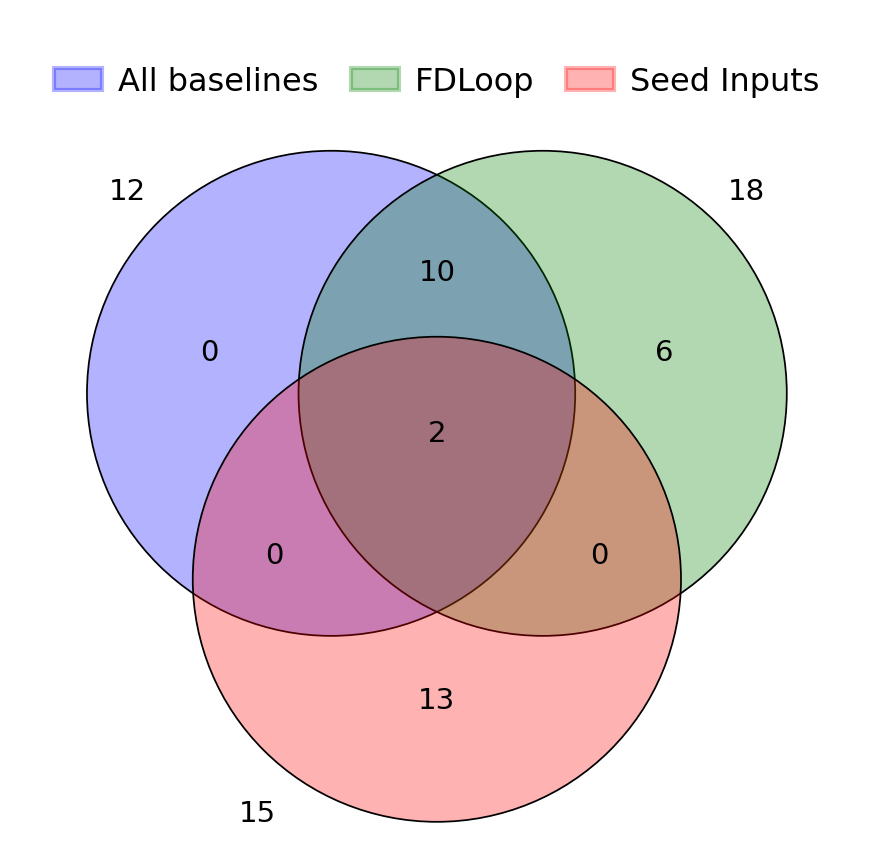}
\label{fig:venn-diagram-2}
\vspace{-0.4cm}
\end{figure}

\subsection{Discussions and Future Outlook}\label{discussion}

%\todo{place of input semantics, size of files, richness, and venn diagram of exceptions}

%\todo{check seed inputs vs generated inputs w.r.t changes in Table 5}

\noindent 
\textbf{\rev{Seed Inputs vs.  Generated Inputs:} }
\rev{
%To elicit opportunities for improvement, we analyze 
%A major observation across all experiments is the disparity between 
Comparing the generated test inputs versus seed inputs, 
%from the comparison experiments (\textbf{RQ2}), 
%across 
%
%all our experiments, 
we found that 
%Our observation is that 
% the 
 seed inputs are often 
richer and more diverse than 
the inputs generated by
%all 
\approach 
%our approach 
and the baselines.  
%In addition, 
%We observed that 
For instance,  seed inputs perform similarly to \approach in achieving runtime, outperforming each other in 10 out of 20 settings (\textit{see} \autoref{tab:seeds-single}  and \autoref{tab:seeds-multiple}).  
%are not asas .   
In addition, 
\autoref{tab:comparison} (\textbf{RQ2}) shows that seed inputs 
%often 
outperform \approach and all baselines in achieving mappings.
% and long runtime.  
We also found that
% It also shows that 
seed inputs are much larger than generated inputs (\textit{see} \autoref{tab:comparison}). 
%More importantly,  
Likewise,  \autoref{fig:venn-diagram-2} shows that 
%there are 
many (13) exceptions are triggered by \textit{only} the seed inputs collected from GitHub. None of these exceptions were triggered by either \approach or 
%any of 
the baselines.  These findings demonstrate the limitations of existing approaches and suggest the need to further improve the quality of automatically generated test inputs.  In future work, we plan to investigate how to address this concern by studying how to better leverage/evolve existing seed inputs
and 
% We also plan to target large files and 
integrate input constraints in our approach.
% to address these concerns. 
%We believe this is due to several reasons, including the lack of input semantics in most approaches and   
}

\smallskip
\noindent 
\textbf{\revise{Additional Feedback:} }
\lrevise{
We plan to investigate the effect of new 
%test or input 
feedback in targeting testing goals. 
For instance,  the better performance of \evogfuzz (CSSValidator) in \textbf{RQ2} suggests that 
%there is room for improving \approach
%when targeting mappings.  It implies that 
incorporating additional test feedback (e.g.,  large input sizes) and leveraging correlations among test feedback 
%, and that targeting larger file sizes 
may improve certain testing goals,  e.g., mappings.  
%In the future,  w
We plan to 
%We leave this investigation 
%of t
investigate the correlation among test feedback and study the effect of additional 
%test or inputs 
feedbacks in future work.  
%for
%The effect of file size on our testing goals require 
%future work. 
}

\smallskip
\noindent %\lrevise{%\skip
\textbf{Efficiency: } % of \approach:}
%\todo{ES: Time Performance of \approach}
\rev{\autoref{tab:time-taken} shows the time performance of \approach, in comparison to the closest baseline, i.e., EvoGFuzz.  Overall,  we observed that \approach is twice as computationally expensive as EvoGFuzz, across all settings.  
Indeed,  we observed that \approach is  computationally more intensive than the baselines.
We attribute the high cost of \approach to its large search space, since it explores multiple (five) test feedbacks than EvoGFuzz (which focuses mostly on exceptions).  
%loops in its 
% aims to target multiple goals and
Overall, this computational cost pays off, since \approach outperforms EvoGFuzz across all settings (\textit{see} \textbf{RQ2} and \autoref{tab:comparison}). 
}

\begin{table}[!tbp]\centering
    \caption{
    \revise{Time taken by \approach when targeting multiple goals versus the closest baseline -- EvoGFuzz 
%     in comparison 
(hh:mm:ss) 
%    \todo{EZ: The last five rows show the raw time to run a single input in the subject programs, we might hide those}
    }}
\begin{tabular}{| l | l | l |}
\hline
\textbf{Time} & \textbf{\approach} & \textbf{EvoGFuzz}\\
\hline
Total time (MinimalJson) & 2:29:36 & 1:56:01 \\
Total time (Genson)      & 2:53:24 & 1:07:35 \\
Total time (Argo)        & 2:38:18 & 1:05:39 \\
Total time (Pojo)        & 2:40:20 & 1:08:23 \\
\hline
Total time               & 10:41:38& 5:17:38 \\
\hline
%Time to run gen. files (MinimalJson) & 0:00:17 & 0:00:08 \\
%Time to run gen. files (Genson)      & 0:00:32 & 0:00:11 \\
%Time to run gen. files (Argo)        & 0:00:18 & 0:00:09 \\
%Time to run gen. files (Pojo)        & 0:00:25 & 0:00:13 \\
%\hline
%Time to run gen. files               & 0:01:32 & 0:00:41 \\
%\hline
%Time to generate final inputs        & 10:40:06 & 5:16:57 \\
%\hline
\end{tabular}
\label{tab:time-taken}
%\vspace{-0.5cm}
\end{table}

\smallskip
\noindent %\skip
\textbf{Testing Goals:}
%\todo{Lower performance or weakness in achieving runtime and coverage goals?}
%Across all experiments, we observed that
Even though  \approach targets each testing goal better than the baselines (\textbf{RQ2}),  we observed that some goals are more difficult to reach or saturate very quickly. 
Notably,  significantly improving code coverage,  mappings and run-time seems to be particularly difficult in our default settings for certain subjects (\textbf{RQ1}). 
% as well as with different parameter configurations \textbf{RQ4}).  
Albeit,  we observed that increasing the number of generations alleviates the difficulty of reaching higher run-time.  Indeed, 
\textbf{RQ4} (\textit{see}  \autoref{tab:sensitivity-num-gens}) shows that increasing the number of input generations leads to a higher runtime.  
However,  different parameter settings do not significantly alleviate the saturation of code coverage and mappings (\textit{see} \textbf{RQ4}).  
All in all, these results suggests that significantly improving both goals,
% code coverage
beyond  saturation point, is challenging and 
%achieving this goal, 
requires further research. 

\smallskip
\noindent %\skip
\textbf{Testing Strategies:}
%\todo{ES: discuss strategies, and weighting and fine-tuning to use cases}
\approach is useful for targeting different, multiple, or no testing goals.
This is due to its evolutionary approach and the flexibility of its fitness function.
It is important to stress that this evolutionary approach allows for developers to employ \approach for different testing goals or policies as desired:  
%More importantly,  i
It allows for a developer to generate test suites that 
% may be interested in achieving 
achieve a specific single testing goal,  multiple testing goals or ignore (an already \rev{achieved}) testing goal. 
%In this work, w
This work has 
%e have 
demonstrated these testing strategies -- single testing goals vs. multiple testing goals (\textbf{RQ1}) and ignoring specific testing goals \revise{(\textbf{RQ3})} -- are 
%can be 
achievable 
%carried out 
by \approach via weight setting in its fitness functions.  
%In addition, we have demonstrated how \approach 's configuration (e.g., number of generations) can help improve its effectivness for some goals (e.g., runtime).  

\def\alreadyfixed{Bug is already fixed in current version}
\begin{table*}[!tbp]\centering
    \caption{Bugs found by test inputs generated by \approach}
\begin{tabular}{| l | l | l |}
\hline
\textbf{Bug exception class} & \textbf{Subject Programs} & \textbf{Remarks}\\
\hline
\scriptsize{\texttt{java.lang.NullPointerException}} & Genson & \alreadyfixed\\
\scriptsize{\texttt{java.lang.ClassCastException}} & Gson & Bug report found in GitHub \cite{bugreport-gson}\\
\scriptsize{\texttt{java.lang.NullPointerException}} & json-flattener & \alreadyfixed\\
\scriptsize{\texttt{java.lang.UnsupportedOperationException}} & json-flattener & \alreadyfixed\\
\scriptsize{\texttt{java.lang.ArrayIndexOutOfBoundsException}} & json2flat & \alreadyfixed\\
\scriptsize{\texttt{java.lang.ArrayIndexOutOfBoundsException}} & JsonToJava & \alreadyfixed\\
%\scriptsize{\texttt{java.lang.NullPointerException}} & jstyleparser & Not reproducible\\
\scriptsize{\texttt{java.lang.StringIndexOutOfBoundsException}} & Pojo & Bug report found in GitHub\cite{bugreport-pojo} \\
% TODO ES: The Rhino IllegalStateException is actually a bug in driver_uninstrumented, not in Rhino!!! Should we report this?
%\todo
%\scriptsize{\texttt{java.lang.IllegalStateException}} & Rhino & Bug of our test driver \\
\hline
\end{tabular}
\label{tab:bugs-found}
%\vspace{-0.5cm}
\end{table*}

\smallskip
\noindent %\skip
\textbf{Bugs Found:}
\autoref{tab:bugs-found} highlights the bugs found during our experiments.  
%During the evaluation of 
\approach revealed several bugs in the subject programs. 
% most notably in \texttt{Gson} and \texttt{Pojo}.
%Most 
Some of the found bugs were either already reported (\texttt{Gson}, \texttt{Pojo}),  or fixed in more recent versions of the respective library (\texttt{Genson},\texttt{ json-flattener}, \texttt{json2flat}, \texttt{JsonToJava}). 
% and a few bugs were not reproducible outside of our test environment 
%driver executable 
%(\texttt{jstyleparser},  \texttt{Rhino}).
As an example,  we found a bug in \texttt{Genson} which is triggered by the following pathological input -- 
\lstinline|{ "\/": {} }|.  
When \texttt{Genson}  processes this JSON input, it 
%leads to  
throws a \texttt{NullPointerException} 
when processed by our test oracle using
\texttt{Genson} version 1.4.
% we used in our evaluation.  
On inspection, we observed that 
%As far as we are aware of, 
this bug has not yet been reported in the bug tracker of the project \cite{gensonbugtracker}.
However, it was not reproducible in recent versions of \texttt{Genson} (e.g., \texttt{Genson} version 1.6).
This suggests that the bug has been fixed, but \approach could have exposed it earlier. 
%We believe the non-reproducible bugs may be due to environmental issues in our settings,  and may only be reproducible outside our environment. 
%\par
%Another faulty file that caused a ClassCastException in Gson: \lstinline|[[[],{}],true]|.

\section{Threats to Validity}\label{threats}

%\todo{scalariazation vs.  many objective approach in evosuite}

%\todo{%
%Discuss the reliability of random seed and the effect of timeouts and determinism of the subject programs execution. 
%%However,  we not that this may not 
%%only 
%%holds if there is there is no time-out during generation.  
%%We also note that the same set of in
%% and the subjects run are full deterministic.  
%}

\noindent
\textbf{Internal Validity:}
\rev{The main threats to internal validity 
of our approach and experiments are the correctness of the implementation of our approach and the reproducibility and consistency of our results. 
To ensure reproducible results, we used a seed for all random factors that were involved in the implementation (especially the input generation and mutation).}
%To ensure old graphs did not change when new results were merged into our result repository, we implemented a Gitlab CI job to automatically build all graohs and documents.
In the implementation phase of our approach, we faced two likely sources of bugs - the grammar probability learning tool and the grammar mutator.
To ensure the correctness of our grammar learner,  
we wrote several unit tests. For instance,  we check learned probabilities against hand-crafted ones to make sure the learned probabilities are actually correct. 
%}
%\vspace{1mm}
%\noindent
%\textbf{Grammar Learning:}
%\revise{For the soundness of our evaluation, it is crucial to learn the new grammar from all input files that were selected in the previous generation.
%However, it is hard to see if that is the case without automatically checking the learned probabilities - especially for complex grammars like JavaScript.
%During the first experiments with CSS and JavaScript, there was a bug in our implementation that resulted in the grammar learning tool only learning from one of the input files instead of all input files.
%Hence, it is important to make sure that the grammar learning tool learns the grammar from all provided input files instead of just the first or last one, which might happen if the counters that are built into the counting grammar are reset for each file.
%This was due to the parser resetting its counters before the run of each input file.
%Since one input file might be complex enough for the grammar probabilities to look reasonable, such a bug might be hard to find in practice.}\par
%\revise{
%To prevent that from happening, we wrote several unit tests that test if all files are considered for learning, or just the first or last one, and some unit tests that check learned probabilities against hand-crafted ones for very simple files, to make sure the learned probabilities are actually correct.}
%Using those test cases, we fixed the bug before running the evaluation.
%
%\vspace{1mm}
%\noindent
%\textbf{Crawled Files:}
%\vspace{1mm}
%\noindent
%\textbf{BNF Grammar Mutator:}
To implement the grammar mutation tool, we implemented an ANTLR v4 grammar for parsing BNF grammars.
%We noticed that the JSON BNF grammar we use to generate JSON files has less syntactic grammar features than the CSS or JavaScript grammar.
%For instance, there are no regular expressions in the JSON grammar, nor parenthesized expressions or optional expansions.}\par
%\revise{
Our final ANTLR BNF grammars supports several syntactic features like regular expressions, parenthesized expressions and optional expansions. 
%of these syntactic constructs while being compatible with grammars that do not contain such features.
To ensure the validity of all generated grammars,  we tested each grammar by parsing the corresponding (CSS, JSON or JavaScript) BNF grammar to discover errors. 
%problems. 
%This is 
%which should be a sufficient test 
%to ensure validity 
%throughout 
%for all generated grammars, s
Since only probabilities change during mutations, the grammar itself remains the same after every mutation.

\rev{Our input grammar settings pose a threat to the validity of \approach and our findings.  For instance,   grammar ambiguity and recursion influence \approach's ability to exactly learn or produce the distribution learned in the initial seed inputs.  Similar to previous works~\cite{9154602, eberlein2020evolutionary}, we handle grammar ambiguity by adapting the ANTLR input grammars, e.g.,  by (re-)writing lexer modes,  shortening lexer tokens and re-writing parser rules.  In addition, our probabilistic generator~\cite{9154602} avoids generating excessively large or unbounded parse trees (e.g., from recursive rules) by setting a threshold that constrains the growth of the parse tree to ensure productions do not exceed a certain number of expansions. This parameter constrains generation to the shortest possible expansion tree once the threshold is reached (if at all).  We acknowledge that this setting may bias probabilistic generation. However,  we note that terminating production is important to avoid unbounded production and generate inputs within a reasonable time budget. }

%\todo{ R3: random selection of subject programs for RQ2-4}

%\todo{R3: generalization to complex programs, e.g., C programs}

\smallskip
\noindent
\textbf{External Validity:}
Even though \approach 
%strategies showed very promising results 
is effective for the subject programs and input formats 
employed in our evaluation, 
%that were shown in the evaluation, it is hard to say if that performance can be 
our results may not generalize to other input formats and programs.
%As we have seen in \autoref{nohold}, the 
%Indeed,  our experimental results (\textbf{RQ1}) show that results may 
%%greatly varied throughout the 
%for different file formats in our evaluation, e.g., the \emph{single goal} strategy had a much worse result for CSS than for the other formats.
%Nevertheless, we attempted t
To mitigate this threat, we have 
% by choosing 
employed 20 different programs and 
three well-known input formats, 
%a  variety of file formats to test,
ranging from a data exchange format (JSON) to programming language formats (CSS, JavaScript).
All three formats are relevant real-world formats that are popularly used in practice by developers and end-users:
JSON is the most popular data exchange format, e.g., JSON-LD is used by almost half (46.4\%) of all websites~\cite{json-ld-usage}.  Similarly,  JavaScript is among the most popular (web) programming languages~\cite{most-used-PL, gitHut-most-used-PL}.  \rev{Despite this setting,  we note that \approach and our findings may not directly apply to more complex structures such as C/C++, Java and Python programs. 
Additionally,  computationally intensive experiments (RQ2-4) were conducted with a randomly selected subset of subjects and a lower number (48) of generations (RQ2) due to time constraints.  We acknowledge that our findings in these settings may not generalize to a larger set of subjects and longer runs. }

%\todo{R1: discuss unit testing of evosuite (dynamosa) versus system level .testing of \approach. Discuss how it is mitigated,  by lifting unit testing to the system tests }

%\todo{R1: Comment on the recursive depth limit of tribble and input from hell.  Also comment on grammar ambiguity: ``Authors do not comment on the input grammars. How, for instance, the complexity of the input grammar would impact the approach.
%Recursively defined grammar rules pose a problem?''}

\smallskip
\noindent
\textbf{Construct Validity:}
To avoid experimeter bias,  we 
%We tried to keep 
employ the same experimental setup 
as closely related works:
%of our experiments similar to other work done in similar topics, 
For instance,  we employ similar 
% the same  
formats (e.g.,  JSON),  programs (e.g., Genson),
%as closely related work
%in other work - JSON, for example, is typically used as a simple format to evaluate the performance of fuzzing tools
%or by 
and metrics (e.g.,  code coverage,  exceptions)~\cite{eberlein2020evolutionary,9154602,8952419,DBLP:journals/corr/abs-1812-00140}. 
% - which is done in a lot of work~\cite{DBLP:journals/corr/abs-1812-00140} - and by using the same subject formats as in other work - JSON, for example, is typically used as a simple format to evaluate the performance of fuzzing tools~\cite{eberlein2020evolutionary,9154602,8952419}.
%By using the same subjects and formats, we tried to avoid having a bias towards input formats that are easier to generate or subject programs that contain faults that are easier to find.
%We also chose to evaluate our approach both on a very simple format (JSON) and on a more complex format (JavaScript) to avoid a bias towards the input format.
%\revise{
To avoid bias in seed inputs selection, 
%To ensure the quality of the seed files for the grammar learning step of the first generation, 
we selected our seed inputs randomly from a large set of crawled inputs.
%This way, we ensure that we do not have any bias towards a kind of input file that follows a specific semantic.
%For instance, during experiments with a JSON crawler, we noticed that there are many repositories that contain a large number of Node.js package files~\cite{npmdocumentation}.
%Although being in JSON format, those files all contain specific patterns like a \texttt{"name"} and a \texttt{"version"} property in the root dictionary that specifies the name and version of the node.js package.
%Having too many of such files with the same pattern would put a bias towards such files into our learned grammar.
%To avoid this bias, w
We crawl a huge set of inputs from GitHub before randomly selecting a smaller set of inputs which we use for grammar learning.  We further experimented with different sets and sizes 
%and types 
of seed inputs (\textit{see} \textbf{RQ4}).  

\rev{The use of weighted sum scalarization in our fitness function may limit the applicability of our approach, especially in settings where the developer does not have testing preferences or understand the relative importance of testing goal(s) or test feedbacks.  Indeed,  many-objective optimization methods (e.g.,  \dynamosa~\cite{panichella2017automated}) may be more appropriate in scenarios where Pareto-optimal solutions for conflicting testing objectives or test feedbacks are required.  In future work, we plan to investigate such scenarios where the mapping between testing goals and test feedbacks may be unknown or conflicting. However, we note that our goal-directed testing setting is better achieved via weighted sum scalarization since we require support for different testing strategies.  As discussed in \autoref{sec:approach},  our weight parameters  allow developers to tune \approach to target multiple testing strategies (single, multiple and ignore goal strategies). }
%This requirement is difficult to achieve with other many-objective optimization methods. }

%To further demonstrate this,  we have compared \approach to \dynamosa in \autoref{evaluation}. }

\section{Related Work}\label{relatedwork}

%\todo{R1: add [J1, J2, C1]. to related works: ``position their work with respect to these works ''}. 

%\vspace{1mm}
\noindent
\textbf{Grammar-Based Fuzzing:} 
Several researchers have proposed \textit{grammar-based test generators} that generate syntactically valid test inputs, such as  \textit{Tribble}~\cite{8952419}, \textit{Gramatron}~\cite{srivastava2021gramatron},  \textit{Superion}~\cite{wang2019superion}, 
\textit{BeDivFuzz}~\cite{nguyen2022bedivfuzz}, 
and  \textit{Grammarinator}~\cite{hodovan2018grammarinator}.  For instance,   \textit{Tribble} is a grammar-based fuzzer that generates inputs to cover certain input structures using an input grammar.   \textit{Tribble} uses a  k-path algorithm which systematically covers syntactic elements for grammar production. 
%k-path input coverage metric  to systematically explore the .   
Similarly,   \textit{Grammarinator}~\cite{hodovan2018grammarinator} is a general purpose grammar-based test generator that employs existing parser grammars for generating valid inputs.   \textit{Grammarinator} further supports mutation-based fuzzing.   
Researchers have also proposed \textit{probabilistic grammar-based fuzzers} which leverage learned input distribution from seed inputs to guide test generation,  namely \textit{Inputs from hell}~\cite{9154602} and  \textit{Skyfire}~\cite{7958599}.   Unlike the aforementioned works on grammar-based test generation, this paper proposes a \textit{directed grammar-based test generators} which focuses on generating  test inputs that are valid, but also maximise a targeted testing goal. 
%Furthermore,  i
In this paper, we employ the random grammar-based test generator of \textit{Tribble}~\cite{8952419} and a probabilistic fuzzer (\textit{Inputs from hell}~\cite{9154602}) as baselines for comparisons (\textit{see} \textbf{RQ2}).

\smallskip
\noindent
\textbf{Evolutionary Grammar-Based Fuzzing:}
Previous works have developed 
%evolutionary grammar-based fuzzing approaches 
techniques that employ evolutionary testing for grammar-based test generation. 
Notably,  
%EvoGFuzz~\cite{eberlein2020evolutionary}  and Nautilius~\cite{Aschermann2019NAUTILUSFF} 
%combine grammar-based test generation with an evolutionary approach. 
\revise{Nautilus~\cite{Aschermann2019NAUTILUSFF}} is one of the first fuzzers to combines grammar-based test generation with an evolutionary approach using feedback from the subject program.  Nautilius  focuses on inputs that trigger a crash or induce new branches.  
Unlike \approach, it does not employ a probabilistic grammar or input distribution for input generation.  However, \approach further target program runtime and input-to-code mappings. 
Likewise,  \emph{LangFuzz}~\cite{180229} proposed a technique for fuzzing \revise{JavaScript Engines~\cite{180229}.} It leverages test feedback to guide test generation by mutating inputs that trigger faults in previous test runs.  Unlike our approach,  LangFuzz's does not learn input distribution from seed inputs,  and employs mutations that are specific to ensuring JavaScript correctness.  \rev{
Researchers have also combined probabilistic grammars and evolutionary testing
to derive input profiles, which serve as a test adequacy criterion for structural testing~\cite{poulding2013optimisation}.  Specifically,  Poulding et al.~\cite{poulding2013optimisation} proposed a probabilistic grammar-based representation for deriving suitable input profiles for  structural testing.  Similar to our probabilistic grammar learning component in \approach, the proposed approach aims to derive probability distributions suitable for software testing, \textit{albeit} for test adequacy rather than test generation.  Unlike \approach, Poulding et al. ~\cite{poulding2013optimisation} aims to derive input profiles that serve as a test adequacy criterion for probabilistic structural testing. 
We note that \approach is not a test adequacy measure. 
Besides,  \approach supports other testing goals (beyond coverage).
}

\rev{
Some  researchers have deployed grammar-based testing to improve evolutionary testing: 
\textsc{G-EvoSuite} is a test generation approach that combines the strength of search-based testing and grammar-based testing to automatically assess programs that process JSON formats~\cite{olsthoorn2020generating}.  
%The search based testing and  grammar-based testing methods of \textsc{G-Evosuite} jointly co-evolve inputs to address the challenges of specifying the entry points for system testing and generating  structured inputs, respectively.   Specifically,  the search based testing component evolves the test case structure and grammar-based testing evolves the input data.  
% \textsc{G-EvoSuite} 
 It primarily employs grammars to inject valid JSON inputs into the initial population of its evolutionary algorithm.  Results show that it outperforms \evosuite, in terms of coverage. Like \textsc{G-EvoSuite},   \approach also employs input grammars, albeit to learn input distribution. Besides,  \approach supports multiple input formats and testing goals.}
\rev{Researchers have also previously combined evolutionary testing and probabilistic grammars for test generation~\cite{kifetew2014combining, kifetew2017generating}. Kifetew et al.~\cite{kifetew2014combining} combined genetic programming and probabilistic grammars to improve system-level coverage testing and demonstrated its effectiveness in improving system coverage  and fault revelation.  In follow-up work, the authors also proposed a grammar annotation scheme that allows developers to add semantic grammar constraints to improve test validity~\cite{kifetew2017generating}.  Results show that the proposed annotation scheme outperforms grammar learning in terms of input validity.  Unlike \approach, these works aim to improve input validity, e.g.,  by reducing 
%by learning or annotating grammar constraints to reduce 
the risks of generating infinite recursion and unrealistic input structures. }
\lrevise{Finally,  \evogfuzz~\cite{eberlein2020evolutionary} is the closest related work to \approach and the most recent evolutionary grammar-based fuzzing technique.  Like \approach, it also combines grammar-based test generation with an evolutionary approach. However,  it is focused on one test feedback (exceptions) and it does not leverage input mutation.  
%Similar to our work, the aforementioned  works employ evolutionary testing and probabilistic grammar learning for test generation. However, unlike 
In contrast to the aforementioned approaches,  \approach supports an arbitrary testing goal,  multiple testing goals and can be tuned to ignore certain testing goals.  In our evaluation,  we compare the performance of \approach versus EvoGFuzz (\textit{see} \textbf{RQ2}).  
}

\smallskip
\noindent
\textbf{Feedback-driven Fuzzing:}
Many fuzzers employ test feedback to drive test generation~\cite{180229,10.1145/3213846.3213874, Kim2019FindingSB, 10.1145/3293882.3339002}.  Notably,  \emph{JQF}~\cite{10.1145/3293882.3339002} is a coverage-guided fuzzing framework designed to fuzz Java programs.  JQF allows users to easily adapt their own \emph{guidance} implementation that guides the fuzzing process to a certain goal, e.g. maximizing code coverage.  \textit{Zest}~\cite{DBLP:journals/corr/abs-1812-00078} is a tool build on the 
the JQF framework~\cite{10.1145/3293882.3339002} which modifies existing test generators to use parameters in their random test generation procedure for feedback-based test generation.  \emph{PerfFuzz}~\cite{10.1145/3213846.3213874} is a  feedback-driven mutation fuzzing approach that aims to find pathological inputs -- test inputs that execute a certain critical section of a program more frequently than other inputs.  
\emph{FairFuzz}~\cite{10.1145/3238147.3238176} is a greybox feedback-driven input generation technique built on top of AFL. It focuses on covering rare branches in the subject program by using random input mutations to ensure diversity.  
Kim et al. ~\cite{Kim2019FindingSB} also proposed a framework for fuzzing file systems which allow for targeting specific goals -- signals from the operating system.
Their feedback loop involves a coverage measurement and a signal that can freely be defined by the programmer. However,  their approach is specifically designed to test file systems.  Our approach aims to be as generalized as possible such that it may be used for any subject program with an input format that has a context-free grammar.
%Simlar to our work, t
These techniques allow to target specific goals albeit via mutation fuzzing or coverage guidance.   
Unlike \approach, 
% (e.g., using AFL),  
%but 
they are not generational, evolutionary or grammar-based approaches.

\smallskip
\noindent 
\textbf{Directed Fuzzing: }
Directed grey-box fuzzing is a technique which generates inputs with the objective of reaching a given set of target program locations efficiently~\cite{huang2022beacon, 10.1145/1321631.1321653}.  Several directed fuzzing approaches have been proposed by researchers~\cite{JI2020102497, 10.1145/3387940.3391457, liang2023multiple, nguyen2020binary, lee2021constraint, 9160891}.  Notably, AFLGo~\cite{bohme2017directed} aimed to efficiently generate interesting inputs, i.e.,  inputs that target patches,  critical system calls or specific code locations.  To improve the efficiency of grey-box fuzzing,  Hawkeye~\cite{chen2018hawkeye} utilizes static analysis for seed prioritization and  power scheduling.  Meanwhile,  CAFL~\cite{lee2021constraint} is a constraint-guided directed greybox fuzzer that aims to satisfy a sequence of constraints.  
%\todo{R1: mention that we use scalarization vs. many-objective optimization in dynamosa}
\revise{
Finally,  \dynamosa~\cite{panichella2017automated}
is one of the closest directed fuzzing methods to our work. 
% \approach.  
%\dynamosa 
It is a multi-objective search method for structural test generation.  
%Similar to \approach, it 
However,  unlike \approach, \dynamosa 
is \textit{only} focused on coverage targets (e.g.,  branches)
and it 
%.  Besides,  \dynamosa 
does not leverage input grammars to drive test generation. 
In this work, we compare the effectiveness of \approach versus \dynamosa (using \evosuite) since both \approach and \dynamosa 
%approaches 
%\approach also 
employ multi-objective search (\textit{see} \textbf{RQ2}). 
}
In contrast to these works,  \approach focuses on targeting and maximizing high-level developer testing goals.  Rather than triggering specific code location(s) (e.g.,  branch(es) or function(s)),  \approach focuses on maximizing coverage,  the number of failures,  complex inputs and runtime. 

%\todo{discuss input/grammar based debugging - ddmax, ddset, fsynth,  alhazen, }
\smallskip
\noindent 
\textbf{Grammar-Based Debugging:} %
\lrevise{%
%On the one hand,  s
Some automated debugging techniques aim to simplify or diagnose failure-inducing inputs~\cite{misherghi2006hdd, dd}.  Notably,  delta debugging (\emph{DD})~\cite{dd} leverages binary search and test experiments to simplify failure-inducing inputs into the minimal subset reproducing a failure. % using a combination of .  
Hierarchical delta debugging (\emph{HDD})~\cite{misherghi2006hdd} improves on DD by employing input grammars.  It leverages input structure to simplify failure-inducing inputs to ensure resulting simplified inputs are syntactically valid. 
%
%, especially for structured input formats (e.g.,  XML).  
Other techniques,  employ input grammars to diagnose and explain faults in programs.  For instance,   \emph{Alhazen}~\cite{alhazen} and  \emph{DDSET}~\cite{ddset} leverages input grammars to explain the circumstances under which a failure occurs. 
% but \approach employs a combination of evolutionary testing and input grammars for test generation.   
}
\lrevise{
Another line of work aims to repair invalid or faulty program inputs~\cite{ddmax, fsynth}.   For instance, \emph{FSynth} employs error feedback from the subject program to conduct input repair without a grammar.  Meanwhile,  (syntactic)  \emph{DDMax} leverages an input grammar to repair invalid inputs.  
Unlike \approach, all of the aforementioned works target input debugging/repair. However,  \approach aims to generate syntactically valid test inputs that target arbitrary testing goals. 
% test generation by combining a combination of evolutionary testing and input grammars.  
In the future, we plan to investigate how \approach can improve program debugging and input debugging.  
}

\section{Conclusion}
\label{sec:conclusion}
This paper proposes a directed grammar-based test generation technique called \approach. The aim of \approach is to enable developers target one or more goal(s) during testing. 
%, e.g.,  generating complex inputs,  triggering new behaviors or exposing failures. 
In particular, we focus on four testing goals which are relevant during software testing activities, namely unique code coverage, input-to-code complexity,  program failures and long run time.  The key insight of \approach is to leverage test feedback to generate goal-specific test inputs. To this end,  \approach employs a combination of evolutionary testing and grammar learning.  \approach iteratively learns relevant input properties from existing inputs to drive the generation of goal-specific inputs. 
It first learns a probabilistic input grammar from sample seed inputs,  then leverages the learned input grammar as a producer to generate new inputs. 
Next, it iteratively evolves the generated inputs towards the testing goal by selecting goal-relevant inputs.  Finally, it employs input mutation and grammar mutation 
% mutates relevant inputs and the learned grammar 
to generate new inputs until the goal-at-hand is achieved.

In  our evaluation, we employ five state-of-the-art grammar-based test generation techniques,  20 open source Java programs and three (3) popular input formats, namely JSON, CSS and JavaScript. 
Our evaluation results demonstrate that \approach outperforms the closest state-of-the-art approach (EvoGfuzz) by up to 77\%. Furthermore, \approach is up to 89\% more effective than the baseline grammar-based test generators (i.e., random, probabilistic and inverse-probabilistic methods).  In our evaluation, \approach effectively targets error-prone behaviors, generates complex inputs,  and produces inputs that trigger long execution time. Overall,  our evaluation demonstrates that FDLOOP is effective for targeting a specific testing goal,  
allows to explore different combination of goals and 
%scales to multiple testing goals and 
it is tunable for ignoring specific goals.  \revise{We} demonstrate that our design decisions contribute to the performance of \approach via an ablation study. 
Finally,  we show that \approach is stable 
% stability of \approach under 
across different 
%its sensitivity to different 
parameter settings through our sensitivity analysis.    
%This work is  
% aims to  
To support replication and reuse, we provide our implementation and experimental data~\cite{artifact}:
\begin{center}
%\url{https://tinyurl.com/DirectedGTG}
\url{https://tinyurl.com/FDLoop-V3}
\end{center}

%\todo{new link for artifact -- FDLOOP}

%\todo{LK: do you want to rename the tool from \approach to "DirectedGTG" given the link above}

\balance

% or


%\ifCLASSOPTIONcompsoc
%  % The Computer Society usually uses the plural form
  \section*{Acknowledgments}
%  \todo{acknowledge \dynamosa and Fitew}
%\todo{LK: We should acknowledge/thank the EvoGFuzz first author who provided their (updated) implementation and checked the evaluation/comparison was correct. }
\revise{
We appreciate Martin Eberlein  
%Yannic Noller and Thomas Vogel and Lars Grunske},
%Our utmost thanks go to , 
and the \evogfuzz team for making their tool chain 
%and evaluation data 
available to us for replication and assessment.  
We also 
%appreciate
acknowledge the guidance of Fitsum Meshesha  Kifetew 
%(\dynamosa team) 
%for 
%guidance 
on the application of \dynamosa.
%comparing to 
%their tool. 
%responding to 
}

%Their support has been exemplary in every aspect.

%\else
%  % regular IEEE prefers the singular form
%\fi
%% Can use something like this to put references on a page
%% by themselves when using endfloat and the captionsoff option.
%\ifCLASSOPTIONcaptionsoff
%  \newpage
%\fi



%\bibliographystyle{IEEEtran}
%\printbibliography
\bibliographystyle{IEEEtran} %ACM-Reference-Format}
\bibliography{reference}

\end{document}